%% file: paper.tex
\documentclass[
	fontsize =11pt , % 10pt , 12 pt ( Schriftgroesse im Dokument )
	paper =a4, % DIN -A4- Format
	draft =false , % Probeversion , falls draft = true
	titlepage=false , % gesonderte Titelseite
	twoside =false , % einseitiges Layout ( sonst doppelseitig )
	open =right , % neue Kapitel beginnen rechts (left , any )
	parskip=half , % Absatzformat full <*+->, half <*+->, no
	DIV =12]{scrartcl}
\usepackage[utf8]{inputenc}
\usepackage[T1]{fontenc}
\usepackage[english]{babel}
\usepackage{amsmath}
\usepackage{amsfonts}
\usepackage{amssymb}
\usepackage{graphicx}
\usepackage{bm}
\usepackage{csquotes}
\usepackage{tikz}
\usepackage{subcaption}
\usepackage{booktabs}
\usepackage{array}
\usepackage{enumitem}
\usepackage{adjustbox}
\usepackage{multirow}
\usepackage{tabularx}
\usepackage{authblk}
\usepackage{pdflscape}
\usepackage{algorithm}
\usepackage{algpseudocode}
\usepackage[markup=nocolor]{changes}
\usepackage{anysize}
\marginsize{2cm}{2.5cm}{2.5cm}{2cm}

\usetikzlibrary{arrows.meta}

\usepackage{hyperref}

\usepackage[backend=biber, 
			style=authoryear-comp, 
			citestyle = authoryear-comp, 
			maxcitenames=2, 
			maxbibnames=20, 
			dashed = false, 
			giveninits=true, 
			url = false, 
			natbib=true, 
			doi = true, 
			eprint =true, 
			bibstyle =authoryear, 
			isbn=false, clearlang=true, 
            uniquename=false,
			uniquelist=false]{biblatex}

\renewbibmacro*{volume+number+eid}{%
	\printfield{volume}%
	\setunit*{\addnbthinspace}% NEW (optional); there's also \addnbthinspace
	\printfield{number}%
	\setunit{\addcomma\space}%
	\printfield{eid}}
\DeclareFieldFormat[article]{number}{\mkbibparens{#1}}

\AtEveryBibitem{\clearfield{month}}
\AtEveryCitekey{\clearfield{month}}
\AtEveryBibitem{\clearfield{day}}
\AtEveryCitekey{\clearfield{day}}
\AtEveryBibitem{\clearfield{note}}
\AtEveryCitekey{\clearfield{note}}

\DefineBibliographyStrings{german}{andothers={\mbox{et~al.}}}

\providecommand{\keywords}[1]
{
  {\small	
  \textbf{\textit{Keywords---}} #1%
  }
}

\NewBibliographyString{toappear}
\DefineBibliographyStrings{english}{%
  toappear = {to appear in:},
  version = {R package version},
}

\renewbibmacro*{in:}{%
  \iffieldundef{pubstate}
    {}
    {\printfield{pubstate}%
     \setunit{\addspace}%
     \clearfield{pubstate}}%
    }

\usepackage[labelformat=simple]{subcaption}

\newcolumntype{C}[1]{>{\centering\let\newline\\\arraybackslash\hspace{0pt}}m{#1}}

\usepackage{listings}
\usepackage{xcolor}

\lstdefinelanguage{myR}{
  morekeywords={if, else, for, while, function, return},
  sensitive=true,
  morecomment=[l]{\#},
  morestring=[b]"
}

\lstdefinestyle{Rclean}{
  language=myR,
  basicstyle=\ttfamily\small,
  keywordstyle=\color{blue},
  commentstyle=\color{green!50!black},
  stringstyle=\color{red!70!black},
  numbers=left,
  numberstyle=\tiny\color{gray},
  stepnumber=1,
  numbersep=5pt,
  showstringspaces=false,
  breaklines=true,
  frame=single,
  columns=fullflexible,
  captionpos=b
}

\newcommand{\R}{\texttt{R}}
\newcommand{\proglang}[1]{\texttt{#1}}
\newcommand{\fct}[1]{\texttt{#1()}}
\newcommand{\pkg}[1]{\texttt{#1}}
\newcommand{\code}[1]{\texttt{#1}}

\usepackage{fancyvrb}
\usepackage{xcolor}

\DefineVerbatimEnvironment{CodeInput}{Verbatim}{fontshape=sl}
\DefineVerbatimEnvironment{CodeOutput}{Verbatim}{}
\newenvironment{CodeChunk}{}{}

\begin{document}

    \title{glmSTARMA -- An R-Package for fitting autoregressive spatio-temporal models following generalized linear models}
    \author[1]{Steffen Maletz}
    \author[2]{Konstantinos Fokianos}
    \author[1]{Roland Fried}
    \affil[1]{Department of Statistics, TU Dortmund University, Germany}
    \affil[2]{Department of Mathematics and Statistics, University of Cyprus, Cyprus}

	\maketitle

\input{sections/abstract.tex}
	\input{sections/introduction.tex}

    \input{sections/framework.tex}

    \input{sections/simulation.tex}

    \input{sections/fitting.tex}

    \input{sections/selection.tex}

    \input{sections/examples.tex}

    \input{sections/summary.tex}

    \section*{Acknowledgements} Financial support by the Deutsche Forschungsgemeinschaft (DFG, German Research Foundation; Project-ID 520388526; TRR 391: Spatio-temporal Statistics for the Transition of Energy and Transport) is gratefully acknowledged.
    
    \section*{Data Availability Statement} The data and code required to reproduce the results presented in this article are publicly available. The code used for the computational experiments and the reproduction of all numerical results is provided in the supplementary repository: \url{https://github.com/stmaletz/glmSTARMA_supplement} The statistical methods are implemented in the \R package \pkg{glmSTARMA}, which is publicly available from CRAN and GitHub: \url{https://cran.r-project.org/package=glmSTARMA}, \url{https://github.com/stmaletz/glmSTARMA}.

    \section*{Disclosure statement}
    The authors report that there are no competing interests to declare.
 
	\newpage
	\appendix
    \numberwithin{equation}{section}
    \renewcommand{\theequation}{A-\arabic{equation}}
	\printbibliography[heading=bibintoc, title={Literature}]

    \newpage
    \include{sections/appendix.tex}

\end{document}

%% file: sections/abstract.tex
\begin{center}
    \sffamily\Large\bfseries Abstract
\end{center}

The R package \texttt{glmSTARMA} implements autoregressive models for spatio-temporal data at fixed locations, with time-invariant spatial dependency structure. We rely on generalized linear models methodology and unify several approaches for the analysis of spatial count time series. Such models allow the (conditional) mean of the response to depend on past observations, lagged  (conditional) expectations, and covariates. The response can be a continuous or a discrete random variable. Additionally, the package develops inference for  double generalized linear models, allowing the dispersion parameter(s) of the marginal distributions to be modeled similarly to the mean process. This is a  new capability which introduces, for example, spatio-temporal volatility models, such as space-time GARCH processes, and count time series models with spatio-temporal overdispersion and underdispersion. We provide functions for model estimation, simulation, inference, and prediction. Its use is illustrated by data examples.

\keywords{copula, covariates, dispersion, double generalized linear models, multivariate time series, space-time data}

\newpage

%% file: sections/introduction.tex
\section{Introduction}

Spatio-temporal data, which capture the evolution of variables over time at fixed locations, are essential in a wide range of scientific disciplines. Economists use these data to model real estate prices \citep{pace_method_2000, otto_spatiotemporal_2018} and analyze macroeconomic relationships \citep{elhorst_spatial_2022}. Environmental scientists employ spatio-temporal models to predict water quality in river networks \citep{clement_spatio-temporal_2006} and study urban heat dynamics \citep{choi_short-term_2021}. Epidemiologists collect and model weekly infection counts from different regions to investigate disease spread \citep{held_statistical_2005, dickson_assessing_2020}. Furthermore, spatio-temporal models are applied in modeling crime patterns \citep{clark_class_2021} and studying traffic flow \citep{kamarianakis_spacetime_2005}. 

Across these diverse applications, the analysis of spatio-temporal data often involves specific assumptions and characteristics tailored to the particular scientific field. 
For example, infection counts or hospitalization rates measured weekly for each district (see, e.g. \cite{maletz_spatio-temporal_2024, martins_space-time_2023}) are integer-valued and often exhibit overdispersion \citep{paul_multivariate_2008}. In meteorological contexts, constraints such as the non-negativity of precipitation, or an excess of zero values representing drought periods, must be taken into account \citep{yang_spatial-temporal_2005}. Even when the data can be approximated by a Gaussian distribution, non-linear dependencies may limit the applicability of linear models such as the Space Time ARMA (STARMA) models introduced by \citet{pfeifer_three-stage_1980}.

In this work, we assume that the data have been observed in equidistant time intervals. Such spatio-temporal data can be viewed as multivariate time series with a spatial structure. Their analysis is subject to the specific scientific field questions. We propose a general modeling framework for spatio-temporal data analysis. The main statistical tool is based on double generalized linear models (DGLM), introduced by \citet{smyth_generalized_1989}, but the approach we employ takes into account autoregressive components. This framework is implemented in the new  \proglang{R} package \pkg{glmSTARMA} and builds on the concept of spatial lagging as suggested by \citet{pfeifer_three-stage_1980}. This enables users to build spatio-temporal models for both the mean and the dispersion parameter of a given marginal distribution.
For example, spatio-temporal GARCH processes \citep{otto_spatial_2025} can be combined with a mean model. For spatio-temporal count data many contributions -- such as those by \citet{jahn_approximately_2023} and \citet{maletz_spatio-temporal_2024} -- are based on the Poisson distribution. Although these simple models account for overdispersion in the marginal distribution to some degree, it is desirable to develop a more flexible class. We will show that the DGLM based approach taken in this work allows the models to be extended, such that they can capture contemporary over- and underdispersion.

Other available \proglang{R} packages for modeling spatio-temporal data  focus on specific model classes. For example, the \pkg{starma} package \citep{cheysson_starma_2016} implements classical linear STARMA models but does not support covariates, which can be included in STARMAX (STARMA with external covariates) processes \citep{stoffer_estimation_1986}. Related to this are the Generalized Network Autoregressive (\pkg{GNAR}) processes, which are implemented in the \proglang{R} package with the same name \citep{knight_generalized_2020}. The \pkg{PNAR} package \citep{armillotta_inference_2024} applies this approach to count data. More precisely, it estimates network time series models with marginal Poisson distributions, closely related to Poisson-STARMA models \citep{maletz_spatio-temporal_2024} and other multivariate count time series models \citep{fokianos_multivariate_2020}. With the \pkg{spGARCH} package \citep{otto_spgarch_2019} it is possible to estimate spatio-temporal GARCH models. Simple univariate autoregressive models can be fitted using the base \proglang{R} function \code{glm} by including lagged values as covariates. The \code{gam} function in the \pkg{mgcv} package \citep{wood_generalized_2017} additionally allows spatial smoothing via Markov random field theory. However, both functions do not enforce parameter constraints that are often necessary for valid inference.

None of the above packages allows simultaneous modeling of spatio-temporal data at the observation or expectation level and at the variance/dispersion level. Even for univariate time series, we are not aware of any software that implements existing time series models for non-normal data with varying dispersion that were proposed by \citet{barretosouza_timevarying_2025, zheng_generalized_2022}.

\begin{table}[t]
    \centering
    \resizebox{\textwidth}{!}{%
    \begin{tabular}{lll}
    \hline
         & Name & Description \\ \hline
       Main Functions  & \code{glmstarma} & Fit model for the mean to given data\\
         & \code{dglmstarma} & Fit mean and dispersion model to given data\\
         & \code{glmstarma.sim} & Simulate observations applying the mean model\\
         & \code{dglmstarma.sim} & Simulate observations applying a mean and dispersion model \\ \midrule
        Help Functions & \texttt{TimeConstant} & Create a time-constant covariate, i.e. $\forall t :\bm{X}_t = \bm{X} \in \mathbb{R}^p$ \\
         & \code{SpatialConstant} & Create a spatial constant covariate, i.e. $\bm{X}_t = x_t\cdot \bm{1}_p,\, x_t \in \mathbb{R}$\\
         & \code{generateW} & Create spatial weight matrices on regular grids. \\ \midrule
       Generic Functions  & \texttt{summary} & Summary of fitted model \\
         & \code{fitted} & Fitted values\\
         & \code{residuals} & Different types of residuals\\
         & \code{AIC} & Akaike Information Criterion\\
         & \code{BIC} & Bayesian Information Criterion\\
         & \code{QIC} & Quasi Information Criterion \\
         & \code{predict} & Predict future values \\
         & \code{vcov} & Variance Covariance Matrix of estimated coefficients \\ \hline
         Data & \code{load\_data} & Download and return preprocessed datasets \\
         & \code{delete\_data} & Remove datasets from the cache \\ \hline
        %& \texttt{delete\_glmSTARMA\_data} & Remove datasets from the cache \\ \hline
    \end{tabular}
    }
    \caption{Key functions of the \texttt{R} package \texttt{glmSTARMA} along with the included datasets.}
    \label{tab:functions}
\end{table}

The framework in \pkg{glmSTARMA} generalizes and unifies these approaches, allowing flexible modeling through covariates, lagged observations, and a latent feedback process for both the mean and the dispersion processes, as we will explain in what follows. The use of spatial lagging enables the analysis of high-dimensional data by using only a small number of free parameters. The package supports a wide range of response distributions. Table~\ref{tab:functions} provides an overview of key functions developed for this R package. The functions \fct{glmstarma} and \fct{dglmstarma} are used to estimate the models described next. Here, \fct{glmstarma} estimates only a spatio-temporal model for the mean, described in Subsection~\ref{sec:mean_model}. The function \fct{dglmstarma} estimates both the mean model and the dispersion model, which is described in Subsection~\ref{sec:dispersion_model}. With the functions \fct{glmstarma.sim} and \fct{dglmstarma.sim}, data can be simulated from the corresponding spatio-temporal models. In Section~\ref{sec:simulation}, we describe the use of these functions and the underlying data-generating process, which also allows for contemporary dependencies between the dimensions, i.e., the locations of the time series. The estimation procedure is explained in Section~\ref{sec:estimation}. Furthermore, the package provides functionality for statistical inference, model diagnostics, and forecasting. We discuss all these issues in Section~\ref{sec:selection}. Section~\ref{sec:examples} contains two data examples. The resulting preprocessed data for these can be loaded with the \fct{load\_data} function of the R package.

%% file: sections/framework.tex
\section{Modeling Framework}

Let $\lbrace \mathbf{Y}_t = (Y_{1, t}, \ldots, Y_{p, t})' \mid t = 0, 1, \ldots \rbrace$ denote a $p$-dimensional time series representing the evolution of a target variable, say $Y$, such as weekly infection counts, observed at $p$ fixed spatial locations over time. Each component process $\lbrace Y_{i,t}, t = 0, 1, \ldots \rbrace, i = 1, \ldots, p$, corresponds to a univariate time series that represents  the  response variable at location $i$ at time $t$.
Assume that  we observe simultaneously two -- not necessarily distinct -- sets of covariate processes, say $\lbrace \bm{X}_{k,t} = (X_{1,k,t}, \ldots, X_{p,k,t})' \rbrace$ for $k = 1, \ldots, m$, and $\lbrace \bm{\tilde{X}}_{k,t} = (\tilde{X}_{1,k,t}, \ldots, \tilde{X}_{p,k,t})' \rbrace$ for $k = 1, \ldots, \tilde{m}$. Introducing  two  covariate series renders the modeling approach we take in this work possible.

The proposed modeling framework is based on the concept of DGLM  introduced by \citet{smyth_generalized_1989}. Specifically, two distinct model components are specified: one for the conditional mean of the response variable $Y$ and another for the dispersion parameter of the assumed distribution. The covariates $\bm{X}_{k,t}$ contribute to the modeling of the mean, while the $\bm{\tilde{X}}_{k,t}$ are used in the dispersion model to explain possible heteroskedasticity or overdispersion in the data. 

Similarly to classical generalized linear models (GLM), DGLMs consist of several key components. For formalization, let $\mathcal{F}_t$ denote the history of the processes up to time $t$. Details will be given later on. In this work, we adapt the DGLM framework as follows:
%For formalization, let $\mathcal{F}_t$ denote the $\sigma$-algebra generated by all observations up to time $t$. 
\begin{enumerate}
    \item \textbf{Distribution}:  
    The univariate conditional distributions $Y_{i,t} \mid \mathcal{F}_{t - 1}$ are assumed to belong to the exponential dispersion family. Their corresponding density can be written as
    \begin{align}
    \label{eq:exponential_family}
    f(y_{i,t} \mid \theta_{i,t}, \phi_{i,t}, \mathcal{F}_{t - 1}) = \exp\left( \frac{\theta_{i,t} y_{i,t} - A(\theta_{i,t})}{\phi_{i,t}} + c(\phi_{i,t}, y_{i,t}) \right),
    \end{align}
    where $\theta_{i,t}$ (respectively,  $\phi_{i,t}$)  denotes the natural  (respectively, dispersion) parameter  of the distribution. The conditional expectation and variance of the observation $Y_{i,t}$ are given by
    \[
    \mu_{i,t} := \mathbb{E}(Y_{i,t} \mid \mathcal{F}_{t - 1}) = \frac{\partial A(\theta_{i,t})}{\partial \theta_{i,t}}, \quad
    \sigma_{i,t}^2 := \text{Var}(Y_{i,t} \mid \mathcal{F}_{t - 1}) = \phi_{i,t} \cdot V(\mu_{i,t}),
    \]
    where $V$ denotes a variance function that depends on the conditional mean $\mu_{i,t}$.

    \item \textbf{Linear Predictors}:  
     \sloppy Define two linear processes $\lbrace \bm{\psi}_t = (\psi_{1,t}, \ldots, \psi_{p,t})' \rbrace$ and $\lbrace \bm{\zeta}_t = (\zeta_{1,t}, \ldots, \zeta_{p,t})' \rbrace$ that will be used  as predictors of the mean and the dispersion process, respectively. Precise definitions are given by equations~\eqref{eq:model} and~\eqref{eq:dispersionmodel} below.

    \item \textbf{Link Functions}:  
    The relationship between the conditional mean, the dispersion parameter, and the linear predictors is established through the link functions $g$ and $\tilde{g}$  that satisfy 
    \[
    g(\mu_{i,t}) = \psi_{i,t}, \quad \tilde{g}(\phi_{i,t}) = \zeta_{i,t}.
    \]
\end{enumerate}

\sloppy The collection of conditional means over space and time is denoted as the mean process  $\lbrace \bm{\mu}_t := (\mu_{1,t}, \ldots, \mu_{p,t})' \rbrace$, and the corresponding collection of dispersion parameters as the dispersion process $\lbrace \bm{\phi}_t := (\phi_{1,t}, \ldots, \phi_{p,t})' \rbrace$.

Early references that deal with modeling mean and dispersion parameters simultaneously, in the context of independent data
are \citet{nelder_quasi-likelihood_1985},  \citet{efron_double_1986}  and  \citet{nelder_extended_1987}, while a Bayesian state space model was employed by \citet{west_dynamic_1985} for modeling time varying generalized linear models. The work by  \citet{jorgensen_theory_1997} offers further insight on the more general concept of exponential dispersion models; see  \citet{smyth_fitting_2002} for their connection with DGLM illustrated by an application to insurance claims data and, more recently, \citet{bonat_extended_2018} who propose a new  class of discrete generalized linear models based on the class of Poisson-Tweedie factorial dispersion models.

The previous assumptions do not specify  the (conditional) multivariate distribution of $\bm{Y}_t \mid \mathcal{F}_{t - 1}$. 
This will be done in detail in Section \ref{sec:simulation}. In practice, various data generating processes  are applicable  to introduce contemporary dependencies between the components $Y_{i,t} \mid \mathcal{F}_{t - 1}$ of the vector 
$\bm{Y}_t \mid \mathcal{F}_{t - 1}$. We use copula-based data-generating processes for the models presented in this paper, with marginal distributions satisfying the assumptions described above; see \citet{fokianos_multivariate_2020} for the case of multivariate count time series.

Table~\ref{tab:distributions} provides an overview of the distributions and link functions currently implemented in the\proglang{R}package, with the corresponding variance functions. In analogy to the \fct{glm} function in R, all functions for simulation and parameter estimation include a \code{family} argument that specifies the distribution and the link function. Each distribution is implemented through a generation function named after the distribution, prefixed by \texttt{v} for ``vector``. The corresponding link function is defined via the \code{link} argument. In addition to specifying the link function, this argument also implicitly defines two transformation functions, $h$ and $\tilde{h}$, which are used in Equations~\eqref{eq:model} and~\eqref{eq:dispersionmodel} to transform past values of the feedback / dispersion process and observations. As we will see later on, the gamma distribution is used to model the dispersion process. Therefore, the link functions for \fct{vgamma} are available for the dispersion model~\eqref{eq:dispersionmodel}.

\begin{landscape}
\begin{table}[p]
\centering
\resizebox{\linewidth}{!}{
\begin{tabular}{llclcccc}
\toprule
\textbf{Data Type} & \textbf{Distribution} & $V(\mu_{i, t})$ &\textbf{Link} & $g(\bm{\mu}_t)$ &$h(\mathbf{\psi}_t)$ & $\tilde{h}(\mathbf{Y}_t)$ & \textbf{NNP} \\
\midrule

\multirow{3}{*}{Gaussian} & \multirow{3}{*}{\shortstack[l]{Normal\\ (\fct{vnormal})}} & \multirow{3}{*}{$1$} 
    & \code{"identity"} & $\bm{\mu}_t$ &$\mathbf{\psi}_t$ & $\mathbf{Y}_t$ &  \\
  & & & \code{"log"}      & $\log(\bm{\mu}_t)$  & $\mathbf{\psi}_t$ & $\log(|\mathbf{Y}_t|)$ &   \\
  & & & \code{"inverse"}  & $1 /\bm{\mu}_t$ & $\mathbf{\psi}_t$ & $1 / \mathbf{Y}_t$ &  \\

\midrule

\multirow{7}{*}{\shortstack[l]{Positive\\ continuous}} & \multirow{3}{*}{\shortstack[l]{Gamma\\ (\fct{vgamma})}} & \multirow{3}{*}{$\mu_{i, t}^2$}  
    & \code{"identity"} & $\bm{\mu}_t$ & $\mathbf{\psi}_t$ & $\mathbf{Y}_t$ & \checkmark \\
  & & & \code{"log"}      & $\log(\bm{\mu}_t)$ & $\mathbf{\psi}_t$ & $\log(\mathbf{Y}_t + c \cdot \bm{1}_p)$ &  \\
  & & & \code{"inverse"}  & $1 / \bm{\mu}_t$ & $\mathbf{\psi}_t$ & $1 / \mathbf{Y}_t$ & \checkmark \\

%\midrule
\cmidrule{2-8}

& \multirow{4}{*}{\shortstack[l]{Inverse-Gauss\\ (\fct{vinverse.gaussian})}} & \multirow{4}{*}{$\mu_{i, t}^3$} 
    & \code{"1/mu\textasciicircum2"}   & $1 / \bm{\mu}_t^2$ & $\mathbf{\psi}_t$ & $1 / \mathbf{Y}_t^2$ & \checkmark  \\
 &  & & \code{"inverse"}  & $1 / \bm{\mu}_t$ & $\mathbf{\psi}_t$ & $1 / \mathbf{Y}_t$ & \checkmark  \\
&  &  & \code{"identity"} & $\bm{\mu}_t$ & $\mathbf{\psi}_t$ & $\mathbf{Y}_t$ & \checkmark  \\
 & &  & \code{"log"}      & $\log(\bm{\mu}_t)$ & $\mathbf{\psi}_t$ & $\log(\mathbf{Y}_t)$ &  \\

\midrule
\multirow{14}{*}{Count Data} & \multirow{4}{*}{\shortstack[l]{Poisson\\ (\fct{vpoisson}, \\ \fct{vquasipoisson})}} & \multirow{4}{*}{$\mu_{i, t}$} 
    & \code{"log"}      & $\log(\bm{\mu}_t)$ & $\mathbf{\psi}_t$ & $\log(\mathbf{Y}_t + c \cdot \mathbf{1}_p)$ &   \\
  & & & \code{"identity"} & $\bm{\mu}_t$ & $\mathbf{\psi}_t$ & $\mathbf{Y}_t$ & \checkmark   \\
 &  & & \code{"sqrt"}     & $\sqrt{\bm{\mu}_t}$ & $\mathbf{\psi}_t$ & $\sqrt{\mathbf{Y}_t}$ & \checkmark  \\
 & &  & \code{"softplus"}  & $c\log(\exp(\bm{\mu}_t / c) - \bm{1}_p)$ & $c \log(\mathbf{1}_p + \exp(\mathbf{\psi}_t / c))$ & $\mathbf{Y}_t$ &  \\

%\midrule
\cmidrule{2-8}

&\multirow{4}{*}{\shortstack[l]{Negative-Binomial\\ (\fct{vnegative.binomial})} } & \multirow{4}{*}{$\mu_{i, t} + \phi_{i, t} \cdot\mu_{i, t}^2$} 
    & \code{"log"}      & $\log(\bm{\mu}_t)$ & $\mathbf{\psi}_t$ & $\log(\mathbf{Y}_t + c \cdot \mathbf{1}_p)$ & \\
 &  & & \code{"identity"} & $\bm{\mu}_t$ & $\mathbf{\psi}_t$ & $\mathbf{Y}_t$ & \checkmark  \\
 &  & & \code{"sqrt"}     & $\sqrt{\bm{\mu}_t}$ & $\mathbf{\psi}_t$ & $\sqrt{\mathbf{Y}_t}$ & \checkmark  \\
 &  & & \code{"softplus"}  & $c\log(\exp(\bm{\mu}_t / c) - \bm{1}_p)$ & $c \log(\mathbf{1}_p + \exp(\mathbf{\psi}_t / c))$ & $\mathbf{Y}_t$ &  \\
%\midrule
\cmidrule{2-8}
& \multirow{4}{*}{\shortstack[l]{Binomial\\ (\fct{vbinomial}, \\ \fct{vquasibinomial})}} & \multirow{4}{*}{$\mu_{i, t}(1 - \mu_{i, t} / n_i)$} 
    & \code{"identity"} & $\bm{\mu}_t / \bm{n}$ & $\bm{\psi}_t$ & $\bm{Y}_t / n$ & \checkmark  \\
 & &  & \code{"softclipping"} & $c \log \left[\frac{\exp(\bm{\mu}_t / c\bm{n}) - \bm{1}_p}{\bm{1}_p - \exp((\bm{\mu}_t - \bm{n}) / c\bm{n} )} \right]$ & 
        $c  \log \left( \frac{1 + \exp(\bm{\psi}_t / c)}{1 + \exp((\bm{\psi}_t - 1)/ c)} \right)$ 
        & $\bm{Y}_t / n$ &  \\
 &  & & \code{"logit"}    & $\log\left(\frac{\bm{\mu}_t}{\bm{n} - \bm{\mu}_t}\right)$ & $(1 + \exp(-\bm{\psi}_t))^{-1}$ & $\bm{Y}_t / n$ &   \\
 &  & & \code{"probit"}   & $\Phi^{-1}(\bm{\mu}_t / \bm{n})$ & $\Phi(\bm{\psi}_t)$ & $\bm{Y}_t / n$ &   \\
\bottomrule
\end{tabular}}
\caption{Summary of distributions and link functions implemented in \pkg{glmSTARMA} along with corresponding transformations of past values of the feedback process and the observations. A checkmark at NNP indicates that the model formula in \eqref{eq:model} or \eqref{eq:dispersionmodel} needs all parameters to be non-negative. The notation  $c > 0$ ($\geq 0$ for Gamma marginals) describes a hyperparameter to be set, with default  value equal to 1, which can be altered via the family argument \code{const}. The vector $\bm{n}$ having  elements from $\mathbb{N}$ in the binomial distribution can be set via the argument \code{size}. If a vector of length 1 is defined, the value is set for all locations.}
\label{tab:distributions}
\end{table}
\end{landscape}

\subsection{Conditional Mean Model}\label{sec:mean_model}

We model the linear process $\lbrace \bm{\psi}_t \rbrace$ in Assumption 2 of the DGLM framework using 
\begin{align}
\label{eq:model_simple}
\bm{\psi}_t = \bm{\delta} 
+ \sum_{i = 1}^q\sum_{\ell = 0}^{a_i} \alpha_{i\ell} \mathbf{W}^{(\ell)}_{\alpha} h\left(\bm{\psi}_{t - i}\right) 
+ \sum_{j = 1}^r \sum_{\ell = 0}^{b_j} \beta_{j\ell} \mathbf{W}^{(\ell)}_{\beta} \tilde{h}\left(\mathbf{Y}_{t - j}\right) 
+ \sum_{k = 1}^m \sum_{\ell = 0}^{c_k} \gamma_{k\ell} \mathbf{W}^{(\ell)}_{\gamma} \mathbf{X}_{k, t}.
\end{align}
This model consists of  an intercept term, past values  of the linear process, functions of  past observations, and effects of the covariate processes. 
Details on the model orders ($q, a_{i}, r, b_{j}, m, c_{k}$) and the weight matrices $\bm{W}_\theta^{(\ell)}$, $\theta\in \{\alpha,\beta,\gamma\}$ will be given later. 
Let $\bm{\vartheta}_{\mu}$ denote the parameter vector  which contains all unknown parameters of \eqref{eq:model_simple}, i.e., the intercept $\bm{\delta}$ and the coefficients $\alpha_{i\ell}$, $\beta_{j\ell}$, and $\gamma_{k\ell}$. Depending on the choice of the link function, it may be necessary to constrain the parameters to be non-negative to ensure that the conditional mean process $\bm{\mu}_t$ conforms  to the distribution implied by~\eqref{eq:exponential_family}.
For instance, when the response is a count variable, then $\bm{\mu}_t >0$; see Table~\ref{tab:distributions}, for all  relevant models and distributions where such constraints are required. 

% Transformations h und \tilde{h}
\paragraph{Transformation of linear process and past observations}
For simplicity, equation~\eqref{eq:model_simple} contains transformations $h$ and $\tilde{h}$, which are applied to the past values of the linear process and the observations. These functions are fixed with the link function, which connects $\bm{\psi}_t$ and $\bm{\mu}_t$. An overview can be found in Table~\ref{tab:distributions}. These transformations allow many models to be used, whereby in most cases we set $h$ equal to the identity function. The function for transforming the observations is chosen such that $\tilde{h}(\bm{Y}_{t})$ has the same scale as $\bm{\psi}_t$, i.e., $\mathbb{E}(\tilde{h}(\bm{Y}_{t}) \mid \mathcal{F}_{t - 1}) \approx \bm{\psi}_t.$

For an example of model \eqref{eq:model_simple} applied to spatio-temporal count data consider the log-linear PSTARMA 
processes; see \citet{maletz_spatio-temporal_2024}. The authors  assume marginal Poisson distributions for $Y_{i, t} \mid \mathcal{F}_{t - 1}$ implemented in the family function \code{vpoisson("log")}. Without covariates, model \eqref{eq:model_simple} becomes
\begin{align}
\label{eq:log_linear}
    \bm{\psi}_t = \bm{\delta} 
+ \sum_{i = 1}^q\sum_{\ell = 0}^{a_i} \alpha_{i\ell} \mathbf{W}^{(\ell)} \bm{\psi}_{t - i} 
+ \sum_{j = 1}^r \sum_{\ell = 0}^{b_j} \beta_{j\ell} \mathbf{W}^{(\ell)} \log\left(\mathbf{Y}_{t - j} + \bm{1}_p\right). 
\end{align}
For the models studied by \citet{jahn_approximately_2023}, the function $h$ corresponds to the inverse link function, such that $h(\bm{\psi}) = \bm{\mu}$, with some abuse of notation. For example, the softplus model can be applied by using the function  \code{vpoisson("softplus")}. If there are no covariates, the softplus model takes the form of
\begin{align}
\label{eq:softplus}
    \bm{\psi}_t = \bm{\delta} 
+ \sum_{i = 1}^q\sum_{\ell = 0}^{a_i} \alpha_{i\ell} \mathbf{W}^{(\ell)} \bm{\mu}_{t - i} 
+ \sum_{j = 1}^r \sum_{\ell = 0}^{b_j} \beta_{j\ell} \mathbf{W}^{(\ell)} \mathbf{Y}_{t - j}.
\end{align}
This is similar to the linear PSTARMA-process of \citet{maletz_spatio-temporal_2024},  which is implemented  via \code{vpoisson("identity")} but with a different link function that implies $\bm{\psi} = \bm{\mu}$. As opposed to the linear model, the  softplus model can account for negative correlation with a cost of an additional hyperparameter $c$ that sets the degree of linearity of the model. By default $c = 1$.

% Spatial Weights
\paragraph{Spatial Weights}
We use so-called spatial weight matrices $\bm{W}^{(\ell)} = (w_{ij}^{(\ell)}) \in \mathbb{R}^{p \times p}$, which describe neighborhoods of different spatial orders $\ell \in \mathbb{N}_0$. For the weights let $w_{ij}^{(\ell)} \in [0, 1]$ and we assume row-normalized matrices, i.e., $\bm{W}^{(\ell)} \bm{1}_p = \bm{1}_p$, for the theoretical properties of the model and for parameter estimation. It is possible to specify different matrices for past values of the feedback term ($\bm{W}^{(\ell)}_{\alpha}$), the observation process ($\bm{W}^{(\ell)}_{\beta}$), and the covariates  ($\bm{W}^{(\ell)}_{\gamma}$).

A neighborhood $\mathcal{N}_i^{(\ell)} \subset \lbrace 1, \ldots, p \rbrace$ of order $\ell$ around location $i$ contains those locations that are considered to be neighbored at spatial order $\ell$ from $i$. Lower-order neighborhoods are intended to include closer locations. The zero-order neighborhood is defined to contain only the location itself: $\mathcal{N}_i^{(0)} = \lbrace i \rbrace$. Neighborhoods of different orders are assumed to be disjoint: $\mathcal{N}_i^{(\ell)} \cap \mathcal{N}_i^{(\tilde{\ell})} = \emptyset$ for $\ell \neq \tilde{\ell}$. By construction of the spatial weights, we assume $w_{ij}^{(\ell)} > 0$ if and only if $j \in \mathcal{N}_i^{(\ell)}$. This implies $\bm{W}^{(0)} = \bm{I}_p$, where $\bm{I}_p$ denotes the $p \times p$ identity matrix.

The specific values of the spatial weights $w_{ij}^{(\ell)}$ can be chosen by the user. A common choice is based on adjacency matrices: two regions are considered adjacent of order one if they share a common border. Higher-order neighbors ($\ell > 1$) can be defined recursively as those reachable by traversing $\ell - 1$ intermediate regions. A simple uniform weighting scheme can be defined as
\begin{align}
    w_{ij}^{(\ell)} = \frac{1}{\# \mathcal{N}_i^{(\ell)}},
\end{align}
where $\# \mathcal{N}_i^{(\ell)}$ denotes the cardinality of the neighborhood set. 
Depending on the application, neighborhoods and matrices can also be defined based on distances, i.e., using $k$-nearest neighbor schemes or all locations within a  distance defined by the user, see \citet[Chap.~7]{moraga_spatial_2023} for further examples. In \proglang{R}, the package \pkg{spdep} \citep{bivand_spdep_2025} contains functions supporting the creation of the weight matrices.

In the \pkg{glmSTARMA} package, the weight matrices are passed as a list to the associated arguments (\code{wlist}, \code{wlist\_past\_mean}, \code{wlist\_covariates}) in the relevant functions, where the $\ell + 1$-th list element corresponds to the spatial weight matrix of order $\ell$. The list elements can be matrices, i.e., objects of class \code{matrix} in base \proglang{R}, or sparse matrices, i.e., objects of class \code{dgCMatrix} from the \pkg{Matrix} \proglang{R} package \citep{bates_matrix_2025}. If no matrices are passed for past values of the feedback process (\code{wlist\_past\_mean}) and the covariates (\code{wlist\_covariates}), i.e., the arguments are equal to \code{NULL}, the same matrices as in \code{wlist} are used.

% The model argument
\paragraph{Model orders}
The model orders in \eqref{eq:model_simple} are passed to the functions of the \pkg{glmSTARMA} package in an argument \code{model} as a named list. 

We distinguish in~\eqref{eq:model_simple} between two types of intercepts: a \emph{homogeneous} intercept, defined  by $\bm{\delta} = \delta_0 \bm{1}_p$, and an \emph{inhomogeneous} intercept, given by an arbitrary vector $\bm{\delta} = (\delta_1, \ldots, \delta_p)'$. The software defaults to a homogeneous intercept. An inhomogeneous intercept is specified by setting \code{intercept = "inhomogeneous"} in the \code{model} list argument. An inhomogeneous intercept can be included cautiously if the model includes time-invariant covariate processes, i.e., when $\bm{X}_{k,t} = \bm{X}_{k,\tilde{t}}$ for all $t, \tilde{t}$. In such cases, identifiability issues may arise, as the intercept and covariate effects may not be uniquely separable.

The maximum spatial orders $a_i$ for the regression on past values of the feedback process can be passed as an integer vector in the list element \code{past\_mean}, where the $i$-th entry belongs to time lag $i$. Similarly, the spatial orders $b_j$ for the past observations are specified as list element \code{past\_obs}. The maximum time lags $q$ and $r$ are determined automatically based on the length of the vector. By default, the maximum spatial order 0 is used for all covariates. Only local covariate effects are then modeled, i.e., the effects do not directly affect neighboring regions. Different maximum spatial orders can be passed in the list element \code{covariates} in the same way as before. The $k$-th entry of the vector then specifies the maximum spatial order for the $k$-th covariate.

In some applications such as modeling periodicities, one does not want to regress on all lagged values up to the maximum lags $q$ and $r$, but only on a small subset. Formally, we can write this using sets $\mathcal{Q}^{(\mu)} \subset \mathbb{N}$ and $\mathcal{R}^{(\mu)} \subset \mathbb{N}$, which contain the time lags to be regressed. These can be passed as integer vectors in list elements \code{past\_mean\_time\_lags} and \code{past\_obs\_time\_lags} of the \code{model} argument. The default setting as in~\eqref{eq:model_simple} corresponds to $\mathcal{Q}^{(\mu)} = \lbrace 1, \ldots, q \rbrace$ and $\mathcal{R}^{(\mu)} = \lbrace 1, \ldots, r \rbrace$.

Moreover, spatial orders below the maximum spatial order can be omitted. Formally, for each time lag $i \in \mathcal{Q}^{(\mu)}$, we define a spatial index set $\mathcal{A}_i^{(\mu)} \subset \mathbb{N}_0$; analogously, for each $j \in \mathcal{R}^{(\mu)}$, we define $\mathcal{B}_j^{(\mu)}$, and for each covariate process $\lbrace \bm{X}_{k, t} \rbrace$, a set $\mathcal{C}_k^{(\mu)}$, $k = 1, \ldots, m$. These sets contain the spatial lags, which are incorporated in the model at a specific time lag or covariate. In this case, instead of integer vectors, binary matrices $\bm{A} = (a_{\ell, i}) \in \lbrace 0, 1 \rbrace^{a \times q}$, $\bm{B} = (b_{\ell, i}) \in \lbrace 0, 1 \rbrace^{b \times r}$ and $\bm{C} = (c_{\ell, i}) \in \lbrace 0, 1 \rbrace^{c \times k}$ must be passed as list elements \code{past\_mean}, \code{past\_obs}, and \code{covariates} in the \code{model} argument, which then define the above sets, i.e., $\mathcal{A}_i^{(\mu)} = \lbrace \ell : a_{\ell + 1, i} = 1 \rbrace$, and the other sets analogously.

Using these sets, we can rewrite model~\eqref{eq:model_simple} as
\begin{align}
\label{eq:model}
\bm{\psi}_t = \bm{\delta} 
+ \sum_{i \in \mathcal{Q}^{(\mu)}}\sum_{\ell \in \mathcal{A}_i^{(\mu)}} \alpha_{i\ell} \mathbf{W}^{(\ell)}_{\alpha} h\left(\bm{\psi}_{t - i}\right) 
+ \sum_{j \in \mathcal{R}^{(\mu)}} \sum_{\ell \in \mathcal{B}_j^{(\mu)}} \beta_{j\ell} \mathbf{W}^{(\ell)}_{\beta} \tilde{h}\left(\mathbf{Y}_{t - j}\right) 
+ \sum_{k = 1}^m \sum_{\ell \in \mathcal{C}_k^{(\mu)}} \gamma_{k\ell} \mathbf{W}^{(\ell)}_{\gamma} \mathbf{X}_{k, t}.
\end{align}

To ensure identifiability, it is important to include regression on past observations, i.e., $\mathcal{R}^{(\mu)} \neq \emptyset$, if regression on past values of the feedback process ($\mathcal{Q}^{(\mu)} \neq \emptyset$) is included in the model. See \citet{maletz_spatio-temporal_2024} for a more detailed discussion. Regression on the feedback process is omitted, if the list element \code{past\_mean} is not defined in the named list or equal to \code{NULL}. If $\mathcal{Q}^{(\mu)} \neq \emptyset$, then it is necessary to initialize values $\bm{\psi}_0, \ldots, \bm{\psi}_{\tau}$ with $\tau = \max \mathcal{Q}^{(\mu)} \cup \mathcal{R}^{(\mu)}$ for parameter estimation. The choices available in the \pkg{glmSTARMA} package are discussed in Section~\ref{sec:estimation}.

\subsection{Conditional Dispersion Model}\label{sec:dispersion_model}

Unlike the conditional mean, the dispersion parameters cannot be inferred directly from the observations. Therefore, consider  pseudo-observations, say $d_{i,t}$, with conditional expectations $\mathbb{E}(d_{i,t} \mid \mathcal{F}_{t-1}) = \phi_{i,t}$ and collect all of them into
a process $\lbrace \bm{d}_t = (d_{1,t}, \ldots, d_{p,t})' \rbrace$.  This is the approach taken by 
\citet{smyth_generalized_1989} who proposes a deviance-based measure for computing $d_{i,t}$, using the (conditional) deviance residuals. Indeed, the $d_{i,t}$ are computed from
the distance between  the log-likelihoods of the actual model to a saturated model, i.e.,
\begin{align}
    \label{eq:deviance}
    d_{i,t} = 2 \cdot \log\Biggl[\frac{f(y_{i,t} \mid  \mu_{i, t} =  y_{i,t}, {\cal F}_{t-1})}{f(y_{i,t} \mid \mu_{i, t} =  \mu_{i,t}, {\cal F}_{t-1}} \Biggr],
\end{align}
where $f(y_{i,t}\mid \mu_{i, t} =  \mu, {\cal F}_{t-1})$ is defined by  \eqref{eq:exponential_family}, conditionally on the past, evaluated at $y_{i,t}$ with the mean $\mu$. The numerator uses $\mu = y_{i,t}$, i.e., a saturated model where the mean is equal to the observation; for more on this approach see \citet{smyth_generalized_1989}.

Using $d_{i,t}$ as observations, define the linear predictor $\bm{\zeta}_t$ for the dispersion process $\bm{\phi}_t$ along the lines of  
the mean model  \eqref{eq:model}, that is, 
\begin{align}
    \label{eq:dispersionmodel}
    \bm{\zeta}_t = \bm{\tilde{\delta}} 
    + \sum_{i \in \mathcal{Q}^{(\phi)}} \sum_{\ell \in \mathcal{A}_{i}^{(\phi)}} \tilde{\alpha}_{i\ell} \bm{\tilde{W}}^{(\ell)}_\alpha h_\phi(\bm{\zeta}_{t-i}) 
    + \sum_{j \in \mathcal{R}^{(\phi)}} \sum_{\ell \in \mathcal{B}_{j}^{(\phi)}} \tilde{\beta}_{j\ell} \bm{\tilde{W}}^{(\ell)}_\beta \tilde{h}_\phi(\bm{d}_{t-j}) 
    + \sum_{k=1}^{\tilde{m}} \sum_{\ell \in \mathcal{C}_{k}^{(\phi)}} \tilde{\gamma}_{k\ell} \bm{\tilde{W}}^{(\ell)}_\gamma \bm{\tilde{X}}_{k,t}.
\end{align}

The unknown parameters, i.e., the intercept $\bm{\tilde{\delta}}$ and the coefficients $\tilde{\alpha}_{i\ell}$, $\tilde{\beta}_{j\ell}$, and $\tilde{\gamma}_{k\ell}$ of \eqref{eq:dispersionmodel} are collected in the vector $\bm{\vartheta}_{\phi}$.
Most existing models in the literature consider a model for the mean, i.e., \eqref{eq:model}, and thus implicitly assume a constant 
dispersion parameter (which can be known or unknown). This is equivalent to \eqref{eq:dispersionmodel} with  $\bm{\zeta}_t = \tilde{\delta}_0 \bm{1}_p$, 
for some constant $\tilde{\delta}_0$. This package contains  two functions, namely  \fct{glmstarma} and \fct{dglmstarma}, where the former fits the model \eqref{eq:model}.

To capture specific dynamics in the dispersion parameters, various link functions are available for the dispersion model. Notably, when modeling normally distributed data, applying the identity link yields a structure that roughly corresponds to a standard GARCH process. Alternatively, utilizing a log-link function mirrors a log-GARCH model. For the log-link, our package defaults to using $\log(\bm{d}_{t-i} + 1)$ instead of the traditional $\log(\bm{d}_{t-i})$. This slight modification is crucial, as we have observed that it prevents numerical instabilities during the parameter estimation procedure.

\paragraph{Remark 1 (Distribution of $d_{i,t}$).}
Under certain assumptions, the conditional distribution of the pseudo-observations $d_{i,t} \mid \mathcal{F}_{t-1}$ approximately follows a Gamma distribution with a fixed dispersion parameter of 2. Following \citet{smyth_generalized_1989}, this approximation is exact for the normal and the inverse Gaussian distributions. For the Poisson distribution, this approximation holds provided that $\mu_{i,t} > 3$. For the binomial case, both $\mu_{i,t} > 3$ and $n - \mu_{i,t} > 3$ should be satisfied for the approximation to hold. For the Gamma distribution, the approximation holds provided that $\phi_{i,t} < 1/3$, although in this case $d_{i,t}$ is known to follow a Digamma distribution exactly, see \citet{smyth_generalized_1989} for more. The link functions of the Gamma distribution are available for the dispersion model in this package, so the model \eqref{eq:dispersionmodel} can be easily fitted; recall Table~\ref{tab:distributions}.

For the Poisson distribution, in simulations we observed that an overall adequate approximation, which yields reliable estimates of the dependence parameters, requires a larger magnitude of observations. Due to the presence of spatial and temporal dependencies, the requirement of $\mu_{i,t} > 3$ does not appear to be satisfied frequently enough, if the global mean of the observations is too low, systematically leading to an underestimation of the autoregressive parameters (see the simulation study in Appendix~\ref{appendix:simulations}). Our results suggest that the approximation becomes sufficiently accurate only when the unbounded mean of the observations reaches a level around 35, but future developments might be able to improve this.

\paragraph{Remark 2 (Count data and overdispersion).}  
For standard Poisson models, the conditional mean and variance are equal, i.e., $\mathbb{E}(Y_{i,t} \mid \mathcal{F}_{t-1}) = \text{Var}(Y_{i,t} \mid \mathcal{F}_{t-1}) = \mu_{i,t}$, implying $\phi_{i,t} = 1$ for all $i,t$. In practice, however, quasi-Poisson models are often used to allow for overdispersion or underdispersion, assuming $\text{Var}(Y_{i,t} \mid \mathcal{F}_{t-1}) = \phi_{i,t} \mu_{i,t}$ with $\phi_{i,t} > 0$. A distribution with this property is the generalized Poisson distribution, see \citet{consul_generalization_1973}. To avoid confusion, the \proglang{R} package contains the two family functions, \fct{vpoisson} and \fct{vquasipoisson}. The former assumes $\phi_{i, t} = 1$ for all $i, t$, while the latter allows variation, which can be modeled via \eqref{eq:dispersionmodel}. The negative binomial case is discussed in the next remark.

The same problem  that occurs with the Poisson distribution applies to the binomial distribution. In this case, the (conditional) variance is given by $V(\mu_{i, t}) = n_i  p_{i, t}  (1 - p_{i, t})$. Over- or under-dispersion can easily occur in practice if the $n_i$ random experiments are not independent \citep{ahn_generation_1995}.  To accommodate this, we set with the family-function \fct{vbinomial} $\phi_{i, t} = 1$ and allow varying dispersions for the \fct{vquasibinomial}-family.

\paragraph{Remark 3 (Negative binomial models).}  
Consider  the following parameterization of the negative binomial distribution.
\begin{align}
    \label{eq:negativbinomial}
    f(y_{i,t} \mid \mathcal{F}_{t-1}) 
    = \frac{\Gamma(\nu_{i,t} + y_{i,t})}{\Gamma(y_{i,t} + 1)\Gamma(\nu_{i,t})}
    \left(\frac{\nu_{i,t}}{\nu_{i,t} + \mu_{i,t}}\right)^{\nu_{i,t}} 
    \left(\frac{\mu_{i,t}}{\nu_{i,t} + \mu_{i,t}}\right)^{y_{i,t}}, \quad y_{i,t} \in \mathbb{N}_0,
\end{align}
where $\Gamma(\cdot)$ is the Gamma function and $\nu_{i,t} > 0$ is a shape parameter. Therefore, $\mathbb{E}(Y_{i,t} \mid \mathcal{F}_{t-1}) = \mu_{i,t}$ and $\text{Var}(Y_{i,t} \mid \mathcal{F}_{t-1}) = \mu_{i,t} + \mu_{i,t}^2 / \nu_{i,t}$. This naturally encompasses conditional overdispersion as $\text{Var}(Y_{i,t} \mid \mathcal{F}_{t-1}) > \mathbb{E}(Y_{i,t} \mid \mathcal{F}_{t-1})$. 
To take advantage of the natural overdispersion property of the negative binomial distribution, we set the GLM-dispersion parameter equal to 1 and instead use the shape parameter $\nu_{i, t}$, or its inverse $\tilde{\phi}_{i, t} = 1 / \nu_{i, t}$. The same trick has been used in the \fct{glm.nb} function from the \pkg{MASS} package \citep{venables_modern_2002} and in the \pkg{tscount} package \citep{liboschik_tscount_2017}.

To calculate the deviance residuals~\eqref{eq:deviance} in case of the negative binomial distribution, we define $d_{i, t}^{\star}$ for this distribution based on Pearson residuals of the Poisson distribution, i.e.,
\begin{align}
    d_{i,t}^{\star} = \frac{(y_{i,t} - \mu_{i,t})^2}{\mu_{i,t}}, 
\end{align}
because the shape parameter is unknown. 
These values are approximately distributed as $(1 + \phi_{i,t} \mu_{i,t}) \cdot \chi^2_1$, with expected value $\phi_{i,t}^{\star} = \mathbb{E}(d_{i,t}^{\star} \mid \mathcal{F}_{t-1})$, which is given by \eqref{eq:dispersionmodel}.
The inverse shape parameter $\tilde{\phi}_{i,t}$ is recovered as
\begin{align}
    \tilde{\phi}_{i,t} = \max\left\lbrace 0, \frac{\phi_{i,t}^{\star} - 1}{\mu_{i,t}} \right\rbrace,
\end{align}
which is the result of solving the conditional expectation of $d_{i,t}^{\star}$ for $\tilde{\phi}_{i, t}$ but for positive values, because 
negative values imply underdispersion, a case which is not  covered by the negative binomial distribution. If $\tilde{\phi}_{i, t} = 0$, the negative binomial distribution reduces to a Poisson distribution.

\paragraph{Remark 4 (Deviance vs. Pearson residuals)}
It is generally possible to define the pseudo-observations $d_{i, t}$ using the (squared) Pearson residuals, i.e., 
\begin{align}
    d_{i,t} = \frac{(y_{i,t} - \mu_{i,t})^2}{V(\mu_{i,t})},
\end{align}
instead of the deviance residuals~\eqref{eq:deviance}, see \citet{wu_dispersion_2016}. However, the deviance residuals have a better theoretical foundation and are therefore used  more often in practice, see \citet{jorgensen_theory_1997}. The unknown parameters of the two models~\eqref{eq:model} and \eqref{eq:dispersionmodel} are orthogonal to each other in the deviance residual case \citep{smyth_generalized_1989}, which facilitates the parameter estimation described in Section~\ref{sec:estimation}.
The \pkg{glmSTARMA} package allows selecting the construction method in the relevant functions via the argument \code{pseudo\_observations}, which can be set to either \code{"deviance"} (default) or \code{"pearson"}. This argument has no effect in case of the negative binomial distribution.

%% file: sections/simulation.tex
\section{Data Generating Process}\label{sec:simulation}

This \R package provides the two functions \fct{glmstarma.sim} and \fct{dglmstarma.sim} to simulate data from  multivariate processes defined by means of \eqref{eq:model} and \eqref{eq:dispersionmodel}. Both functions are similar in structure but differ in how the dispersion parameter is handled, as explained next.

\subsection{Main Function}

The function \fct{glmstarma.sim} assumes a constant dispersion by default, i.e., $\phi_{i,t} = 1$ for all $i$ and $t$. Custom dispersion values can be supplied by the user via the \texttt{dispersion} argument within the family-functions; see Table~\ref{tab:distributions}. The \code{dispersion} argument can be set to a scalar, a numeric vector of length $p$, or a $p \times T$ matrix, where $T$ denotes the number of time points to simulate. The argument \code{ntime} specifies the length of the time series $T$, while \code{model} defines the model orders for the mean structure, as introduced in the previous sections. The true parameters of the mean model, denoted by $\boldsymbol{\vartheta}$, are provided via the argument \code{parameters}.

In Section~\ref{appendix:simulation_example} of the Appendix, we illustrate with two examples how the functions can be used to simulate data from these processes.

The function \fct{dglmstarma.sim} allows us to incorporate the dispersion model in the simulation explicitly. In addition to the arguments mentioned above, it includes:
\begin{itemize}
  \item \code{dispersion\_model} to specify the model orders of the dispersion component,
  \item \code{pseudo\_observations} to define how pseudo-observations are constructed, and
  \item \code{parameters\_dispersion} to specify the true parameters of the dispersion model~\eqref{eq:dispersionmodel}.
\end{itemize}

Both \code{parameters} and \code{parameters\_dispersion} follow the same structure as \code{model} and \code{dispersion\_model}, respectively. They are named lists that must contain the following elements:
\begin{itemize}
  \item \code{intercept}: a vector with the intercept parameter(s),
  \item \code{past\_obs}: a matrix $\mathbf{B} = ({\beta})_{i\ell}$, where the $(\ell + 1)$-th row corresponds to the parameters of spatial lag $\ell$, and the columns refer to the temporal lags,
  \item \code{past\_mean}: parameters $\alpha_{i\ell}$ in the same matrix format, referring to past means, and
  \item \code{covariates}: parameters $\gamma_{k, \ell}$ in the same matrix format, referring to the coefficients of the covariates. The columns refer to the covariates and the rows to the spatial lag.
\end{itemize}
Note that parameter entries that do not match the model orders defined in \code{model} or \code{dispersion\_model} will be ignored, i.e., by setting them to  zero. If  covariates or regression on past values of the feedback/dispersion process are not included  in the models, the corresponding list elements can be omitted.

Covariates are supplied as a (named) list using the arguments \code{covariates} and (if applicable) \code{dispersion\_covariates}, with each element of the list representing one covariate process. These must be consistent with the total number of simulated time points, i.e., they must cover the same observation times as the time series to be simulated, i.e., \code{ntime}. The covariates are supplied as a matrix of dimension $p \times T$ or as a result of the auxiliary functions \fct{SpatialConstant} or \fct{TimeConstant}. The first function accepts a vector of length $T$ as input and creates a covariate of the form $\bm{X}_t = x_t \bm{1}_p$, i.e., it varies only over time, but takes the same values for all locations. The second function accepts a vector of length $p$ as input and creates a covariate of the form $\bm{X}_t = \bm{X} \in \mathbb{R}^p$. These covariates can vary with location, but are constant in time.

An optional argument \code{n\_start} in the simulation functions controls the number of initial burn-in observations used to initialize the process. During this burn-in phase, time-variant covariates are not taken into account. By default, the burn-in phase consists of 100 time points.

\subsection{Model Application and Simulation Procedure}

The simulation process consists of two phases: a burn-in phase and a main sampling phase during which the output data are generated. In the burn-in phase, the first $\max \mathcal{Q}^{(\mu)} \cup \mathcal{Q}^{(\phi)} \cup \mathcal{R}^{(\mu)} \cup \mathcal{R}^{(\phi)}$ observations are drawn independently from the distribution specified in the \code{family} argument. Recall that, for $\nu \in \{\mu,\phi\}$, the sets $\mathcal{Q}^{(\nu)}$ and $\mathcal{R}^{(\nu)}$ denote the time lags included in the feedback and observation processes of the mean and dispersion models, respectively. The initial values for $\mu_{i,t}$ and $\phi_{i,t}$ are set using the approximate expectation under assumption of stationarity of the process, see eq.~\eqref{eq:expectation} in the Appendix. The observations are generated using the methods described in the following subsection.

After the initial observations, the model equations~\eqref{eq:model} and~\eqref{eq:dispersionmodel} are iteratively applied to generate new values of $\bm{\mu}_t$ and $\bm{\phi}_t$, which are then used to simulate new observations. Covariate effects are ignored during the burn-in phase.
After the burn-in phase, the simulation runs for further $T$ time points, as specified by \code{ntime}, and covariates are incorporated into the model dynamics provided that they have been included in the model.

\subsection{Sampling from Joint Distribution}
Note that the model assumptions in Section~2 do not imply conditional independence of the components in $\bm{Y}_t$ conditionally on $\mathcal{F}_{t - 1}$. To simulate contemporary dependencies among components of $\bm{Y}_t$, this package allows the use of copulas. The distribution functions listed in Table~\ref{tab:distributions} support the arguments \code{copula} and \code{copula\_param}. The main idea follows \citet{fokianos_multivariate_2020}; the marginals are of some known form (for example Poisson, negative binomial), and the copula function determines the joint dependence. This enables us to use established methods based on the property of keeping the marginals conditionally on the past within a specific distribution family. Examples of this approach have been reported by \citet{armillotta_count_2024} and \citet{maletz_spatio-temporal_2024} in the context of count data.

The \code{copula} argument accepts one of the following character strings: \code{"normal"}, \code{"t"}, \code{"clayton"}, \code{"frank"}, \code{"gumbel"}, or \texttt{"joe"}, each corresponding to a copula implemented in the \pkg{copula} package (see \citet{yan_enjoy_2007}). The scalar \code{copula\_param} specifies the associated copula parameter. By default, i.e., when no copula is specified, observations are generated assuming conditional independence using the standard random number generators. If a copula is specified, one of the sampling algorithms described below is applied, depending on the chosen distribution.

In the Appendix in Section~\ref{appendix:simulation_example} we show two examples on how to simulate data from these processes using the functions of the package.

\paragraph{Inversion Method}
For most implemented distributions (Normal, Poisson with $\phi_{i,t} = 1$, Negative Binomial, Gamma, Binomial), the inversion method can be used \citep[see e.g.][Chapter 2]{devroye_non-uniform_1986}. A random vector $\bm{u}_{0,t} = (u_{0,t,1}, \ldots, u_{0,t,p})'$ is drawn from the specified copula. Due to the properties of copulas, the components are uniformly distributed on $[0, 1]$ but not necessarily independent. The observation $\bm{Y}_t$ is then generated element-wise as
\[
Y_{i,t} = F^{-1}(u_{0,t,i}; \mu_{i,t}, \phi_{i, t}),
\]
where $F^{-1}$ denotes the quantile function of the corresponding distribution with mean $\mu_{i,t}$ and dispersion parameter $\phi_{i, t}$. In cases of non-unique quantile functions, the smallest value for which the cumulative distribution function of the corresponding distribution exceeds the value $u_{0,t,i}$ is used.

\paragraph{Michael-Schucany-Haas Method}

For the inverse Gaussian distribution, the inversion method is challenging due to the lack of a simple closed-form quantile function. Therefore, we use the Michael-Schucany-Haas method \citep{michael_generating_1976} for the generation of random numbers. First, two independent random vectors, say  $\bm{U}_{1,t}$ and $\bm{U}_{2,t}$ are drawn from a copula. Then, the observation $Y_{i,t}$ is computed as follows:
\begin{enumerate}
  \item Define \( Z_{i,t} = \Phi^{-1}(U_{1,t,i}) \) so that $Z_{i,t} \sim \mathcal{N}(0,1)$.
  \item Let $\tilde{Z}_{i,t} = Z_{i,t}^2$ and compute
  \[
  \tilde{Y}_{i,t} = \mu_{i,t} + \frac{\mu_{i,t}^2 \tilde{Z}_{i,t}}{2\phi_i} - \frac{\mu_{i,t}}{2\phi_i} \sqrt{4\mu_{i,t}\phi_i \tilde{Z}_{i,t} + \mu_{i,t}^2 \tilde{Z}_{i,t}^2}.
  \]
  \item Then set
  \[
  Y_{i,t} = 
  \begin{cases}
    \tilde{Y}_{i,t}, & \text{if } U_{2,t,i} \leq  \displaystyle \frac{\mu_{i,t}}{\mu_{i,t} + \tilde{Y}_{i,t}}, \\
    \displaystyle \frac{\mu_{i,t}^2}{\tilde{Y}_{i,t}}, & \text{otherwise}.
  \end{cases}
  \]
\end{enumerate}

\paragraph{Poisson Process}
For the special case of the Poisson distribution without over- or underdispersion (i.e., without a dispersion model~\eqref{eq:dispersionmodel}), an alternative data-generating process based on \citet{fokianos_multivariate_2020} is available. This method is implemented only in \texttt{glmstarma.sim} and can be used in the \fct{vpoisson}-family with the argument \code{sampling\_method = "poisson\_process"}.
See \citet{fokianos_multivariate_2020}  and \citet{fokianos_multivariate_2024} for details.

\paragraph{Quasi-Poisson-Models}
For quasi-Poisson models with time-varying dispersion, simulated via the function \fct{dglmstarma.sim}, the sampling method can be specified using the argument \code{sampling\_method} within the \fct{vquasipoisson} family. The available options are \code{"build\_up"}, \code{"chop\_down"}, \code{"branching"}, and \code{"negbin"}.

The first three options generate samples from a suitably parameterized generalized Poisson distribution, as defined by \citet{consul_generalization_1973}. The implementation follows the approach used in the \proglang{R} package \pkg{RNGforGPD} \citep{li_rngforgpd_2021}. The \fct{"negbin"} method, on the other hand, uses the negative binomial distribution as defined in Equation~\eqref{eq:negativbinomial}, with
\[
\nu_{i, t} = \frac{\mu_{i, t}}{\phi_{i, t} - 1}.
\]

Note that the \code{"branching"} and \code{"negbin"} methods do not simulate underdispersion (i.e., $\phi_{i, t} < 1$). In such cases, sampling defaults to the standard Poisson distribution.

\paragraph{Quasi-Binomial-Models}
Two different methods are employed for simulating quasi-binomial models, depending on whether the data exhibit 
overdispersion or underdispersion. For the observations we use the notation
\[
Y_{i, t} \mid \mathcal{F}_{t - 1} \sim \text{QuasiBin}(n_i, \pi_{i, t}, \phi_{i, t})
\]
with conditional mean and variance 
\[
\mathbb{E}(Y_{i, t} \mid \mathcal{F}_{t - 1}) = n_i \pi_{i, t} ~~
\text{Var}(Y_{i, t} \mid \mathcal{F}_{t - 1}) = \phi_{i, t} n_i \pi_{i, t}(1 - \pi_{i, t}).
\]

In case of overdispersion (i.e., $\phi_{i, t} \geq 1$), following the lines of \citet{ahn_generation_1995}, we generate $n_i + 1$ independent Bernoulli random variables $Z_j \sim \text{Bin}(1, \pi_{i, t})$ for $j = 0, 1, \ldots, n_i$, and $n_i$ independent Bernoulli variables $\tilde{Z}_j \sim \text{Bin}(1, \rho)$ for $j = 1, \ldots, n_i$, where
$
\rho = \sqrt{{(\phi_{i, t} - 1)}/{(n_i - 1)}}.
$
Then, we define $
\tilde{Y}_j = \tilde{Z}_j Z_0 + (1 - \tilde{Z}_j) Z_j, \quad j = 1, \ldots, n_i,
$
and compute $Y_{i, t} = \sum_{j = 1}^{n_i} \tilde{Y}_j$.
In the case of underdispersion (i.e., $\phi_{i, t} < 1$), a normal approximation is used, that is, 
\[
\tilde{Y}_{i, t} \sim \mathcal{N}(n_i \pi_{i, t}, \phi_{i, t} n_i \pi_{i, t}(1 - \pi_{i, t})),
\]
and $Y_{i, t}$ is set to the integer in $\{0, 1, \ldots, n_i\}$ closest to $\tilde{Y}_{i, t}$.

While the first method works exactly with desired mean and variance, the normal approximation faces problems when the mean is very close to the boundaries $0$ and $n_i$.

%% file: sections/fitting.tex
\section{Parameter Estimation}\label{sec:estimation}

Similarly to the data simulation functionality, the \proglang{R} package \pkg{glmSTARMA} provides two functions for model fitting within the framework established in this work: \fct{glmstarma} estimates a model for the conditional mean only, while \fct{dglmstarma} estimates both a mean and a dispersion model. In both cases, the parameters are estimated using a quasi-maximum likelihood (QMLE) approach. In \fct{dglmstarma}, the estimation proceeds iteratively, alternating between fitting the mean model~\eqref{eq:model} and the dispersion model~\eqref{eq:dispersionmodel}, as is common in the estimation of double generalized linear models; see \citet{smyth_generalized_1989}. The estimation is performed by default under stability constraints; see Eq.~\eqref{eq:stationarity} in the Appendix. 

\subsection{Quasi-Maximum-Likelihood Estimation}
The unkown parameters $\bm{\vartheta}_{\mu}$ of the mean model \eqref{eq:model} are estimated by maximizing the following conditional quasi log-likelihood function
\begin{align}
    \label{eq:loglike}
    \ell(\bm{\vartheta}_{\mu}) = \sum_{t = \tau}^T \sum_{i = 1}^p \log(f(y_{i,t} \mid \theta_{i,t}, \phi_{i,t}, {\cal F}_{t-1})) 
    = \sum_{t = \tau}^T \sum_{i = 1}^p \left( \frac{\theta_{i,t}(\bm{\vartheta}_{\mu}) y_{i,t} - A(\theta_{i,t}(\bm{\vartheta}_{\mu}))}{\phi_{i,t}} + c(\phi_{i,t}, y_{i,t}) \right),
\end{align}
with $\tau = \max\mathcal{Q}^{(\mu)} \cup \mathcal{R}^{(\mu)} \cup \mathcal{Q}^{(\phi)} \cup \mathcal{R}^{(\phi)}$, i.e., the largest time lag used in the models~\eqref{eq:model} and \eqref{eq:dispersionmodel}.
For estimation of the parameters $\bm{\vartheta}_{\phi}$ of the dispersion model~\eqref{eq:dispersionmodel} a quasi-likelihood, using a Gamma density with fixed dispersion parameter 2, is calculated on the pseudo-observations $d_{i, t}$ instead of $y_{i, t}$ and $\bm{\vartheta}_{\phi}$ instead of $\bm{\vartheta}_{\mu}$; see the next subsection for more details on the estimation procedure. Although the log-likelihood function is constructed assuming conditional independence among the components of $\bm{Y}_t \mid \mathcal{F}_{t - 1}$, it still takes into account temporal and spatial correlations because of eq.~\eqref{eq:model} and \eqref{eq:dispersionmodel}. In case

The corresponding (quasi) score function of the mean model is
\begin{align}
    \label{eq:score}
    S_T(\bm{\vartheta}_{\mu}) = \sum_{t = \tau}^T \left[ \frac{\partial}{\partial \bm{\vartheta}_{\mu}} \bm{\psi}_t \right]' \mathbf{D}_t (\mathbf{Y}_t - g^{-1}(\bm{\psi}_t)),
\end{align}
where $\mathbf{D}_t = \mathrm{diag}((g^{-1})'(\psi_{i,t}) / \sigma_{i,t}^2, i = 1, \ldots, p) = \mathrm{diag}(1 / g'(g^{-1}(\psi_{i,t})) \sigma_{i,t}^2, i = 1, \ldots, p)$ and $\sigma_{i,t}^2 = \phi_{i, t} \cdot A''_i(\theta_{i,t}(\bm{\vartheta}_{\mu}))$. The (quasi) Fisher information matrix is given by the expected second derivative of~\eqref{eq:loglike}:
\begin{align}
    \label{eq:information}
    \mathbf{G}_T(\bm{\vartheta}_{\mu}) = \sum_{t = \tau}^T \left[ \frac{\partial}{\partial \bm{\vartheta}_{\mu}} \bm{\psi}_t \right]' \mathbf{\tilde{D}}_t \left[ \frac{\partial}{\partial \bm{\vartheta}_{\mu}} \bm{\psi}_t \right],
\end{align}
with $\mathbf{\tilde{D}}_t = \mathrm{diag}(((g^{-1})'(\psi_{i,t}))^2 / \sigma_{i,t}^2, i = 1, \ldots, p)$. A derivation of this matrix is provided in Appendix~\ref{sec:information}. The derivatives $\left[ \partial \bm{\psi}_t / \partial \bm{\vartheta}_{\mu} \right]'$ in~\eqref{eq:score} and~\eqref{eq:information} are computed component-wise. Here, $\bm{X} \circ \bm{Y}$ denotes the element-wise (Hadamard) product of the matrices $\bm{X}, \bm{Y} \in \mathbb{R}^{p \times q}$:
\begin{equation}
	\label{eq:recursions}
	\begin{aligned}
        \frac{\partial \bm{\psi}_t'}{\partial \delta_0} &= \mathbf{1}_p + \sum_{j = 1}^q \sum_{\ell = 0}^{a_j} \alpha_{j\ell} \mathbf{W}^{(\ell)} \left[ h'(\bm{\psi}_{t-j}) \circ \frac{\partial \bm{\psi}_{t-j}'}{\partial \delta_0} \right], \\
        \frac{\partial \bm{\psi}_t'}{\partial \alpha_{jo}} &= \mathbf{W}^{(o)} h(\bm{\psi}_{t-j}) + \sum_{i = 1}^q \sum_{\ell = 0}^{a_i} \alpha_{i\ell} \mathbf{W}^{(\ell)} \left[ h'(\bm{\psi}_{t-j}) \circ \frac{\partial \bm{\psi}_{t-j}'}{\partial \alpha_{jo}} \right], \\
        \frac{\partial \bm{\psi}_t'}{\partial \beta_{io}} &= \mathbf{W}^{(o)} \tilde{h}(\mathbf{Y}_{t-i}) + \sum_{j = 1}^q \sum_{\ell = 0}^{a_j} \alpha_{j\ell} \mathbf{W}^{(\ell)} \left[ h'(\bm{\psi}_{t-j}) \circ \frac{\partial \bm{\psi}_{t-j}'}{\partial \beta_{io}} \right], \\
        \frac{\partial \bm{\psi}_t'}{\partial \gamma_{ko}} &= \mathbf{W}^{(o)} \mathbf{X}_{k,t} + \sum_{j = 1}^q \sum_{\ell = 0}^{a_j} \alpha_{j\ell} \mathbf{W}^{(\ell)} \left[ h'(\bm{\psi}_{t-j}) \circ \frac{\partial \bm{\psi}_{t-j}'}{\partial \gamma_{ko}} \right].
	\end{aligned}
\end{equation}

\subsection{Iterative Estimation of Mean and Dispersion Models}
Similarly to the DGLMs, we use an iterative procedure that alternates between estimating the mean model \eqref{eq:model} and the dispersion model \eqref{eq:dispersionmodel}. This approach exploits the separability of the two submodels while accounting for their interdependence through the variance structure.

In each iteration, we maximize a quasi-log-likelihood using numerical optimization methods. For the mean model, we employ the density function specified by the \code{family} argument. For the dispersion model, we use the density function of the Gamma distribution with its dispersion parameter fixed at 2, evaluated on pseudo-observations $d_{i,t}$ derived from the Pearson residuals of the mean model. The quasi-log-likelihood for the dispersion model follows the same structural form as the mean model, with $\bm{Y}_t$ replaced by $\bm{d}_t$, linear processes $\bm{\psi}_t$ replaced by $\bm{\zeta}_t$, and all other model components adapted accordingly.

To ensure monotonic improvement of the log-likelihood~\eqref{eq:loglike} after each submodel fit, and numerical stability, our implementation incorporates two key enhancements beyond the standard
algorithm. First, we monitor the joint log-likelihood after updating each submodel. If an update leads to a decrease in the overall likelihood, we apply a step-halving procedure: the parameter update is scaled by successive factors of $0.5$ until the likelihood increases or a maximum number of halving steps is reached. This backtracking strategy prevents divergent behavior and ensures that each accepted update improves the model fit.

Second, we employ dual convergence criteria. The iteration terminates when either (i) the change in the combined parameter vector falls below a specified tolerance, say $\varepsilon$, indicating parameter stability, or (ii) the relative improvement in log-likelihood between successive iterations becomes negligible.
%suggesting that further iterations yield diminishing returns. 
This dual-criterion approach balances computational efficiency with robust convergence detection.

Algorithm~\ref{alg:estimation} provides a detailed description of the iterative estimation procedure, including the likelihood monitoring and step-halving mechanism. This enhanced iterative scheme ensures robust convergence while maintaining computational efficiency.

\begin{algorithm}[t]
\caption{Iterative Estimation Procedure with Likelihood Monitoring}
\label{alg:estimation}
\begin{algorithmic}[1]
\State Initialize iteration counter: $k \gets 0$
\State Set constant dispersion: $\phi_{i,t}^{(0)} \gets \phi$ for all $i, t$
\State Compute initial joint log-likelihood: $\ell^{(0)}$
\Repeat
    \State \textbf{Mean Model Update:}
    \State \quad Estimate mean model via QMLE using current $\phi_{i,t}^{(k)}$: $\bm{\tilde\vartheta}_{\mu}^{(k)}$
    \State \quad Compute joint log-likelihood based on mean model estimates: $\tilde\ell_{\mu}^{(k)}$
    \If{$\tilde\ell_{\mu}^{(k)} < \ell^{(k)}$}
        \State \quad Apply step-halving to $\bm{\tilde\vartheta}_{\mu}^{(k)}$ until $\tilde\ell_{\mu}^{(k)} \geq \ell^{(k)}$
    \EndIf
    \State \quad Accept update: $\bm{\hat\vartheta}_{\mu}^{(k)} \gets \bm{\tilde\vartheta}_{\mu}^{(k)}$
    \State Update joint log-likelihood: $\ell^{(k)} \gets \ell_{\mu}^{(k)}$
    \State
    \State \textbf{Dispersion Model Update:}
    \State \quad Compute pseudo-observations: $\lbrace \bm{d}_t^{(k)} \rbrace$ from current mean model
    \State \quad Estimate dispersion model on $\lbrace \bm{d}_t^{(k)} \rbrace$ via QMLE: $\bm{\tilde\vartheta}_{\phi}^{(k)}$
    \State \quad Update dispersion: $\tilde\phi_{i,t}^{(k)}$ based on $\bm{\tilde\vartheta}_{\phi}^{(k)}$
    \State \quad Compute joint log-likelihood based on disperson estimates: $\tilde\ell_{\phi}^{(k)}$
    \If{$\tilde\ell_{\phi}^{(k)} < \ell^{(k)}$}
        \State \quad Apply step-halving to $\bm{\tilde\vartheta}_{\phi}^{(k)}$ until $ \tilde\ell_{\phi}^{(k)} \geq \ell^{(k)}$
    \EndIf
    \State \quad Accept update: $\bm{\hat\vartheta}_{\phi}^{(k)} \gets \bm{\tilde\vartheta}_{\phi}^{(k)}$, $\phi_{i,t}^{(k+1)} \gets \tilde\phi_{i,t}^{(k)}$
    \State
    \State Set joint log-likelihood: $\ell^{(k+1)} \gets \tilde\ell_{\phi}^{(k)}$
    \State $k \gets k + 1$
\Until{$\|(\bm{\hat\vartheta}_{\mu}^{(k)}, \bm{\hat\vartheta}_{\phi}^{(k)}) - (\bm{\hat\vartheta}_{\mu}^{(k-1)}, \bm{\hat\vartheta}_{\phi}^{(k-1)})\|_2 < \varepsilon$ \textbf{or} $|\ell^{(k)} - \ell^{(k-1)}|/|\ell^{(k-1)}| < \delta$}
\State \Return $(\bm{\hat\vartheta}_{\mu}, \bm{\hat\vartheta}_{\phi})$
\end{algorithmic}
\end{algorithm}

Estimation behavior can be adjusted using the \code{control} argument, which accepts a named list of control parameters passed internally to the function \fct{glmstarma.control} and \fct{dglmstarma.control}. Further details about the control parameters are given in Section~\ref{appendix:control} of the Appendix.

%% file: sections/selection.tex
\section{Model Selection and Assessment}\label{sec:selection}
This \R package provides several tools for model assessment and selection. The \fct{summary} function computes standard errors for the estimated parameters and derives asymptotic $p$-values based on the normal approximation of the quasi-maximum likelihood estimator (QMLE), as detailed in Section~\ref{sec:theory}. For models with parameters restricted to non-negative values, see Table~\ref{tab:distributions}, the normal approximation is invalid under the null hypothesis of a true parameter equal to zero. Following the theory of GARCH processes \citep{francq_garch_2019}, we apply a heuristic correction by halving the $p$-values. Simulation studies in \citet{maletz_spatio-temporal_2024} support  the validity of this approach.

The \fct{residuals} function extracts residuals from fitted models. Three types of residuals are available: standard residuals $r_{i, t} = y_{i, t} - \mu_{i, t}$, Pearson residuals $r_{i,t}^{P} = (y_{i, t} - \hat{\mu}_{i, t})/\sqrt{V(\hat{\mu}_{i, t})}$, and deviance residuals $r_{i, t}^D = d_{i, t}$ (corresponding to pseudo-observations). Optionally, Pearson and deviance residuals can be scaled by the estimated dispersion parameter.

For forecasting in space and time, the \fct{predict} function generates future values by iteratively applying the model equations~\eqref{eq:model_simple} and~\eqref{eq:dispersionmodel}. When new observations are provided, rolling predictions can be computed; otherwise, the estimated conditional expectation substitutes for unobserved values.

\subsection{Information Criteria}
To facilitate model selection, including the choice of covariates, temporal orders, and spatial dependencies, several information criteria are computed during estimation: the Akaike Information Criterion (AIC) \citep{akaike_new_1974}, the Bayesian Information Criterion (BIC) \citep{schwarz_estimating_1978}, and the Quasi Information Criterion (QIC) \citep{pan_akaikes_2001}.

\paragraph{General Structure.}
All criteria follow the form
\begin{equation}\label{eq:ic_general}
\text{IC} = -2\tilde{\ell}(\bm{\vartheta}) + \text{Penalty},
\end{equation}
where $\tilde{\ell}(\bm{\vartheta})$ denotes the scaled (quasi-)log-likelihood and the penalty term varies by criterion. To ensure comparability between models with different temporal orders, the log-likelihood~\eqref{eq:loglike} is scaled by the effective number of observations:
\begin{equation}\label{eq:scaled_loglik}
\tilde{\ell}(\bm{\vartheta}) = \frac{T}{T - \max(\mathcal{Q}^{(\mu)} \cup \mathcal{R}^{(\mu)} \cup \mathcal{Q}^{(\phi)} \cup \mathcal{R}^{(\phi)})} \ell(\bm{\vartheta}),
\end{equation}
where $T$ denotes the total number of time points.

\paragraph{Quasi-Likelihood Models.}
For quasi models, where the mean structure of the Poisson or binomial distribution is paired with a dispersion model, a proper likelihood function may not exist or cannot be uniquely defined. We therefore employ adjusted profile likelihoods \citep{smyth_adjusted_1999, mccullagh_simple_1990}. For the quasi-Poisson case, the quasi log-likelihood for one observation with mean $\mu_{i, t}$ and dispersion-parameter $\phi_{i, t}$ is
\begin{equation*}
\begin{aligned}
	{\ell}(y_{i,t}; \mu_{i, t}, \phi_{i, t}) &= 
-\frac{1}{2} \biggl[
\log(\phi_{i, t})
+ 2y_{i, t}
+ 2 \log\Gamma(y_{i, t} + 1) \\
&\qquad \left. - 2y_{i, t} \log(y_{i, t})
+ \frac{y_{i, t} \log\left(\frac{y_{i, t}}{\mu_{i, t}}\right) - (y_{i, t} - \mu_{i, t})}{\phi_{i, t}}
\right],
\end{aligned}
\end{equation*}
with the convention $0\log(0) := 0$.

For the quasi-binomial case with observations $y_{i, t} \in \{0, \dotsc, n_i\}$, $\mu_{i, t} = n_i \pi_{i, t}$ and dispersion $\phi_{i, t}$, the quasi-log-likelihood is
\begin{equation}\label{eq:quasi_binomial_loglik}
\begin{split}
\ell(y_{i, t}; \mu_{i, t}, \phi_{i, t}, n_i) 
&= -\frac{1}{2} \Bigg[
    \log(\phi_{i, t})
    + 2 \log\binom{n_i}{y_{i, t}}
    - 2 y_{i, t} \log(y_{i, t})
    - 2 (n_i - y_{i, t}) \log(n_i - y_{i, t}) \\
&\qquad + \frac{1}{\phi_{i, t}} \bigg(
    2y_{i, t} \log\left( \frac{y_{i, t}}{\mu_{i, t}} \right)
    + 2(n_i - y_{i, t}) \log\left( \frac{n_i - y_{i, t}}{n_i - \mu_{i, t}} \right)
\bigg)
\Bigg].
\end{split}
\end{equation}

\paragraph{Penalty Terms.}
The penalty terms for the three information criteria differ in how they penalize model complexity. For the AIC, the penalty is given by $2k$, where $k$ denotes the total number of parameters in the mean and dispersion models. The BIC employs a stronger penalty of $k \log(Tp)$, where $p$ is the number of spatial units, thereby favoring more parsimonious models as the sample size increases. The QIC accounts for model complexity through the sandwich variance estimator of the QMLE and is defined as
\begin{equation}\label{eq:qic_penalty}
\text{Penalty}_{\text{QIC}} = 2 \left( \mathrm{tr}(\bm{G}_{\mu}^{-1} \bm{H}_{\mu}) + \mathrm{tr}(\bm{G}_{\phi}^{-1} \bm{H}_{\phi}) \right),
\end{equation}
where $\bm{G}_{\mu}$ and $\bm{H}_{\mu}$ are the sandwich components for the mean model (see Equation~\eqref{eq:variance_matrices}), and $\bm{G}_{\phi}$ and $\bm{H}_{\phi}$ are defined analogously for the dispersion model.

%% file: sections/examples.tex
\section{Usage and Examples}\label{sec:examples}
The most recent stable version of the \pkg{glmSTARMA} package is available on the Comprehensive R Archive Network (CRAN). Versions with newer and possibly untested features are available on GitHub \url{https://github.com/stmaletz/glmSTARMA}. Since most of the program code of the R package is written in \proglang{C++}, installing the GitHub version requires suitable compilers, e.g., \code{Rtools} for Windows (\url{https://cran. r-project.org/bin/windows/Rtools/}) and for MacOS \code{XCode} and \code{GNU Fortran compiler}; see \url{https://mac.r-project.org/tools/}. After installation, the package can be loaded via \code{library("glmSTARMA")}.

The help pages, accessible through \code{?glmstarma} and \code{?dglmstarma}, provide a detailed summary of the  two key functions and their arguments. In this Section we provide two data examples to illustrate the usage of the package. The preprocessed data sets corresponding to these examples can be downloaded from the GitHub repository with the \fct{load\_data} function. To comply with CRAN policies, the data sets could not be added directly to the \R package. 
The function saves the data sets in a user directory when the function is called for the first time. The available data sets are \code{chickenpox}, \code{rota}, and \code{SST}. We demonstrate the application of the functions using the last two data sets as examples in the following subsections. Two examples for simulation of data are discussed in the Appendix in Subsection~\ref{appendix:simulation_example}.
The following analyses should not be seen as exhaustive data analyses, but rather as illustrative examples of software usage.

\subsection{Rotavirus Infections in Germany}

As a first example, we analyze weekly reported rotavirus infections in $p = 411$ urban and rural districts in Germany from 2001 to 2024, comprising $T = 1,252$ weekly observations. An earlier version of this dataset (limited to 2018) was examined in our previous work \citep{maletz_spatio-temporal_2024}. Note that Germany's 2021 territorial reform merged the city of Eisenach into Wartburg district, reducing the total number of districts by one.

The data were obtained from the Robert Koch Institute's SurvStat@RKI 2.0 platform (\url{https://survstat.rki.de/}) and can be loaded in preprocessed form via the package command \code{load\_data("rota")}. The dataset includes linearly interpolated weekly population sizes for each region. Additional preprocessing details are available on the help page (\code{?rota}). Population data were derived from the Federal Statistical Office of Germany (Destatis).

\begin{figure}[t]
	\begin{subfigure}{0.42\textwidth}
		\includegraphics[width= \textwidth, keepaspectratio]{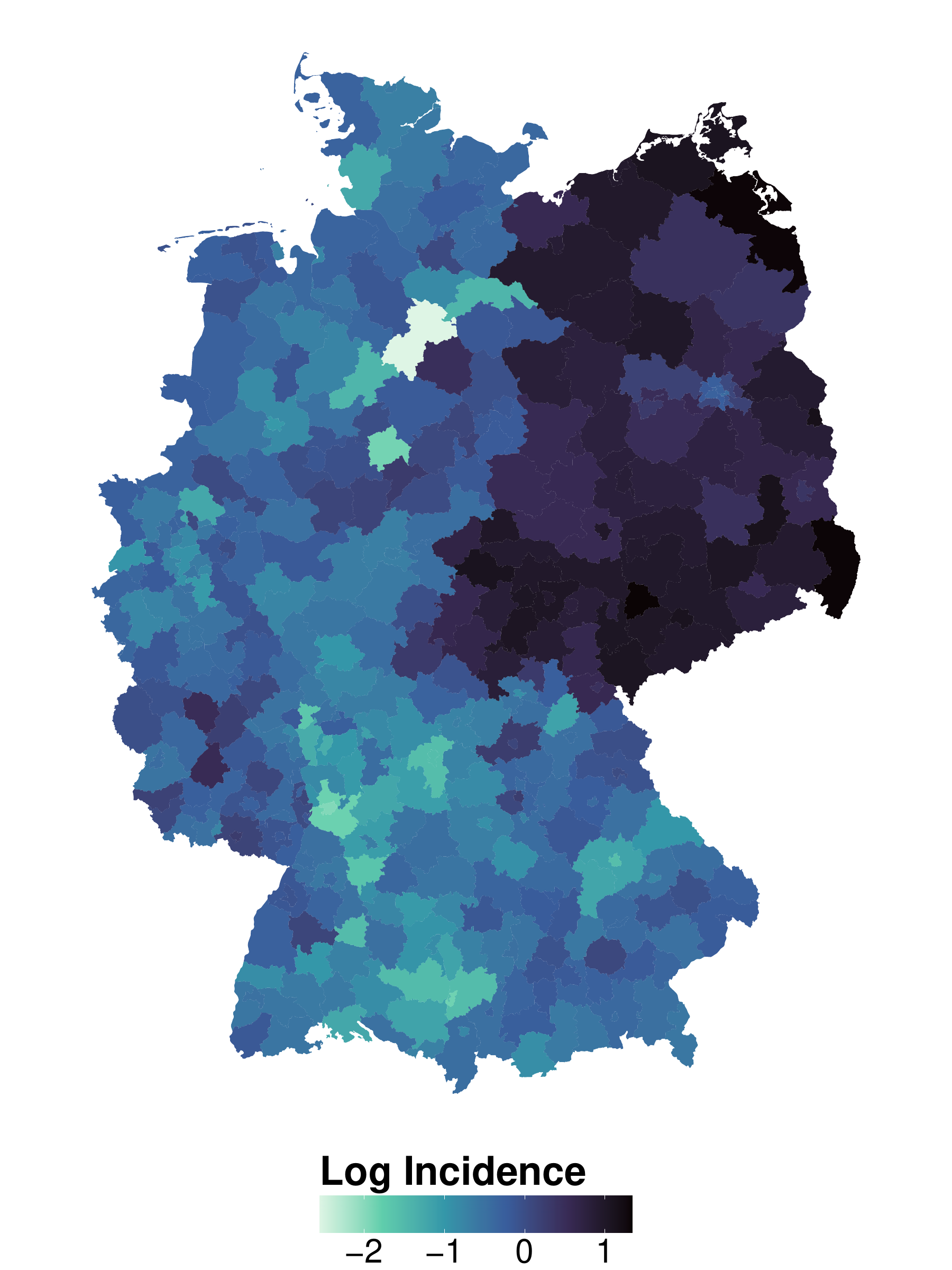}
		\caption{Log Mean Incidence of Rota Virus infections in Germany}
		\label{fig:rota_incidence}
	\end{subfigure}
	\hfill 
	\begin{subfigure}{0.42\textwidth}
		\includegraphics[width= \textwidth, keepaspectratio]{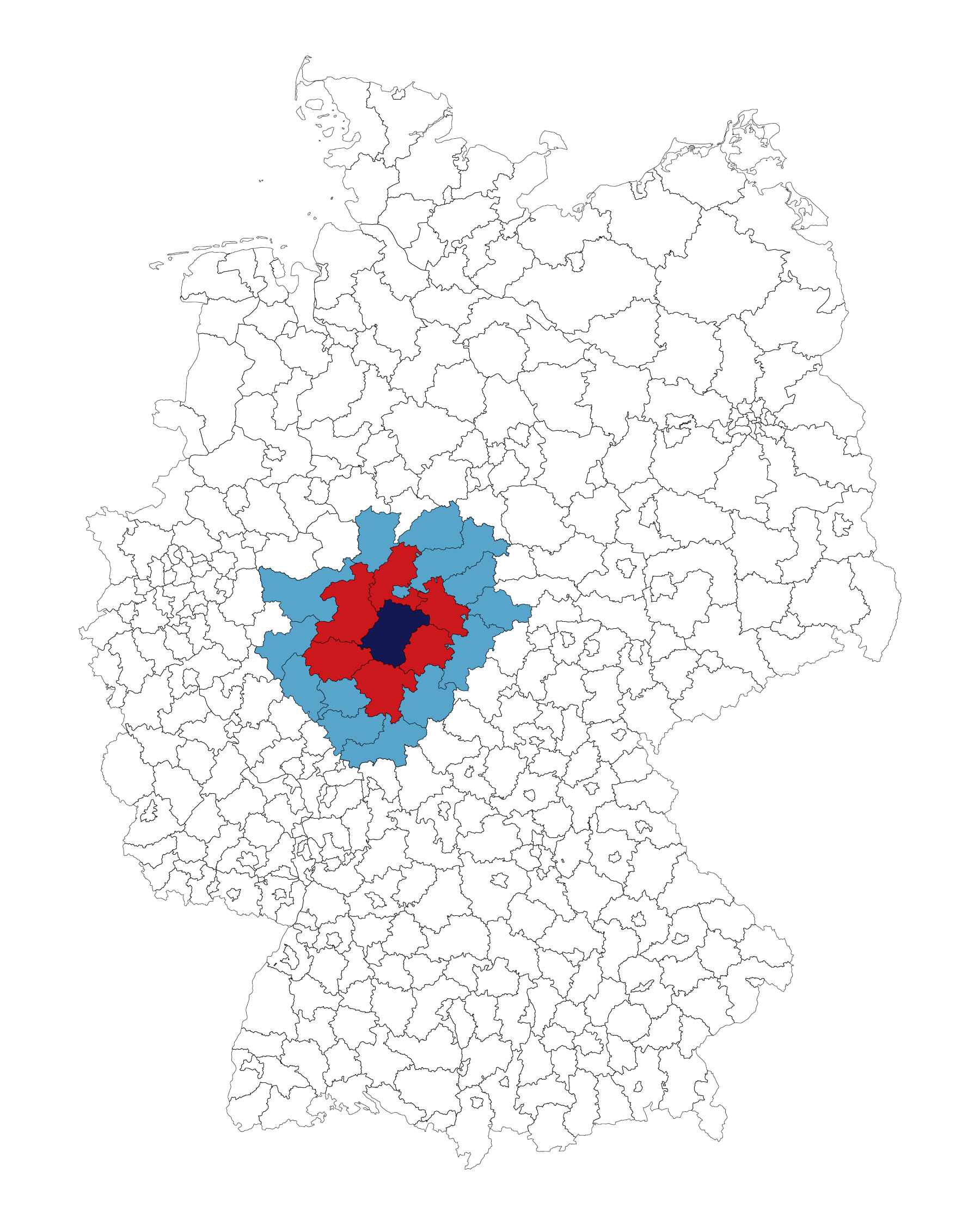}
		\caption{Neighbors up to order 2 of Schwalm Eder Kreis county.}
		\label{fig:neighbors_germany}
	\end{subfigure}
	\caption{German cities and districts.}
\end{figure}

Figure~\ref{fig:rota_incidence} displays the log-transformed mean incidence rates (infections per 100,000 inhabitants) in all regions, revealing persistently higher case numbers in former East German territories. Since July 2013, rotavirus vaccination has been officially recommended in Germany. We now reproduce the log-linear model from \citet{maletz_spatio-temporal_2024} while extending the vaccination effect to distinguish between eastern and western regions. Spatial neighborhoods are defined as in the original study: first-order neighbors share a common border, while second-order neighbors are reachable by traversing exactly one intermediate region (without being direct neighbors). Figure~\ref{fig:neighbors_germany} illustrates these neighborhood definitions for a sample region.

The original model assumes marginal Poisson distributions with linear process $\bm{\psi}_t = \bm{\mu}_t$ and the following specification:
\begin{equation}
\bm{\psi}_t = \delta_0 \bm{1}_p + \sum_{i = 1}^4 \sum_{\ell = 0}^2 \beta_{i, \ell} \bm{W}^{(\ell)}\log(\bm{Y}_{t - i} + \bm{1}_p) + \sum_{k = 1}^5 \gamma_k \bm{X}_{k, t}.
\end{equation}
We modify the global vaccination effect to estimate separate effects for eastern and western Germany. The model includes these covariates:
\begin{itemize}
    \item \textbf{Population}: log-transformed, linearly interpolated population size (per 100,000) for each region
    \item \textbf{GDR}: time-constant binary indicator for regions in former East Germany
    \item \textbf{Seasonality}: $\bm{X}_t = \cos\left(\frac{2 \pi}{52} t\right)$ and $\bm{X}_t = \sin\left(\frac{2 \pi}{52} t\right)$ for annual fluctuations
    \item \textbf{Vaccination}: interaction between the GDR indicator and a step function (1 post-vaccination recommendation)
\end{itemize}
These covariates are specified using the following code:
\begin{CodeChunk}
\begin{CodeInput}
R>  covariates <- list(
+       population = log(population_germany),
+       gdr = TimeConstant(1 * (gdr_feature > 0)),
+       season_cos = SpatialConstant(cos(2 * pi / 52 * 1:1252)),
+       season_sin = SpatialConstant(sin(2 * pi / 52 * 1:1252)),
+       vaccine_west = (gdr_feature == 0) %*% t(seq(ncol(rota)) >= 654),
+       vaccine_east = (gdr_feature > 0) %*% t(seq(ncol(rota)) >= 654))
\end{CodeInput}
\end{CodeChunk}
The model is estimated using the code shown below, with output displaying coefficient estimates for each component (intercept, autoregressive terms, covariates). Autoregressive coefficients and covariates are presented as matrices where columns represent time lags or specific covariates (e.g. \code{t\_4} for time-lag-4 effects), and rows represent spatial lags (e.g. \code{s\_1} for first-order neighbors). The results match closely those in \citet{maletz_spatio-temporal_2024}, suggesting stable dependency structures in the extended dataset. The regional vaccination effects reveal a 23.5\% reduction in western Germany ($1 - \exp(-0.2687)$) versus only 2\% in eastern regions.
\begin{CodeChunk}
\begin{CodeInput}
R> glmstarma(rota, list(past_obs = rep(2, 4)), wlist = W_germany,
+           covariates = covariates, family = vpoisson("log"))
\end{CodeInput}
\begin{CodeOutput}
Call:
glmstarma(ts = rota, model = list(past_obs = rep(2, 4)), wlist = W_germany,
    family = vpoisson("log"), covariates = covariates)

Estimated Coefficients:
Intercept:
[1] -0.754

Autoregressive Coefficients:
     t_1      t_2      t_3      t_4
s_0  0.4710   0.2163   0.1116   0.0529
s_1  0.1034   0.0026   0.0007  -0.0006
s_2  0.0238   0.0018   0.0004  -0.0003

Covariate Coefficients:
     population  gdr  season_cos  season_sin  vaccine_west  vaccine_east
s_0    0.3213   0.2655   0.0632      0.4467     -0.2687       -0.0197
\end{CodeOutput}
\end{CodeChunk}
\paragraph{Dispersion Handling}
The default Poisson assumption fixes the dispersion parameter at $\phi_{i,t} = 1$ for all $i,t$, although count data often exhibit overdispersion. Alternative count data families (\fct{vquasipoisson} or \fct{vnegative.binomial}) accommodate this:
\begin{itemize}
    \item \fct{vquasipoisson} estimates the variance as $\mathrm{Var}(Y_{i,t} \mid \mathcal{F}_{t-1}) = \phi \mu_{i,t}$
    \item \fct{vnegative.binomial} uses $\mathrm{Var}(Y_{i,t} \mid \mathcal{F}_{t-1}) = \mu_{i,t} + \phi \mu_{i,t}^2$ (with $\phi=0$ for Poisson as a special case)
\end{itemize}
All three families yield identical point estimates, differing only in standard error calculations. The sandwich estimator further reduces these differences. For this dataset, the quasipoisson dispersion is estimated at 1.804, while the negative binomial dispersion is 1.068. Model comparison via AIC favors the negative binomial family (1,429,264) over Poisson (1,608,721) and quasipoisson (1,498,655).

\paragraph{Software Comparison}
Our package design follows the user-friendly interface of \pkg{tscount} \citep{liboschik_tscount_2017} for univariate count time series, extending its INGARCH and log-linear models to the spatio-temporal domain. Alternative \R packages for spatio-temporal count data include:

\begin{itemize}
    \item \pkg{PNAR} \citep{armillotta_inference_2024}: Implements linear/log-linear Poisson Network Autoregression models, which are special cases of our framework (achievable via \code{vpoisson("identity")} or \code{vpoisson("log")}) for a fixed finite  number of spatial points. Includes test statistics for testing linearity, but restricts covariates to time-constant terms.

    \item \pkg{surveillance} \citep{meyer_spatio-temporal_2017}: Implements the \code{hhh4} framework, decomposing the conditional expectation into endemic and epidemic components:
    \begin{equation}
    \mu_{i,t} = e_{i,t}\nu_{i,t} + \lambda_{i,t} Y_{i,t-1} + \kappa_{i,t} \sum_{j \neq i} w_{ij}^{(1)} Y_{j,t-1}.
    \end{equation}
    While structurally similar to our linear models, \code{hhh4} natively supports time-varying parameters, but limits spatial dependencies to a single lag/matrix. \citet{armillotta_inference_2024} note that its standard errors may be underestimated due to ignored contemporaneous dependencies.
\end{itemize}

%%%%%%%%%%%%%%%%%%%%%%%%%%%%%%%%%%%%%%%%
\subsection{Sea Surface Temperature Anomalies}
We consider Sea Surface Temperature Anomaly (SST) data in the Pacific Ocean. The data describe (monthly) deviations (in °C) from climatology over the period from January 1970 to December 2002 and were measured on a regular grid from $120^{\circ}$E to $290^{\circ}$E ($70^{\circ}$W) longitude and $29^{\circ}$S to $29^{\circ}$N latitude in $2^{\circ}$ increments. The complete data set is included in the \R package \pkg{STRbook} \citep{wikle_spatio-temporal_2019} under the name \code{SST\_df}. We follow \citet{holleland_stationary_2020} and consider the rectangular area $29^\circ$S to $29^{\circ}$N latitude and $160^{\circ}$E to $240^{\circ}$E ($120^{\circ}$W), which does not contain land masses as there are no measurements available.

\begin{figure}[t]
    \centering
    \includegraphics[width=\linewidth, keepaspectratio]{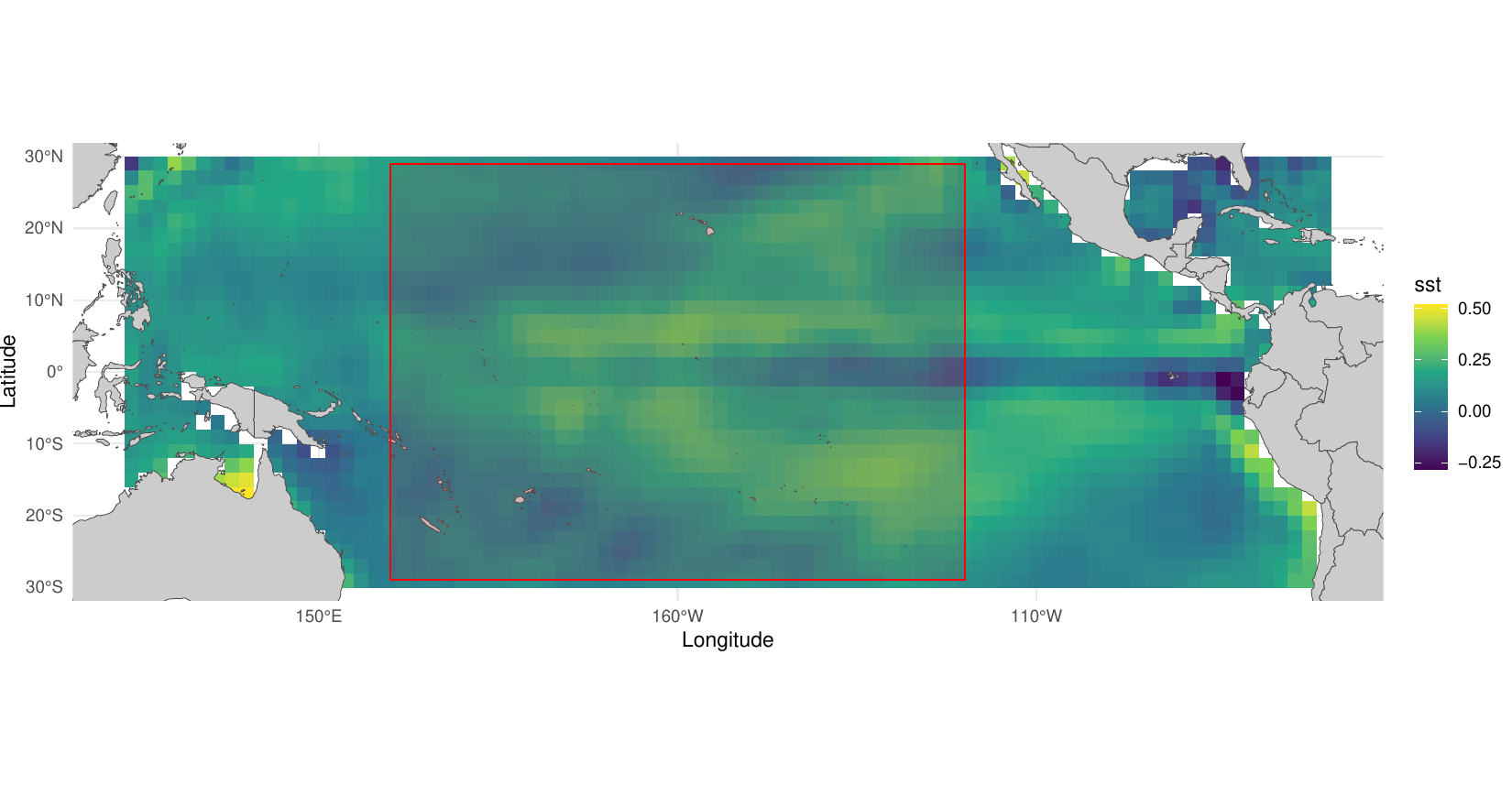}
    \caption{Pixel-wise mean of sea surface temperature anomalies (in °C) in the Pacific. The red rectangle, $29^\circ$S to $29^{\circ}$N latitude and $160^{\circ}$E to $240^{\circ}$E ($120^{\circ}$W), marks the area modeled in this data example.}
    \label{fig:sst_area}
\end{figure}

Figure~\ref{fig:sst_area} shows the area we consider for modeling, as well as the coordinate-wise mean values of the SST variable. In total, we use monthly observations of this $30 \times 41$ grid ($p = 1230$) for the time period from January 1970 to December 2002 ($T = 396$). We observe minor temperature anomalies on average near the equator, which increase until $6^{\circ}$N and $6^\circ$S and then decrease again. In the eastern direction, i.e., with increasing longitude, we also observe higher anomalies than in the western part of the area under consideration. Based on these observations, we construct several covariates that we account for in the modeling: (1) a linear temporal trend with values between 0 (beginning of the observation period) and 1 (end) to capture a potential increase in temperature, (2) longitude/360 to describe higher temperatures further east, (3) $\cos(2\pi/12 \cdot t)$ and $\sin(2\pi/12 \cdot t)$ to capture potential seasonality, and (4) two covariates $\min\lbrace |\text{latitude}|, 6 \rbrace / 90$ and $\max\lbrace |\text{latitude}| - 6, 0 \rbrace / 90$ to describe the increasing temperature anomalies up to the 6th degree and the decline beyond that in a piecewise linear manner.

With the following \R code we specify these covariates for use in the modeling functions. We utilize the helper function \fct{SpatialConstant} to define a covariate that varies only temporally. This is particularly intended for temporal trends or seasonality. The helper function \fct{TimeConstant} defines a covariate that remains constant over the entire observation period and varies only spatially. Such covariates can describe effects that are generated exclusively by the location of the measurement site.
\begin{CodeChunk}
\begin{CodeInput}
R> times <- seq(from = as.Date("1970-01-01"), 
+               to = as.Date("2002-12-01"), by = "m")
R> times <- format(times, "%b %Y")
R> covariates <- list(
+    trend = SpatialConstant(seq(times) / length(times)),
+    longitude = TimeConstant(locations$lon / 360),
+    season_cos = SpatialConstant(cos(2 * pi / 12 * seq(times))),
+    season_sin = SpatialConstant(sin(2 * pi / 12 * seq(times))),
+    abs_lat_inc = TimeConstant(pmin(abs(locations$lat), 6) / 90),
+    abs_lat_dec = TimeConstant(pmax(abs(locations$lat) - 6, 0) / 90))
\end{CodeInput}
\end{CodeChunk}

We consider a model under the marginal normality assumption, i.e., $Y_{i, t} \mid \mathcal{F}_{t - 1} \sim \mathcal{N}(\mu_{i, t}, \phi_{i, t})$ and $\mathrm{Var}(Y_{i, t} \mid \mathcal{F}_{t - 1}) = \phi_{i, t}$. We first use the mean model
$$ \bm{\mu}_{t} = \delta_0\bm{1}_{1230} + \left(\beta_0 \bm{I}_{1230} + \beta_{\mathrm{N}}\bm{W}^{(\mathrm{N})} + \beta_{\mathrm{E}}\bm{W}^{(\mathrm{E})} + \beta_{\mathrm{S}}\bm{W}^{(\mathrm{S})} + \beta_{\mathrm{W}}\bm{W}^{(\mathrm{W})} \right) \bm{Y}_{t - 1} + \sum_{k = 1}^6 \gamma_{k} \bm{X}_{k, t}, $$
where $\bm{W}^{(\ell)}, \ell \in \lbrace \mathrm{N}, \mathrm{E}, \mathrm{S}, \mathrm{W} \rbrace$ indicates row-wise for each location exactly the adjacent pixel/coordinate point to the north, east, south, and west, respectively. In each row, there is exactly one 1 and otherwise only 0s. For coordinates located at the edges of the grid shown in Figure~\ref{fig:sst_area}, such a neighbor does not necessarily exist, i.e., some rows contain only zeros. This violates the condition $\bm{W}^{(\ell)} \bm{1}_p = \bm{1}_p$ mentioned earlier in this work. However, due to the very high number of locations $p = 1230$ and the relatively small number of boundary points (138), the impact of this is negligible. The cyclic neighborhood matrices used by \citet{holleland_stationary_2020} would, in our opinion, introduce a bias into the analysis, as they would treat locations in the far north of the grid as direct neighbors of points on the southern edge of the grid.

We model the variance process $\bm{\phi}_t$ on a log-linear scale to capture negative dependencies as well. We assume $\bm{\zeta}_t = \log(\bm{\phi}_t)$ and
\begin{equation*}
\begin{aligned}
\bm{\zeta}_t
={}& \delta_0\bm{1}_{1230} \\
&+ \Bigl(
    \beta_0 \bm{I}_{1230}
    + \beta_{\mathrm{N}}\bm{W}^{(\mathrm{N})}
    + \beta_{\mathrm{E}}\bm{W}^{(\mathrm{E})}
    + \beta_{\mathrm{S}}\bm{W}^{(\mathrm{S})}
    + \beta_{\mathrm{W}}\bm{W}^{(\mathrm{W})}
    \Bigr)
\log\!\left((\bm{Y}_{t-1}-\bm{\mu}_t)^2+\bm{1}_p\right) \\
&+ \sum_{k=1}^{6}\gamma_k\bm{X}_{k,t},
\end{aligned}
\end{equation*}
i.e., we assume the same model order as in the mean model and use the same covariates already specified above. The model structure of the dispersion model is closely related to that of a spatio-temporal log-GARCH model with additional covariates \citep{otto_spatial_2025}. Model estimation is performed using the code below.
\begin{CodeChunk}
\begin{CodeInput}
R>  fit1 <- dglmstarma(SST, mean_model = list(past_obs = 4), 
+                    dispersion_model = list(past_obs = 4),
+                    mean_family = vnormal(), 
+                    dispersion_link = "log",
+                    wlist = W_directed, 
+                    mean_covariates = covariates, 
+                    dispersion_covariates = covariates)
\end{CodeInput}
\end{CodeChunk}

With \code{summary(fit1)}, we can display a summary of the output. For the parameters of the mean model, we detect spatial dependence. The parameter associated with $\bm{W}^{(\mathrm{E})}$, labelled  \code{past\_obs\_\{s\_2, t\_1\}} in the output, is estimated to be significantly larger than the other spatial dependence parameters, indicating a stronger dependence from the eastern direction than from other directions. This can be explained by the northern and southern equatorial currents, which are driven by trade winds and transport water masses in a westward direction, resulting in stronger spatial dependence from the east. Additionally, dependencies from other directions are also significant, which can be explained by other currents that are, however, dominated by the equatorial currents on a global scale relative to the grid section. Furthermore, the mean model shows a positive temporal trend, which is likely attributable to rising water temperatures due to climate change. The other covariates in the mean model are not significant.

In the dispersion model, we also detect spatio-temporal autoregressive dependencies. These are primarily driven by dependencies on past values of the pseudo-observations from the north (\code{past\_obs\_\{s\_1, t\_1\}}) and south (\code{past\_obs\_\{s\_3, t\_1\}}), as well as from the location itself (\code{past\_obs\_\{s\_0, t\_1\}}). Spatio-temporal dependencies from the east and west are not significant at the 5\% level in the dispersion model. Additional to the autoregressive dependencies, a negative trend is observable in the dispersion, suggesting that the variance decreases over time. Additionally, there is an annual seasonality in the variance. Longitude does not appear to have an effect on either the mean or the variance. Latitude leads to a decrease in variance up to the 6th degree north or south and then to an increase again. This suggests a higher concentration of high or low anomalies in this region, possibly an artefact generated by equatorial currents.
\begin{CodeChunk}
\begin{CodeInput}
R>  summary(fit1)
\end{CodeInput}

\begin{CodeOutput}
Call:
dglmstarma(ts = SST, mean_model = list(past_obs = 4), 
	dispersion_model = list(past_obs = 4), 
	mean_family = vnormal(), wlist = W_directed, 
	mean_covariates = covariates, 
	dispersion_covariates = covariates)

==================================================
            Coefficients of Mean Model
==================================================
                     Estimate Std. Error z value Pr(>|z|)    
(Intercept)         -0.102614   0.056218  -1.825  0.06796 .  
past_obs_{s_0, t_1}  0.150587   0.021989   6.848 7.47e-12 ***
past_obs_{s_1, t_1}  0.146773   0.014122  10.393  < 2e-16 ***
past_obs_{s_2, t_1}  0.262654   0.010661  24.636  < 2e-16 ***
past_obs_{s_3, t_1}  0.152767   0.015036  10.160  < 2e-16 ***
past_obs_{s_4, t_1}  0.111290   0.013089   8.503  < 2e-16 ***
trend_{s_0}          0.106865   0.033320   3.207  0.00134 ** 
longitude_{s_0}      0.126819   0.080194   1.581  0.11378    
season_cos_{s_0}    -0.007730   0.007681  -1.006  0.31422    
season_sin_{s_0}    -0.011069   0.009553  -1.159  0.24657    
abs_lat_inc_{s_0}    0.269895   0.264460   1.021  0.30747    
abs_lat_dec_{s_0}   -0.089160   0.058985  -1.512  0.13064    
---
Signif. codes:  0 ‘***’ 0.001 ‘**’ 0.01 ‘*’ 0.05 ‘.’ 0.1 ‘ ’ 1

Marginal Distribution: gaussian
Link: identity

==================================================
         Coefficients of Dispersion Model
==================================================
                      Estimate Std. Error z value Pr(>|z|)    
(Intercept)          -1.030314   0.117310  -8.783  < 2e-16 ***
past_obs_{s_0, t_1}   0.561574   0.124929   4.495 6.95e-06 ***
past_obs_{s_1, t_1}   0.206228   0.080298   2.568   0.0102 *  
past_obs_{s_2, t_1}   0.022348   0.075720   0.295   0.7679    
past_obs_{s_3, t_1}   0.201855   0.079095   2.552   0.0107 *  
past_obs_{s_4, t_1}   0.007037   0.071090   0.099   0.9211    
trend_{s_0}          -0.677960   0.063748 -10.635  < 2e-16 ***
longitude_{s_0}       0.127095   0.225730   0.563   0.5734    
season_cos_{s_0}      0.148817   0.026095   5.703 1.18e-08 ***
season_sin_{s_0}      0.032206   0.026243   1.227   0.2197    
abs_lat_inc_{s_0}   -17.334921   1.279691 -13.546  < 2e-16 ***
abs_lat_dec_{s_0}     2.316219   0.242909   9.535  < 2e-16 ***
---
Signif. codes:  0 ‘***’ 0.001 ‘**’ 0.01 ‘*’ 0.05 ‘.’ 0.1 ‘ ’ 1
Link: log
Dispersion Parameter of dispersion model fixed at 2

Number of coefficients of mean model: 12
Number of coefficients of dispersion model: 12

Quasi-Log-Likelihood: -210434
AIC: 420916
BIC: 421182.3
QIC: 424241.9
\end{CodeOutput}
\end{CodeChunk}
\paragraph{Discussion}
Another R package for modeling spatio-temporal data under the assumption of marginal normal distributions is the \pkg{starma} package \citep{cheysson_starma_2016}, which provides an implementation of the STARMA processes of the same name by \citet{pfeifer_three-stage_1980}. The STARMA-GARCH combination introduced by \citet{holleland_stationary_2020} is implemented in the \R package \pkg{starmagarch} and available on GitHub (\url{https://github.com/holleland/starmagarch/}).

Both models describe the data solely through autoregressive dependence structures, meaning that location-specific differences cannot be easily described through covariates. Furthermore, the data must be detrended beforehand. The theoretical properties of these models are based, among other things, on spatial stationarity assumptions. The applicability of these models therefore requires several complex preprocessing steps, which additionally compromise interpretability.

The \pkg{starma} package also works exclusively with dense spatial weight matrices. This is particularly problematic in high-dimensional cases such as in this application, when the weight matrices are only sparsely populated. This drastically increases computation time. On an Apple MacBook Air with M2 processor, the computation of the DGLM using the \fct{dglmstarma} call above takes approximately 50 seconds. The estimation of a first-order STAR process with the same neighborhood matrices, i.e., $$ \bm{Y}_{t} = \left(\beta_0 \bm{I}_{1230} + \beta_{\mathrm{N}}\bm{W}^{(\mathrm{N})} + \beta_{\mathrm{E}}\bm{W}^{(\mathrm{E})} + \beta_{\mathrm{S}}\bm{W}^{(\mathrm{S})} + \beta_{\mathrm{W}}\bm{W}^{(\mathrm{W})}\right) \bm{Y}_{t - 1} + \bm{\varepsilon}_t$$ takes approximately 250 seconds with the \pkg{starma} package, making it more suitable for low-dimensional data.

A more practical alternative is provided by the Generalized Network Autoregression (GNAR) models \citep{knight_modelling_2016}, as well as their extension to include exogenous covariates (GNARX) \citep{nason_quantifying_2022}. Both models are implemented in the \R package \pkg{GNAR} \citep{knight_generalized_2020}.

In the functions of this \R package, users only need to provide a graph that defines the spatial neighbors. Higher-order neighbors are automatically calculated if required. On the one hand, this has the advantage that users only need to consider the first-order dependence structure. On the other hand, users may not be aware of what types of spatial dependencies are implied when they use this feature unknowingly. Furthermore, it is not possible to estimate different dependencies for different directions, as in the example above, since only a single network can be passed.

The GNAR framework provides the possibility to model time-varying networks, which is currently not possible within our framework. Additionally, GNAR models can easily estimate individual autoregressive dependence parameters for each location. In the \pkg{glmSTARMA} package this can only be realized by passing a large list of neighborhood matrices $\bm{W}^{(i)}, i = 1, \ldots, p$, where the $i$-th diagonal element of $\bm{W}^{(i)}$ equals 1 and all other values in the matrix equal 0. The spatial dependence matrices are then appended to this list, and the spatial orders must be chosen accordingly.

For a comparison, we now estimate a simple GNARX model of the form
$$ \bm{Y}_t = \beta_0 \bm{Y}_{t - 1} + \beta_1 \bm{W} \bm{Y}_{t - 1} + \sum_{k = 1}^7 \gamma_k \bm{X}_{k, t} + \bm{\varepsilon}_t, $$
where $\bm{W}$ is the row-normalized adjacency matrix and $\bm{\varepsilon}_t \sim \mathcal{N}(0, \sigma^2 \bm{I})$ are independent and identically distributed. A coordinate is considered adjacent, as in the example above, if it is located directly to the north, east, south, or west of a point. The covariates are identical to those from the example above. The additional seventh covariate is the intercept, which is not automatically fitted.

Additionally, we estimate a similar mean model with \fct{glmstarma} under the assumption of marginal normal distributions, i.e.,
$$ \bm{\mu}_t = \delta_0 \bm{1}_p + \beta_0 \bm{Y}_{t - 1} + \beta_1 \bm{W} \bm{Y}_{t - 1} + \sum_{k = 1}^6 \gamma_k \bm{X}_{k, t}. $$
The estimation of the GNARX model takes approximately 15 seconds on a MacBook Air M2, while the \code{glmstarma} model requires approximately 1.7 seconds. Table~\ref{tab:sst_estimate} contains the parameter estimates of both models, as well as the standard errors calculated by supplementary calculations. Both models estimate the parameters of the covariates almost identically. Differences arise in $\beta_0$ and $\beta_1$. We enforce a stationary solution in \fct{glmstarma}, while the \pkg{GNAR} package estimates the parameters using the \fct{lm} function. Furthermore, the standard errors we calculate are larger than those of the GNARX model, which is because we do not assume independent and identically distributed errors, but also account for contemporaneous dependencies.

\begin{table}[t]
	\centering
	\resizebox{\textwidth}{!}{%
	\begin{tabular}{llrrrrrrrrr}
		\toprule
		\multicolumn{2}{c}{Model} & Intercept & $\beta_0$ & $\beta_1$ & $\gamma_1$ & $\gamma_2$ & $\gamma_3$ & $\gamma_4$ & $\gamma_5$ & $\gamma_6$ \\
		\midrule
		\multirow{2}{*}{GNARX} & Est. & $-0.0899$ & $-0.1603$ & $0.9749$ & $0.1080$ & $0.0874$ & $-0.0069$ & $-0.0084$ & $0.4124$ & $-0.0964$ \\
		& Std. Err. & $0.0053$ & $0.0100$ & $0.0104$ & $0.0020$ & $0.0085$ & $0.0008$ & $0.0008$ & $0.0419$ & $0.0075$ \\
		\midrule
		\multirow{2}{*}{\code{glmstarma}} & Est. & $-0.0889$ & $-0.0937$ & $0.9062$ & $0.1086$ & $0.0867$ & $-0.0068$ & $-0.0086$ & $0.4025$ & $-0.0974$ \\
		& Std. Err. & $0.0406$ & $0.0356$ & $0.0393$ & $0.0185$ & $0.0711$ & $0.0069$ & $0.0070$ & $0.2674$ & $0.0541$ \\
		\bottomrule
	\end{tabular}
	}
	\caption{Parameter estimates and standard errors of GNARX- and a \code{glmstarma} process.}
	\label{tab:sst_estimate}
\end{table}

%% file: sections/summary.tex
\section{Conclusion}

While existing R packages for spatio-temporal modeling provide essential tools, they often rely on restrictive assumptions, such as contemporaneous independence or strictly linear dependencies, that may not be justified in real-world applications. These simplifications often lead to underestimated standard errors and biased model selection, undermining the reliability of applied research. Our package addresses these limitations by offering a unified, user-friendly framework designed for flexible spatio-temporal modeling under diverse distributional assumptions. By supporting both linear and non-linear structures with customizable spatial and temporal lags, the package allows caputuring complex dependencies that are often overlooked by more rigid software implementations.

A key innovation of our approach is the inclusion of time-varying dispersion parameters, a feature that allows the model to adapt to fluctuating volatility within the data over time. To ensure statistical integrity, we utilize robust sandwich estimators that provide reliable inference even when traditional assumptions fail. This focus on methodological rigor is paired with a modular design, ensuring that the package remains adaptable as analytical needs evolve.

%% file: sections/appendix.tex
% Theoretical properties
\section*{Appendix}
\markboth{Appendix}{Appendix}

In the following sections of the Appendix, we present further examples and discuss results that expand and supplement the main part. First,  in Section~\ref{sec:asymptotics}, we discuss theoretical properties of the model framework and parameter estimation. In Section~\ref{appendix:control} we explain additional control parameters of the estimation functions from Subsection~\ref{sec:estimation} from the main part. In Section~\ref{appendix:simulation_example}, we illustrate usage of the simulation functions of the package using two examples. We support our claims about the theoretical properties with simulation results in Section~\ref{appendix:simulations}.

\section{Asymptotics and Stability}\label{sec:asymptotics}

In time series analysis, stationary and ergodic processes are often required to establish consistent and (asymptotically) normally distributed parameter estimates. Introducing time-varying deterministic covariate processes directly violates the stationarity of the process to be modeled. Nevertheless, we conjecture that consistent and asymptotically normal distributed parameter estimates can be obtained under weak assumptions. In Subsection~\ref{appendix_subsec:stability} we discuss the stability of processes. Subsection~\ref{sec:theory} discusses our conjecture about the asymptotic results of the parameter estimation procedure.

\subsection{Some Heuristics on Model Stability}\label{appendix_subsec:stability}

To the best of our knowledge, there are currently no studies in the field of multivariate time series analysis that jointly model the mean and the dispersion as both time- and dimension-dependent. Therefore, the theoretical properties of such models have yet to be investigated in more detail.
In the univariate case, it is common practice to combine an ARMA process with a GARCH process; see \citet[Sec. 7]{francq_garch_2019}, or generalizations thereof within the GARMA-GARCH framework; see \citet{zheng_generalized_2022}. Specifically, for count data, univariate INGARCH processes with time-varying dispersion have been studied by \citet{barretosouza_timevarying_2025} under the negative binomial assumption. 

A necessary condition for the stationarity of the process $\lbrace \bm{Y}_t \rbrace$ is the stationarity of $\lbrace\bm{\mu}_t\rbrace$ and $\lbrace\bm{\phi}_t \rbrace$. This is generally only guaranteed if (a) no covariates are present, or (b) all covariate processes in \eqref{eq:model} and \eqref{eq:dispersionmodel} are time-invariant, or (c) (strictly) stationary; see \citet{aknouche_count_2021}. Furthermore, the existence of higher order moments might be necessary.

To ensure the stability of the process, it is important to prevent the individual components of the processes of~\eqref{eq:model} and \eqref{eq:dispersionmodel} from exploding, i.e., suddenly growing in an uncontrolled manner. This means that the functions $h_{\mu}$, $\tilde{h}_{\mu}$ and $h_{\phi}$, $\tilde{h}_{\phi}$ in \eqref{eq:model} and \eqref{eq:dispersionmodel}, used to transform past values of the feedback- or dispersion process and past (pseudo-)observations, should be \emph{well-behaved}.

We conjecture that an appropriate assumption for this is that the functions are twice continuously differentiable and do not increase too fast. Heuristically, the functions $h_{\mu}$ and $h_{\phi}$ shall fulfil $|h(x)| \leq c|x| + d$ for constants $c, d > 0$, which implies linear growth at most. The functions $\tilde{h}_{\mu}$ and $\tilde{h}_{\phi}$ shall behave asymptotically the same as the link functions $g_{\mu}$ and ${g}_{\phi}$, which we express as $|\tilde{h}(g^{-1}(x))| \leq \tilde{c}|x| + \tilde{d}$ for some $\tilde{c}, \tilde{d} > 0$. Not that this is fulfilled by all functions implemented in the package; see Table~\ref{tab:distributions}.

Recall \eqref{eq:model} but without covariate terms. We assume that with the previous assumptions the condition
\begin{align}
	\label{eq:stationarity}
	\sum_{i \in \mathcal{Q}^{(\mu)}} \sum_{\ell \in \mathcal{A}_i^{(\mu)}}|\alpha_{i\ell}| + \sum_{j \in \mathcal{R}^{(\mu)}} \sum_{\ell \in \mathcal{B}_j^{(\mu)}}|\beta_{j\ell}| < 1,
\end{align}
together with the identical condition in terms of $\tilde{\alpha}_{i, \ell}$ and $\tilde{\beta}_{i, \ell}$  of~\eqref{eq:dispersionmodel}, are sufficient to guarantee stationarity and ergodicity, as well as the existence of all moments of $\lbrace \bm{Y}_t \rbrace$. Similar conditions hold for univariate time-series models, such as GARCH and INGARCH processes, as well as spatio-temporal models for counts \citep{armillotta_count_2024, maletz_spatio-temporal_2024} and generalized network models \citep{knight_generalized_2020}.

However, these conditions might be too strict. For the Softplus and Softclipping processes, see \citet{jahn_approximately_2023}, weaker bounds can be achieved but at the price of assuming  constant dispersion parameters, i.e.,
\begin{align}
	\label{eq:stationarity_jahn}
	\sum_{i \in \mathcal{Q}} \sum_{\ell \in \mathcal{A}_i}\max\lbrace 0, \alpha_{i\ell}\rbrace + \sum_{j \in \mathcal{R}} \sum_{\ell \in \mathcal{B}_j}\max \lbrace 0, \beta_{j\ell} \rbrace < 1 \quad \text{and} \quad \sum_{i \in \mathcal{Q}} \sum_{\ell \in \mathcal{A}_i} |\alpha_{i\ell}| < 1.
\end{align}

Under stationarity assumptions, the expectation of the process $\bm{\psi}_t$ is approximately given by
\begin{align}
	\label{eq:expectation}
	\mathbb{E}(\bm{\psi}_t) \approx \left(\bm{I}_p - \sum_{i \in \mathcal{Q}} \sum_{\ell \in \mathcal{A}_i}\alpha_{i\ell}\bm{W}_{\alpha}^{(\ell)} + \sum_{j \in \mathcal{R}} \sum_{\ell \in \mathcal{B}_j}\beta_{j\ell} \bm{W}_{\beta}^{(\ell)} \right)^{-1}\bm{\delta}^{\star}
\end{align}
and $\mathbb{E}(\bm{Y}_{t}) \approx g^{-1}(\mathbb{E}(\bm{\psi}_t))$. Here, $\bm{\delta}^{\star}$ denotes the intercept plus the expected values of the covariate effects, if any. These formulas also hold analogously for $\mathbb{E}(\bm{\zeta}_t)$ and $\mathbb{E}(\bm{d}_t)$.

\subsection{Theoretical Properties}\label{sec:theory}
\citet{smyth_generalized_1989} argues that the parameter vectors $\bm{\vartheta}_{\mu}$ and $\bm{\vartheta}_{\phi}$ are orthogonal and can therefore be estimated separately. We assume that this property carries over to our context, even though \citet{smyth_generalized_1989} assumes independent observations.
%However, these arguments assume independent observations. We assume that this property carries over to our context, even though \citet{smyth_generalized_1989} assumes independent observations.

Consistent parameter estimation requires certain conditions. First, the model parameters must satisfy the conditions \eqref{eq:stationarity} or \eqref{eq:stationarity_jahn}. Second, covariate processes must be bounded with finite variance. Specifically, for deterministic covariates, we require the existence of $c > 0$ such that $\Vert\mathbf{X}_{k, t}\Vert_2^2 < c$ for all $t$ and $k$. For stochastic covariates, we must have $\mathbb{E} \left[\Vert\mathbf{X}_{k, t}\Vert_2^2 \right] < c$ for all $t$ and $k$.

Additionally, as the sample size increases, the information provided by the covariates must grow sufficiently fast, compare, for example, \citet{fahrmeir_consistency_1985} or \citet{chen_strong_1999}. This can be guaranteed by the condition
\begin{align}
	\label{eq:information_gain}
	\frac{\overline{\sigma_T}^{0.5} \log(\overline{\sigma_T})^{0.5\varepsilon}}{\underline{\sigma_T}} \to 0 \quad (T \to \infty),
\end{align}
where $\overline{\sigma_T}$ and $\underline{\sigma_T}$ are the largest and smallest eigenvalue of the matrix 
\[
\bm{F}_T = \sum_{t = 1}^T \mathbf{\tilde{X}}_t^\top \mathbf{\tilde{X}}_t, \quad \text{with} \quad
\mathbf{\tilde{X}}_t = \left(\mathbf{1}_p, \mathbf{W}^{(0)}\mathbf{X}_{1, t}, \ldots, \mathbf{W}^{(s_m)}\mathbf{X}_{m, t} \right).
\]
This condition should be satisfied for the covariates of both the mean model \eqref{eq:model} and the dispersion model \eqref{eq:dispersionmodel}. Under these assumptions, we conjecture that the estimators $\bm{\hat{\vartheta}}_{\mu}$ and $\bm{\hat{\vartheta}}_{\phi}$ are asymptotically normal:
\begin{align}
	\label{eq:asymptotics}
	\sqrt{T}(\bm{\hat{\vartheta}} - \bm{\vartheta}_0) \to \mathcal{N}(0, \mathbf{G}^{-1} \mathbf{H} \mathbf{G}^{-1}) \quad \text{as} \quad T \to \infty,
\end{align}
where $\bm{\vartheta}_0$ denotes the true parameter vector for the mean or the dispersion model. The asymptotic variance takes a sandwich form, with matrices (for the mean model):
\begin{align}
	\label{eq:variance_matrices}
	\mathbf{G}(\bm{\vartheta}) &= \mathbb{E}\left[ \frac{\partial \bm{\psi}_t^\top}{\partial \bm{\vartheta}} \mathbf{\tilde{D}}_t(\bm{\vartheta}) \frac{\partial \bm{\psi}_t}{\partial \bm{\vartheta}} \right], \qquad
	\mathbf{H}(\bm{\vartheta}) = \mathbb{E}\left[\frac{\partial \bm{\psi}_t^\top}{\partial \bm{\vartheta}} \mathbf{D}_t(\bm{\vartheta}) \mathbf{\Sigma}_t(\bm{\vartheta}) \mathbf{D}_t(\bm{\vartheta}) \frac{\partial \bm{\psi}_t}{\partial \bm{\vartheta}} \right],
\end{align}
which can be approximated empirically using $\bm{\hat{\vartheta}}$. For the dispersion model, the analogous matrices are computed using $\bm{\zeta}_t$ instead of $\bm{\psi}_t$ and the pseudo-observations $\bm{d}_t$. Recall from Section~\ref{sec:estimation} that $$\mathbf{D}_t = \mathrm{diag}((g^{-1})'(\psi_{i,t}) / \sigma_{i,t}^2, i = 1, \ldots, p) = \mathrm{diag}(1 / g'(g^{-1}(\psi_{i,t})) \sigma_{i,t}^2, i = 1, \ldots, p),$$  
with $\sigma_{i,t}^2 = \phi_{i, t} \cdot A''_i(\theta_{i,t}(\bm{\vartheta}^{(\mu)})) = \phi_{i, t} V(\mu_{i, t})$ the conditional variance estimation and
$$\mathbf{\tilde{D}}_t = \mathrm{diag}(((g^{-1})'(\psi_{i,t}))^2 / \sigma_{i,t}^2, i = 1, \ldots, p).$$
In the R package the two matrices $\mathbf{G}(\bm{\vartheta})$ and $\mathbf{H}(\bm{\vartheta})$ are replaced by their empirical counterparts, that is $\bm{G}(\bm{\vartheta})$ is estimated by the expected hessian, see the next subsection, and $\bm{H}$ is calculated based on the score.

These  asymptotic results hold only if the true parameters lie in the interior of the parameter space. If inference is based on such results and some parameters lie on the boundary (e.g., due to non-negativity constraints, cf. Table~\ref{tab:distributions}), then test statistics need to be adjusted \citep[cf.][]{francq_garch_2019}. For example, in one-sided tests such as $\text{H}_0: \vartheta_i = 0$ versus $\text{H}_1: \vartheta_i \neq 0$, doubling the local significance level suffices. The \fct{summary} function adjusts the reported $p$-values accordingly.

\subsubsection{Derivation of the  information matrix}\label{sec:information}
The Hessian is obtained with the common derivation rules as $\bm{H}_T(\bm{\theta}) = \frac{\partial^2 \ell}{\partial \bm{\theta} \partial \bm{\theta}^{\mathrm
		T}} = \frac{\partial S_T(\bm{\theta})}{\partial \bm{\theta}^{\mathrm{T}}}$. For the Hessian we define $\bm{\tilde{D}}_t = \mathrm{diag}\left\lbrace\left[g'(g^{-1}(\psi_{i, t}))\right]^2 g^{-1}(\psi_{i, t}), i = 1, \ldots, n\right\rbrace$.

\begin{align*}
	\bm{H}_T(\bm{\theta}) &= \sum_{t = 1}^T \sum_{i = 1}^n \frac{\partial}{\partial \bm{\theta}^{\mathrm{T}}} \left[ \frac{\partial \psi_{i, t}}{\partial \bm{\theta}} \frac{y_{i, t} - g^{-1}(\psi_{i, t})}{g'(g^{-1}(\psi_{i, t})) g^{-1}(\psi_{i, t})}\right] \\
	&= \underbrace{\sum_{t = 1}^T \sum_{i = 1}^n \frac{\partial^2 \psi_{i, t}}{\partial \bm{\theta} \partial \bm{\theta}^{\mathrm{T}}} \frac{y_{i, t} - g^{-1}(\psi_{i, t})}{g'(g^{-1}(\psi_{i, t})) g^{-1}(\psi_{i, t})}}_{(\star)} + \sum_{t = 1}^T \sum_{i = 1}^n \frac{\partial \psi_{i, t}}{\partial \bm{\theta}} \frac{\partial}{\partial\bm{\theta}^{\mathrm{T}}} \left[\frac{y_{i, t} - g^{-1}(\psi_{i, t})}{g'(g^{-1}(\psi_{i, t})) g^{-1}(\psi_{i, t})}\right] \\
	&= (\star) + \sum_{t = 1}^T \sum_{i = 1}^n \frac{\partial \psi_{i, t}}{\partial \bm{\theta}} \frac{\frac{\partial}{\partial\bm{\theta}^{\mathrm{T}}} (y_{i, t} - g^{-1}(\psi_{i, t}))\cdot g'(g^{-1}(\psi_{i, t}))g^{-1}(\psi_{i, t}) }{\left[g'(g^{-1}(\psi_{i, t}))g^{-1}(\psi_{i, t})\right]^2} \\
	&\qquad- \sum_{t = 1}^T \sum_{i = 1}^n \frac{\partial \psi_{i, t}}{\partial \bm{\theta}} \frac{(y_{i, t} - g^{-1}(\psi_{i, t})) \frac{\partial}{\partial\bm{\theta}^{\mathrm{T}}} \left(g'(g^{-1}(\psi_{i, t}))g^{-1}(\psi_{i, t})\right)}{\left[g'(g^{-1}(\psi_{i, t}))g^{-1}(\psi_{i, t})\right]^2} \\
	&= (\star) - \sum_{t = 1}^T \sum_{i = 1}^n \frac{\partial \psi_{i, t}}{\partial \bm{\theta}} \cfrac{\frac{g'(g^{-1}(\psi_{i, t}))g^{-1}(\psi_{i, t})}{g'(g^{-1}(\psi_{i, t}))}}{\left[g'(g^{-1}(\psi_{i, t}))g^{-1}(\psi_{i, t})\right]^2} \frac{\partial \psi_{i, t}}{\partial \bm{\theta}^{\mathrm{T}}} \\
	&\qquad- \sum_{t = 1}^T \sum_{i = 1}^n \frac{\partial \psi_{i, t}}{\partial \bm{\theta}} \cfrac{(y_{i, t} - g^{-1}(\psi_{i, t}))\left[\frac{g''(g^{-1}(\psi_{i, t}))}{g'(g^{-1}(\psi_{i, t}))}g^{-1}(\psi_{i, t}) + \frac{g'(g^{-1}(\psi_{i, t}))}{g'(g^{-1}(\psi_{i, t}))}\right]}{\left[g'(g^{-1}(\psi_{i, t}))g^{-1}(\psi_{i, t})\right]^2} \frac{\partial \psi_{i, t}}{\partial \bm{\theta}^{\mathrm{T}}} \\
	&= (\star) - \sum_{t = 1}^T \sum_{i = 1}^n \frac{\partial \psi_{i, t}}{\partial \bm{\theta}} \frac{1}{\left[g'(g^{-1}(\psi_{i, t}))\right]^2 g^{-1}(\psi_{i, t})} \frac{\partial \psi_{i, t}}{\partial \bm{\theta}^{\mathrm{T}}} \\
	&\qquad- \sum_{t = 1}^T \sum_{i = 1}^n \frac{\partial \psi_{i, t}}{\partial \bm{\theta}} \cfrac{(y_{i, t} - g^{-1}(\psi_{i, t}))\left[g''(g^{-1}(\psi_{i, t}))g^{-1}(\psi_{i, t}) + g'(g^{-1}(\psi_{i, t}))\right]}{\left[g'(g^{-1}(\psi_{i, t}))\right]^3 \left[g^{-1}(\psi_{i, t})\right]^2} \frac{\partial \psi_{i, t}}{\partial \bm{\theta}^{\mathrm{T}}} \\
	&= \sum_{t = 1}^T \sum_{i = 1}^n \frac{\partial^2 \psi_{i, t}}{\partial \bm{\theta} \partial \bm{\theta}^{\mathrm{T}}} \frac{y_{i, t} - g^{-1}(\psi_{i, t})}{g'(g^{-1}(\psi_{i, t})) g^{-1}(\psi_{i, t})} \\
	&\qquad- \sum_{t = 1}^T \sum_{i = 1}^n \frac{\partial \psi_{i, t}}{\partial \bm{\theta}} \cfrac{(y_{i, t} - g^{-1}(\psi_{i, t}))\left[g''(g^{-1}(\psi_{i, t}))g^{-1}(\psi_{i, t}) + g'(g^{-1}(\psi_{i, t}))\right]}{\left[g'(g^{-1}(\psi_{i, t}))\right]^3 \left[g^{-1}(\psi_{i, t})\right]^2} \frac{\partial \psi_{i, t}}{\partial \bm{\theta}^{\mathrm{T}}} \\
	&\qquad- \sum_{t = 1}^T \frac{\partial \bm{\psi}_t'}{\partial \bm{\theta}} \bm{\tilde{D}}^{-1}_t \frac{\partial \bm{\psi}_t'}{\partial \bm{\theta}^{\mathrm{T}}}.
\end{align*}

By taking expectation, we receive the conditional information matrix as
\begin{align*}
	\bm{G}_T(\bm{\theta}) = \mathbb{E}\left[ \bm{H}_T (\bm{\theta}) \right] =  -\sum_{t = 1}^T \frac{\partial \bm{\psi}_t'}{\partial \bm{\theta}} \bm{\tilde{D}}^{-1}_t \frac{\partial \bm{\psi}_t'}{\partial \bm{\theta}^{\mathrm{T}}}
\end{align*}

\section{Control Parameters}\label{appendix:control}

Estimation behavior, see Section~\ref{sec:estimation}, can be adjusted using the \code{control} argument, which accepts a named list of control parameters passed internally to the function \fct{glmstarma.control} and \fct{dglmstarma.control} 

By default, the log-likelihood function in \eqref{eq:loglike} is maximized using the SLSQP algorithm \citep{kraft_algorithm_1994} from the \pkg{nloptr} package \citep{ypma_nloptr_2022}, subject to potential non-negativity constraints on the parameters and stability constraints \eqref{eq:stationarity} or \eqref{eq:stationarity_jahn}. The latter constraints can be disabled by setting \code{constrained = FALSE}. 
%\kostas{But this depends on the model!!}

An alternative optimization method can be specified via the \code{method} argument. If this is set to \code{"optim"}, the BFGS algorithm \citep{fletcher_function_1964} is employed for unconstrained models, whereas the L-BFGS algorithm \citep{liu_limited_1989} with a lower bound of zero is used when non-negativity constraints are in force. Both algorithms utilize the \R internal \fct{optim} implementation via the \pkg{roptim} package \citep{pan_roptim_2022}, executed in \proglang{C++}. Note, however, that these algorithms do not enforce stability constraints.

Models involving feedback terms (i.e., $\mathcal{Q}^{(\mu)} \neq \emptyset$ or $\mathcal{Q}^{(\phi)} \neq \emptyset$) require initial values for the feedback process, i.e., values for $\bm{\psi}_0, \ldots, \bm{\psi}_{\max\mathcal{Q}^{(\mu)} - 1}$. These can be provided via the \code{init\_link} argument. The default, \code{"first\_obs"}, sets $\bm{\psi}_t = \tilde{h}(\bm{Y}_t)$. Alternatives include \code{"mean"}, which uses $\bm{\psi}_t =  (T + 1)^{-1} \sum_{t = 0}^{T}\tilde{h}(\bm{Y}_t)$, and \code{"transformed\_mean"}, which applies the link function after averaging: $\bm{\psi}_t =  \tilde{h}\left( (T + 1)^{-1} \sum_{t = 0}^{T}\bm{Y}_t \right)$. The \code{"zero"} option sets $\bm{\psi}_t = \bm{0}_p$. Alternatively, a matrix of initial values can be passed, where the $(t + 1)$-st column corresponds to $\bm{\psi}_t$. The initialization for the dispersion process follows the same logic and is specified via \code{init\_dispersion}. The default here is \code{"zero"}, since \code{init\_link = "first\_obs"} implies $d_{i, t} = 0$ for $t = 0, \ldots, \max \mathcal{Q}$.

\sloppy Initial values for parameter estimation are  specified via the list elements \code{parameter\_init} or \code{parameter\_init\_dispersion}. These can be given as named lists, analogous to the parameter specification in simulations (see Section~\ref{sec:simulation}). Alternatively, \code{"zero"} (default) and \code{"random"} are available. If no parameter constraints are imposed, all dependence parameters $\alpha_{i\ell}, \beta_{j\ell}$ and covariate parameters $\gamma_{k\ell}$ are initialized to zero. If non-negativity constraints apply, they are initialized with a value slightly above zero, i.e.,
\[
\alpha_{i\ell} = \beta_{j\ell} = \frac{1}{|\mathcal{Q}| + |\mathcal{R}| + 10}, \quad \gamma_{k\ell} = 1.
\]
If \code{"random"} is selected, parameters are initialized randomly within the region defined by \eqref{eq:stationarity}. The intercept is initialized based on the initial values of the other parameter values using
\begin{align}
    \bm{\delta} = \frac{1}{T}\left(1 - \sum_{i \in \mathcal{Q}}\sum_{\ell \in \mathcal{A}_i}\alpha_{i\ell} - \sum_{j \in \mathcal{R}}\sum_{\ell \in \mathcal{B}_j}\beta_{j\ell}\right)\sum_{t = 1}^T \tilde{h}(\bm{Y}_t).
\end{align}
For homogeneous models, the mean of this vector is used. In the \fct{dglmstarma} function, these values serve only for the first estimation step. In general, the parameter estimates from iteration $k - 1$ in Algorithm~\ref{alg:estimation} are used as starting values in iteration $k$.

Additional control parameters influencing convergence speed and memory usage are documented on the help page of \fct{glmstarma.control} and \fct{dglmstarma.control}.

\section{Simulation Examples}\label{appendix:simulation_example}

In this section we illustrate, with two examples, how data can be simulated from the model framework. We consider observations on a regular $10 \times 10$ grid (so $p=100$) and include neighbours up to spatial order~2. The neighbourhood structure used throughout the examples is shown in Figure~\ref{fig:neighbors_example}. Panel~\ref{fig:grid} highlights, for a representative location (dark), the first-order neighbours (green), and the second-order neighbours (teal). All neighbours within the same order receive equal weight. The helper function \fct{generateW} in the package constructs the corresponding weight matrices for this neighbourhood structure. The weight matrices used here are obtained by \code{generateW("rectangle", dim = 100, maxOrder = 2, width = 10)}. The call returns a list of matrices $\bm{W}^{(\ell)} \in \mathbb{R}^{p \times p}$ for $\ell = 0, 1, 2$, which are row-normalized. Panel~\ref{fig:neighbors_w1} visualizes the weights of the first-order neighborhood, i.e.,\ $\bm{W}^{(1)}$, and Panel~\ref{fig:neighbors_w2} shows the weights for the second-order neighbourhood, i.e.,\ $\bm{W}^{(2)}$. The matrix $\bm{W}^{(0)}$ is the $100\times100$ identity matrix.

\definecolor{ord0}{RGB}{20,24,82}   % Zentrum
\definecolor{ord1}{RGB}{248,203,68}  % Ordnung 1
\definecolor{ord2}{RGB}{88,165,204} % Ordnung 2

\newcommand{\N}{10}    % 10x10 Gitter
\newcommand{\C}{4}     % Zentrum (0-basiert)
\newcommand{\CELL}{0.45} % Skalierung

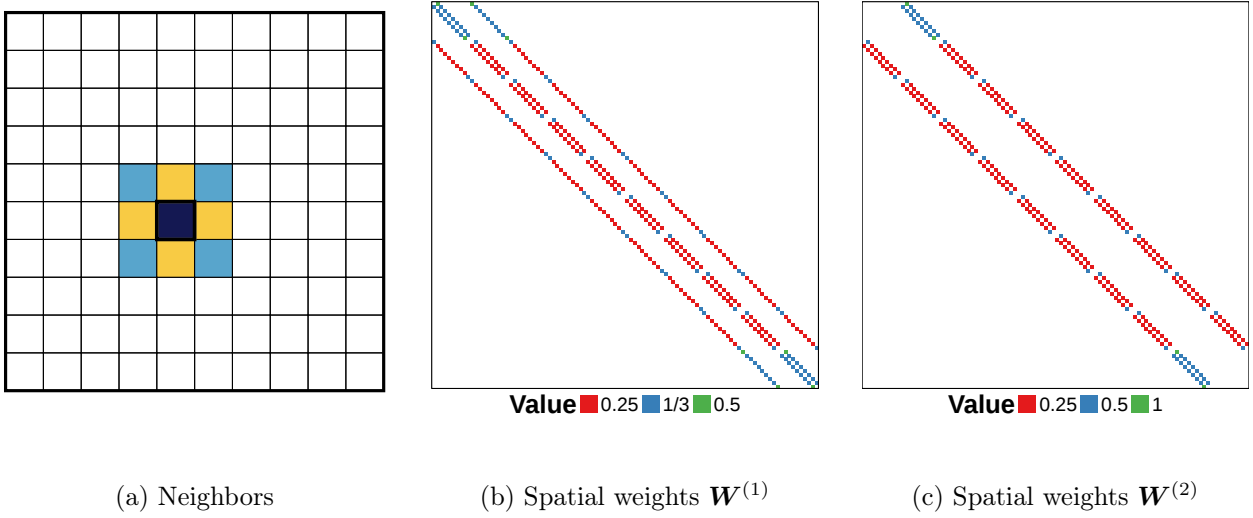
\begin{figure}[t]
	\centering
	\begin{subfigure}[b]{0.31\textwidth}
		\centering
		% Hier ggf. scale anpassen, damit es in die 0.32\textwidth passt
		\begin{tikzpicture}[scale=0.5] 
			\foreach \x in {0,...,9}{
				\foreach \y in {0,...,9}{
					\pgfmathtruncatemacro{\dx}{abs(\x-\C)}
					\pgfmathtruncatemacro{\dy}{abs(\y-\C)}
					\pgfmathtruncatemacro{\man}{\dx+\dy}
					
					\pgfmathtruncatemacro{\drawcell}{0}
					\ifnum\man<2 \pgfmathtruncatemacro{\drawcell}{1} \else
					\ifnum\man=2 \ifnum\dx=1 \pgfmathtruncatemacro{\drawcell}{1} \fi \fi
					\fi
					
					\ifnum\drawcell=1 \fill[ord\man] (\x,\y) rectangle ++(1,1); \fi
					\draw[thin] (\x,\y) rectangle ++(1,1);
				}
			}
			\draw[very thick] (0,0) rectangle (10,10);
			\draw[line width=1.2pt, white] (\C,\C) rectangle ++(1,1);
			\draw[line width=1.4pt] (\C,\C) rectangle ++(1,1);
		\end{tikzpicture}
		\vspace{0.5cm}
		\caption{Neighbors}
		\label{fig:grid}
	\end{subfigure}
	\hfill
	% Grafik 1 (W1)
	\begin{subfigure}[b]{0.31\textwidth}
		\centering
		\includegraphics[width=\linewidth, keepaspectratio]{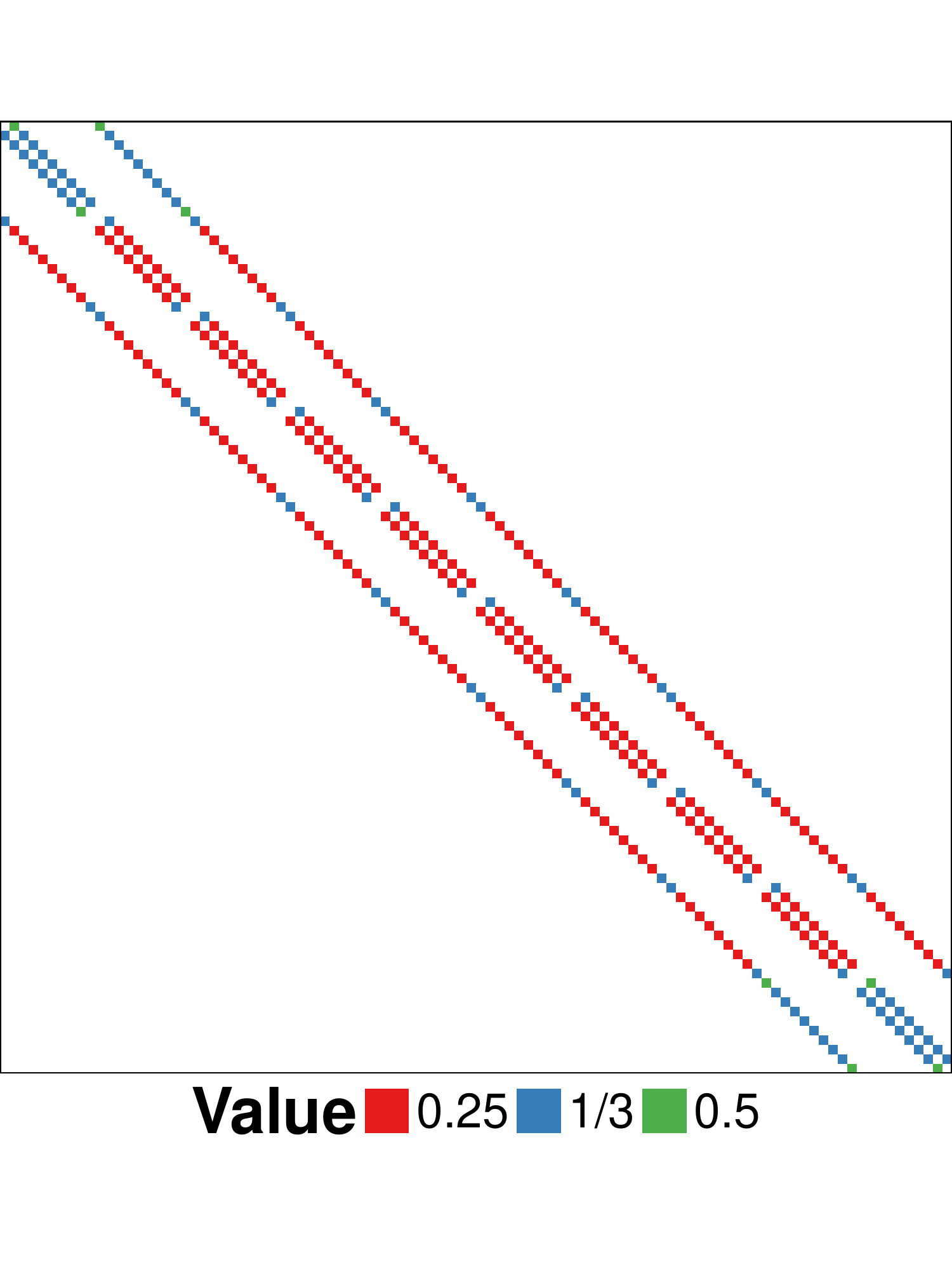}
		\caption{Spatial weights $\bm{W}^{(1)}$}
		\label{fig:neighbors_w1}
	\end{subfigure}
	\hfill
	% Grafik 2 (W2)
	\begin{subfigure}[b]{0.31\textwidth}
		\centering
		\includegraphics[width=\linewidth, keepaspectratio]{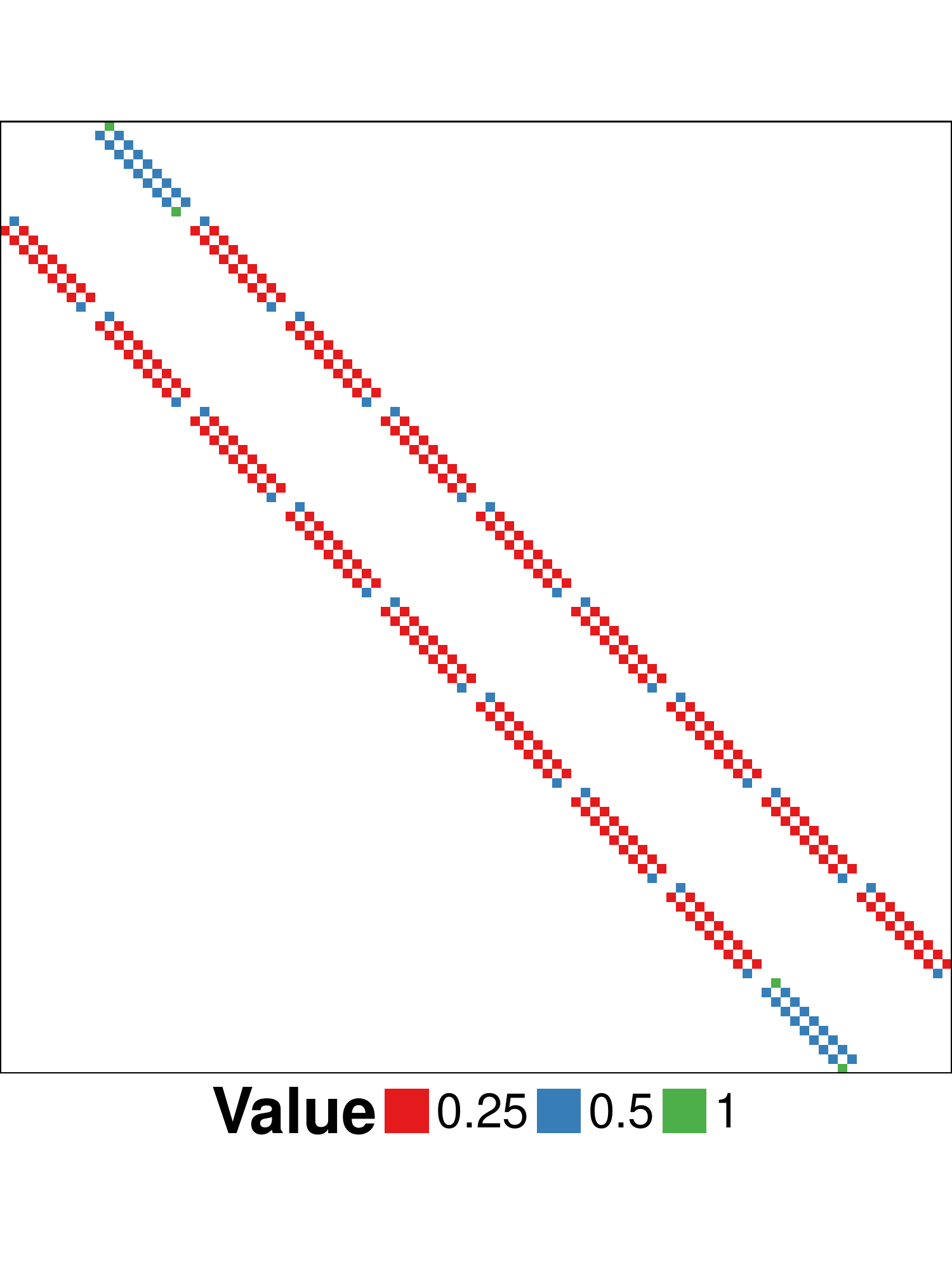}
		\caption{Spatial weights $\bm{W}^{(2)}$}
		\label{fig:neighbors_w2}
	\end{subfigure}
	\caption{Visualization of the neighborhood generated by \code{generateW} on a uniform $10 \times 10$ rectangular grid up to second-order neighbors}
	\label{fig:neighbors_example}
\end{figure}

\subsection{Linear PSTARMA Process}
In the first example we generate data on the grid from the linear spatio-temporal model of \citet{maletz_glmstarma_2026}. Because that model specifies only the conditional mean of the observations, we use the function \fct{glmstarma.sim}. Contemporaneous dependence is induced using the data-generating mechanism of \citet{fokianos_multivariate_2020} with a Frank copula (parameter $2$). We simulate a process of length $T=150$ according to the model
\[
\bm{\mu}_t = 3\cdot\bm{1}_{100} + 0.2\,\bm{\mu}_{t-1}
+ \big(0.3\,\bm{I}_{100} + 0.2\,\bm{W}^{(1)} + 0.1\,\bm{W}^{(2)}\big)\bm{Y}_{t-1}
+ 0.1\,\bm{Y}_{t-7}.
\]
The \R code used for this simulation is given below. The function returns a named list containing the simulated observations (element \code{observations}), the values of the linear predictor $\bm{\psi}_t$ (element \code{link\_values}), the model orders (element \code{model}), and the true parameters (element \code{parameters}).
\begin{CodeChunk}
\begin{CodeInput}
R>  set.seed(42)
R>  sim <- glmstarma.sim(
+    150,
+    parameters = list(
+      intercept = 1,
+      past_mean = 0.2,
+      past_obs  = cbind(c(0.3, 0.2, 0.1),
+                        c(0.1, 0, 0))),
+    model = list(
+      intercept = "homogeneous",
+      past_mean= 0,
+      past_obs = c(2, 0),
+      past_obs_time_lags = c(1, 7)),
+    family = vpoisson("identity", copula = "frank", copula_param = 2, 
+                      sampling_method = "poisson_process"),
+    wlist = W
+  )
\end{CodeInput}
\end{CodeChunk}

Because the intercept is homogeneous, the unconditional global mean can be computed as
\[
\mu = \frac{1}{1 - 0.2 - 0.3 - 0.2 - 0.1 - 0.1} = 10.
\]
The  mean of the simulated observations, e.g.,\ \code{mean(sim\$observations)}, is approximately $10.0264$, which is in agreement with the theoretical value. Using the true conditional means, the average of the Pearson residuals,
\code{mean((sim\$observations - sim\$link\_values)\^{}2 / sim\$link\_values)}, is approximately $0.9963$.

The marginal distribution of the observations in this example is Poisson by default. To introduce overdispersion, the marginal Poisson family can be replaced. For instance, random draws from a generalized Poisson distribution are obtained by using the \fct{vquasipoisson} family with an additional \code{dispersion} argument instead of \fct{vpoisson}. Similarly, marginal negative binomial observations can be generated by selecting the \fct{vnegative.binomial} family.

\subsection{Continous-data Process}
We now simulate data on the same spatial grid with marginal normal distributions,
\[
Y_{i,t}\mid\mathcal{F}_{t-1}\sim\mathcal{N}\big(\mu_{i,t},\phi_{i,t}\big).
\]
Contemporaneous dependence is introduced via the probability integral transform applied to a sample from a Joe copula with parameter $1.5$. We assume a log-link between the conditional mean process $\{\bm{\mu}_t\}$ and the linear predictor $\{\bm{\psi}_t\}$, i.e.\ $\bm{\psi}_t=\log(\bm{\mu}_t)$, with
\[
\bm{\psi}_t = 0.1\cdot\bm{1}_{100}
+ \big(0.3\,\bm{I}_{100} + 0.2\,\bm{W}^{(1)} + 0.1\,\bm{W}^{(2)}\big)\bm{Y}_{t-1}
+ 0.3\sin\!\left(\frac{2\pi}{52}t\right)\bm{1}_{100}
- 0.2\cos\!\left(\frac{2\pi}{52}t\right)\bm{1}_{100}.
\]

In addition, we specify a spatio-temporally varying variance vector $\bm{\phi}_t$ that follows a GARCH-like dynamics with seasonal component and a spatial gradient that increases with the column index of the grid in Panel~\ref{fig:grid}. Let $\bm{X} = \big(1,\ldots,10\big)'\otimes\bm{1}_{10}$ denote the column-index covariate replicated across rows. Then
\[
\bm{\phi}_t = 2\cdot\bm{1}_{100}
+ \big(0.1\,\bm{I}_{100} + 0.05\,\bm{W}^{(2)}\big)\bm{\phi}_{t-1}
+ 0.3\big(\bm{Y}_{t-1}-\bm{\mu}_{t-1}\big)^2
+ 2\exp\!\left(\sin\!\left(\frac{2\pi}{26}t\right)\right)\bm{1}_{100}
+ 0.5\,\bm{X}.
\]
This specification yields a dispersion process that varies in both space and time. For simulation of such a model the function \fct{dglmstarma.sim} is provided. The covariates are generated using the following code:
\begin{CodeChunk}
	\begin{CodeInput}
R> covariates_mean <- list(
+    sine   = SpatialConstant(sin(2 * pi / 52 * seq(250))),
+    cosine = SpatialConstant(cos(2 * pi / 52 * seq(250))))

R> covariates_dispersion <- list(
+    sine   = SpatialConstant(exp(sin(2 * pi / 26 * seq(250)))),
+    column = TimeConstant(c(1:10 %x% rep(1, 10))))
	\end{CodeInput}
\end{CodeChunk}
We then set the orders and parameters of the mean model via
\begin{CodeChunk}
	\begin{CodeInput}
R>  mean_model <- list(
+    intercept  = "homogeneous",
+    past_obs   = matrix(c(1, 1, 1), ncol = 1),
+    covariates = c(0, 0))

R>  mean_parameters <- list(
+    intercept  = 0.1,
+    past_obs   = matrix(c(0.3, 0.2, 0.1), ncol = 1),
+    covariates = matrix(c(0.3, -0.2), nrow = 1))
\end{CodeInput}
\end{CodeChunk}
Similar we define the dispersion model:
\begin{CodeChunk}
	\begin{CodeInput}
R>  dispersion_model <- list(
+    intercept  = "homogeneous",
+    past_mean  = matrix(c(1, 0, 1), ncol = 1),
+    past_obs   = 0,
+    covariates = c(0, 0))

R>  dispersion_parameters <- list(
+    intercept  = 2,
+    past_mean  = matrix(c(0.1, 0, 0.05), ncol = 1),
+    past_obs   = matrix(0.3),
+    covariates = matrix(c(3, 0.5), nrow = 1))
\end{CodeInput}
\end{CodeChunk}
 Finally, we simulate the from the process using the following \R code:
\begin{CodeChunk}
\begin{CodeInput}
R>  sim <- dglmstarma.sim(
+    250,
+    mean_parameters,
+    dispersion_parameters,
+    mean_model,
+    dispersion_model,
+    mean_family = vnormal("log", copula = "joe", copula_param = 1.5),
+    dispersion_link      = "identity",
+    wlist                = W,
+    mean_covariates      = covariates_mean,
+    dispersion_covariates = covariates_dispersion)
\end{CodeInput}
\end{CodeChunk}

\section{Simulation study}\label{appendix:simulations}
In this section, we assess the theoretical properties stated in Section~\ref{sec:asymptotics} with simulations. We adopt the same spatial grid structure as in Section~\ref{appendix:simulation_example}, but restrict spatial interactions to first-order neighbors.

\subsection{Construction}\label{appendix:construction}

We consider models for two types of marginal distributions: (i) count data and (ii) marginally Gaussian (normal) data. For the count-data experiments we generate observations from the \fct{vquasipoisson} family with a log link; for the normal-data experiments we use the identity link. The contemporary dependence between locations is induced by a Frank copula with parameter $2$. Data are generated in several configurations for the conditional mean and for the dispersion process.

For the conditional mean, we use two alternative dynamic specifications:
\begin{align}
\bm{\psi}_t &= \delta_0\cdot \bm{1}_{100} + (0.4 \bm{I}_{100} + 0.2\bm{W}^{(1)}) \,h(\bm{Y}_{t - 1}) + \gamma \,\bm{X}_t,
	\label{eq:simulation_mean_without_feedback} \tag{M-1} \\[4pt]
	\bm{\psi}_t &= \delta_0\cdot \bm{1}_{100} + (0.2 \bm{I}_{100} + 0.1\bm{W}^{(1)}) \,\bm{\psi}_{t - 1} 
	+ (0.2 \bm{I}_{100} + 0.1\bm{W}^{(1)}) \,h(\bm{Y}_{t - 1}) + \gamma \,\bm{X}_t, \label{eq:simulation_mean_with_feedback} \tag{M-2}
\end{align}
where for the count-data experiments we set $\delta_0 = 0.6$, $\gamma = 0.9$, and $
h(\bm{Y}_{t - 1}) = \log(\bm{Y}_{t - 1} + \bm{1}_{100}),$
while for the normally distributed experiments we set $\delta_0 = 5$, $\gamma = 2$, and $h(\bm{Y}_{t - 1}) = \bm{Y}_{t - 1}.$

The covariate process $\{\bm{X}_t\}$ used in all simulation runs was generated once (and then reused across scenarios) with the following R code:
\begin{CodeChunk}
\begin{CodeInput}
R>  set.seed(42)
R>  covariates <- t(replicate(
+    100,
+    arima.sim(n = 1000,
+              list(ar = c(0.89, -0.3), ma = c(-0.1, 0.28)), 
+              rand.gen = runif)))
R>  covariates <- covariates / max(covariates)
R>  covariates[covariates < 0] <- 0
\end{CodeInput}
\end{CodeChunk}

We also consider several dispersion processes. The dispersion parameter vector \(\bm{\phi}_t\) is linked to a linear process \(\bm{\zeta}_t\) through the log link (i.e.\ \(\bm{\phi}_t = \exp(\bm{\zeta}_t)\)). The following alternatives are used:
\begin{align}
	\bm{\phi}_t &= 2 \cdot \bm{1}_{100}, \label{eq:simulation_dispersion_constant} \tag{D-1} \\[4pt]
    \bm{\zeta}_t &= 0.5 \cdot\bm{1}_{100} + (0.5\bm{I}_{100} + 0.2\bm{W}^{(1)}) \,\log(\bm{d}_{t - 1} + \bm{1}_p).
	\label{eq:simulation_dispersion_without_feedback} \tag{D-2} \\[4pt]
	\bm{\zeta}_t &= 0.5 \cdot\bm{1}_{100} + (0.15\bm{I}_{100} + 0.05\bm{W}^{(1)}) \,\bm{\zeta}_{t - 1} 
	+ (0.3\bm{I}_{100} + 0.2\bm{W}^{(1)}) \,\log(\bm{d}_{t - 1} + \bm{1}_p), 
	\label{eq:simulation_dispersion_with_feedback} \tag{D-3}
\end{align}

Each combination of the mean specifications \eqref{eq:simulation_mean_without_feedback}--\eqref{eq:simulation_mean_with_feedback} with the dispersion specifications \eqref{eq:simulation_dispersion_constant}--\eqref{eq:simulation_dispersion_with_feedback} is used once to generate data. Table~\ref{tab:settings} summarises the data-generating configurations and the corresponding fitted models. When data are generated under a time-varying dispersion process, we estimate both (i) the correctly specified model with time-varying dispersion and (ii) a misspecified model that assumes a single global (time-constant) dispersion parameter; this allows us to examine robustness and the effect of misspecification. When the data-generating process has a global, time-constant dispersion \eqref{eq:simulation_dispersion_constant}, we fit only the variable-dispersion model in order to study the asymptotic behavior of the overparameterized model under the null hypothesis of no temporal dispersion dynamics.

Each scenario is simulated for \(T \in \{50, 100, 250, 500, 1000\}\) and repeated over \(1000\) independent replications. Results are reported in the next Subsection. For models that include a feedback mechanism, i.e. \eqref{eq:simulation_mean_with_feedback} and \eqref{eq:simulation_dispersion_with_feedback}, we initialize this during parameter estimation using the first observations or pseudo-observations.

\begin{table}[tb]
	\centering
	\resizebox{\linewidth}{!}{
	\begin{tabular}{|cc|cc|cc|}
		\hline \multirow{2}{*}{True Model} & \multirow{2}{*}{Fitted Models} & \multicolumn{2}{c}{Normal} & \multicolumn{2}{|c|}{Generalized Poisson} \\
		& & Boxplots & QQ-Plots & Boxplots & QQ-Plots\\ \hline
        \eqref{eq:simulation_mean_without_feedback} + \eqref{eq:simulation_dispersion_constant} & \eqref{eq:simulation_mean_without_feedback} + \eqref{eq:simulation_dispersion_without_feedback} & - & - & - & - \\

        \eqref{eq:simulation_mean_without_feedback} + \eqref{eq:simulation_dispersion_without_feedback} & \eqref{eq:simulation_mean_without_feedback} + \eqref{eq:simulation_dispersion_constant} \& \eqref{eq:simulation_mean_without_feedback} + \eqref{eq:simulation_dispersion_without_feedback} & Fig.~\ref{fig:boxplot_normal_without_without} & Fig.~\ref{fig:qqplot_normal_without_without} & Fig.~\ref{fig:boxplot_poisson_without_without} & Fig.~\ref{fig:qqplot_poisson_without_without} \\ 

        \eqref{eq:simulation_mean_without_feedback} + \eqref{eq:simulation_dispersion_with_feedback} & \eqref{eq:simulation_mean_without_feedback} + \eqref{eq:simulation_dispersion_constant} \& \eqref{eq:simulation_mean_without_feedback} + \eqref{eq:simulation_dispersion_with_feedback} & Fig.~\ref{fig:boxplot_normal_without_with} & Fig.~\ref{fig:qqplot_normal_without_with} & Fig.~\ref{fig:boxplot_poisson_without_with} & Fig.~\ref{fig:qqplot_poisson_without_with} \\

		\eqref{eq:simulation_mean_with_feedback} + \eqref{eq:simulation_dispersion_constant} & \eqref{eq:simulation_mean_with_feedback} + \eqref{eq:simulation_dispersion_without_feedback} & - & - & - & - \\

        \eqref{eq:simulation_mean_with_feedback} + \eqref{eq:simulation_dispersion_without_feedback} & \eqref{eq:simulation_mean_with_feedback} + \eqref{eq:simulation_dispersion_constant} \& \eqref{eq:simulation_mean_with_feedback} + \eqref{eq:simulation_dispersion_without_feedback} & Fig.~\ref{fig:boxplot_normal_with_without} & Fig.~\ref{fig:qqplot_normal_with_without} & Fig.~\ref{fig:boxplot_poisson_with_without} & Fig.~\ref{fig:qqplot_poisson_with_without} \\
        
		\eqref{eq:simulation_mean_with_feedback} + \eqref{eq:simulation_dispersion_with_feedback} & \eqref{eq:simulation_mean_with_feedback} + \eqref{eq:simulation_dispersion_constant} \& \eqref{eq:simulation_mean_with_feedback} + \eqref{eq:simulation_dispersion_with_feedback} & Fig.~\ref{fig:boxplot_normal_with_with} & Fig.~\ref{fig:qqplot_normal_with_with} & Fig.~\ref{fig:boxplot_poisson_with_with} & Fig.~\ref{fig:qqplot_poisson_with_with} \\
		\hline
	\end{tabular}}
	\caption{Overview of data-generating models and estimated models on the generated data. For data generation, one of the two mean models (\eqref{eq:simulation_mean_without_feedback} -- without feedback mechanism, \eqref{eq:simulation_mean_with_feedback} -- with feedback mechanism) was combined with one of the dispersion models (\eqref{eq:simulation_dispersion_constant} -- constant dispersion, \eqref{eq:simulation_dispersion_without_feedback} -- dispersion model without feedback mechanism, \eqref{eq:simulation_dispersion_with_feedback} -- dispersion model with a feedback mechanism). For settings with variating dispersion, the correctly specified model was estimated, and one model was estimated assuming constant dispersion.}
	\label{tab:settings}
\end{table}

\subsection{Results}\label{appendix:results}
Table~\ref{tab:settings} provides an overview of the figures containing the simulation results for the respective settings. Each figure includes boxplots of the parameter estimates for both the mean and dispersion models. For the mean model, the estimates under a correctly specified dispersion are compared with those obtained under a misspecified constant-variance structure. In both cases, the estimates improve with increasing time series length; however, misspecification comes at the cost of higher variability in the parameter estimates. Incorporating a feedback mechanism into at least one model component reduces this difference.

\begin{figure}[p]
	\centering
	\begin{subfigure}{\textwidth}
		\centering
		\includegraphics[width=\textwidth, keepaspectratio]{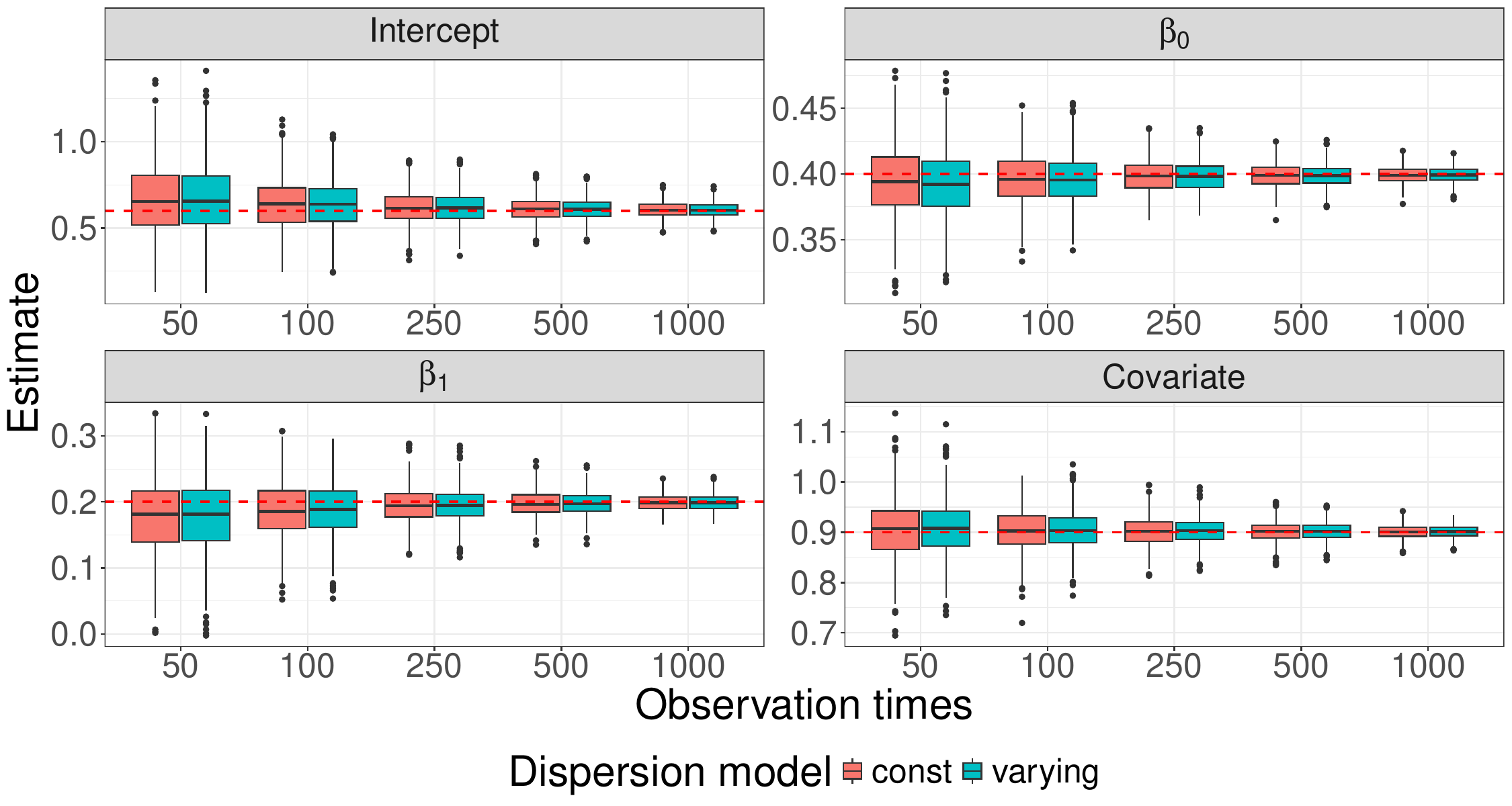}
		\caption{Parameter estimates of the mean model~\eqref{eq:simulation_mean_without_feedback} under the constant dispersion specification~\eqref{eq:simulation_dispersion_constant} and the varying dispersion specification~\eqref{eq:simulation_dispersion_without_feedback}.}
	\end{subfigure}
	
	\vspace{3em}
	
	\begin{subfigure}{\textwidth}
		\centering
		\includegraphics[width=\textwidth, keepaspectratio]{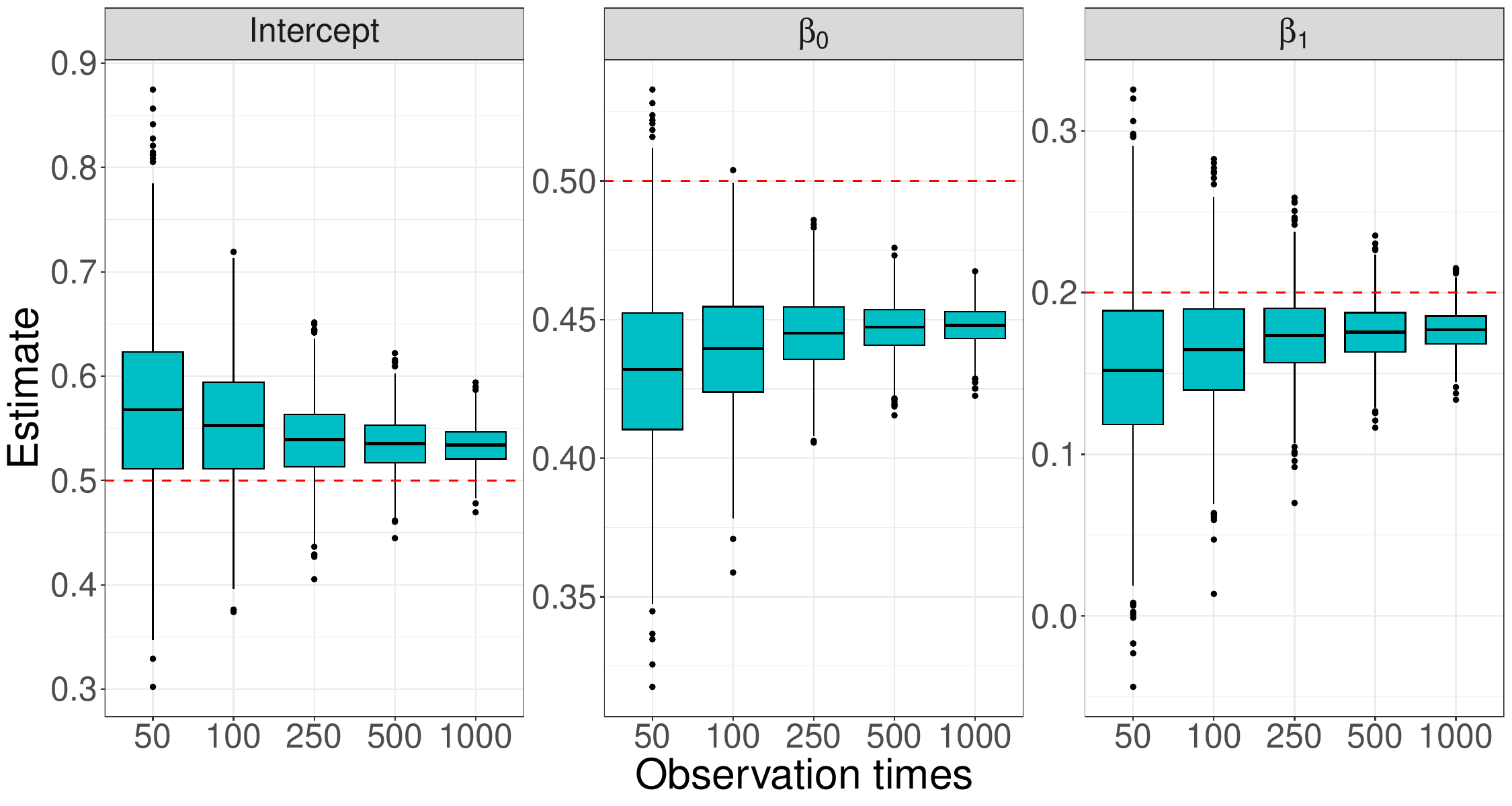}
		\caption{Parameter estimates of the varying dispersion model~\eqref{eq:simulation_dispersion_without_feedback} combined with the mean model~\eqref{eq:simulation_mean_without_feedback}.}
        \label{subfig:boxplot_poisson_without_without}
	\end{subfigure}
	\caption{Box plots of parameter estimates under alternative dispersion specifications.  Panel~(a) shows estimates of the mean model under constant and varying dispersion, whereas panel~(b) displays estimates of the varying dispersion model. The data-generating process with generalized Poisson marginals corresponds to \eqref{eq:simulation_mean_without_feedback} and~\eqref{eq:simulation_dispersion_without_feedback}, i.e., both models without feedback mechanism.}
	\label{fig:boxplot_poisson_without_without}
\end{figure}

\begin{figure}[p]
	\centering
	\begin{subfigure}{\textwidth}
		\centering
		\includegraphics[width=\textwidth, keepaspectratio]{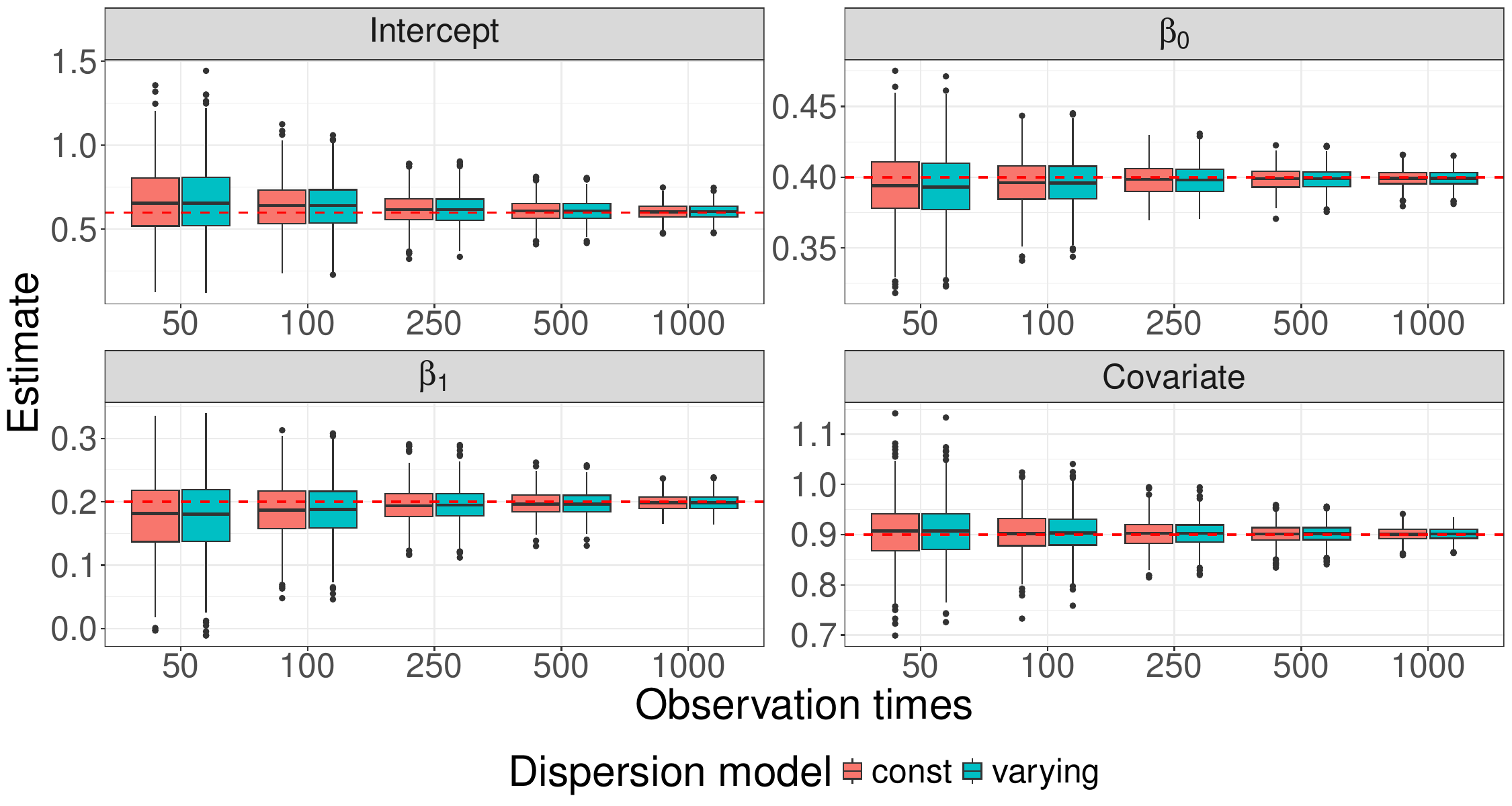}
		\caption{Parameter estimates of the mean model~\eqref{eq:simulation_mean_without_feedback} under the constant dispersion specification~\eqref{eq:simulation_dispersion_constant} and the varying dispersion specification~\eqref{eq:simulation_dispersion_with_feedback}.}
	\end{subfigure}
	
	\vspace{3em}
	
	\begin{subfigure}{\textwidth}
		\centering
		\includegraphics[width=\textwidth, keepaspectratio]{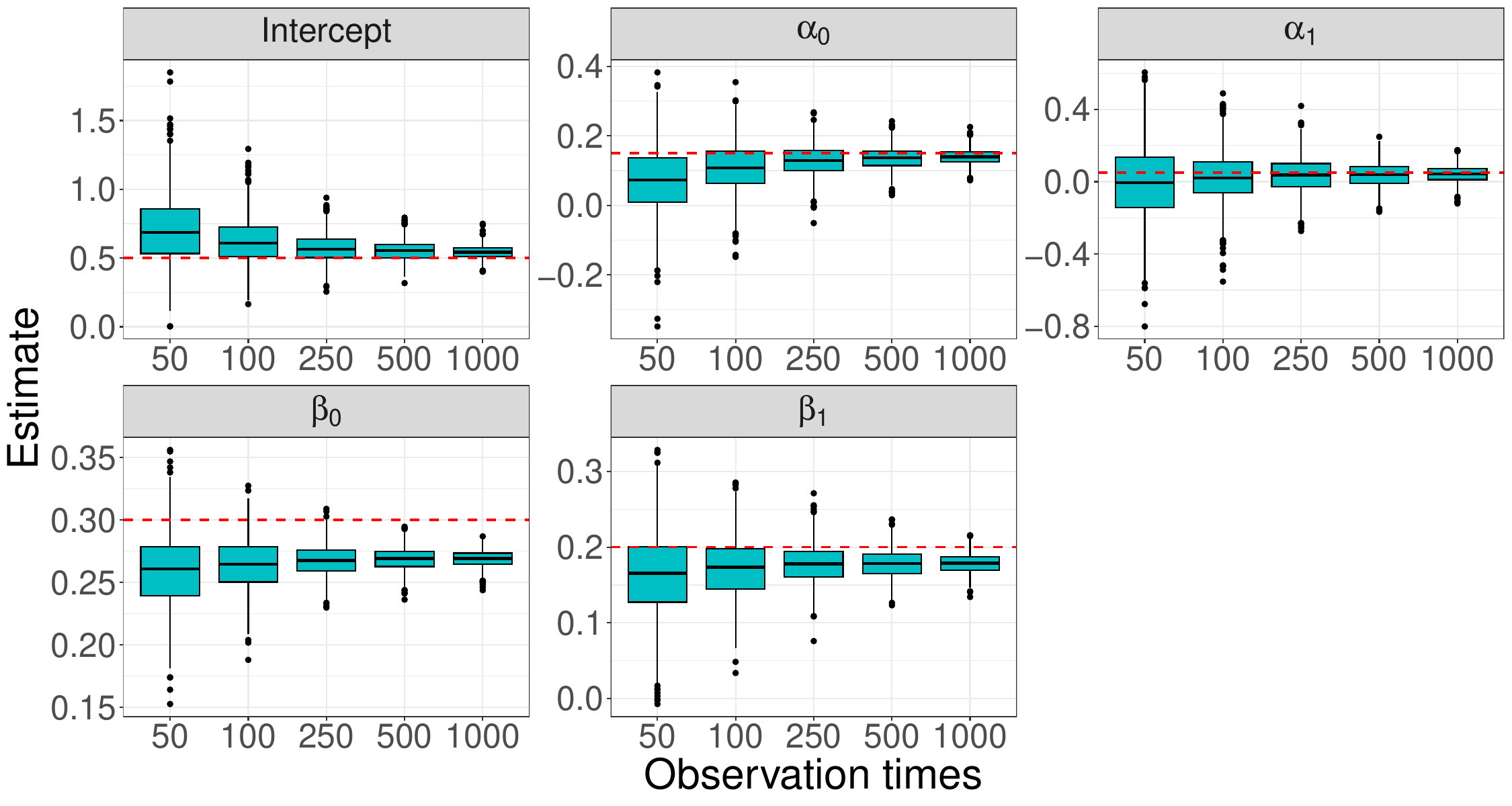}
		\caption{Parameter estimates of the varying dispersion model~\eqref{eq:simulation_dispersion_with_feedback} combined with the mean model~\eqref{eq:simulation_mean_without_feedback}.}
        \label{subfig:boxplot_poisson_without_with}
	\end{subfigure}
	\caption{Boxplots of parameter estimates under alternative dispersion specifications.  Panel~(a) shows estimates of the mean model under constant and varying dispersion, whereas panel~(b) displays estimates of the varying dispersion model. The data-generating process with generalized Poisson marginals corresponds to \eqref{eq:simulation_mean_without_feedback} and~\eqref{eq:simulation_dispersion_with_feedback}, i.e., the mean model does not include a feedback mechanism while the dispersion process does.}
	\label{fig:boxplot_poisson_without_with}
\end{figure}

\begin{figure}[p]
	\centering
	\begin{subfigure}{\textwidth}
		\centering
		\includegraphics[width=\textwidth, keepaspectratio]{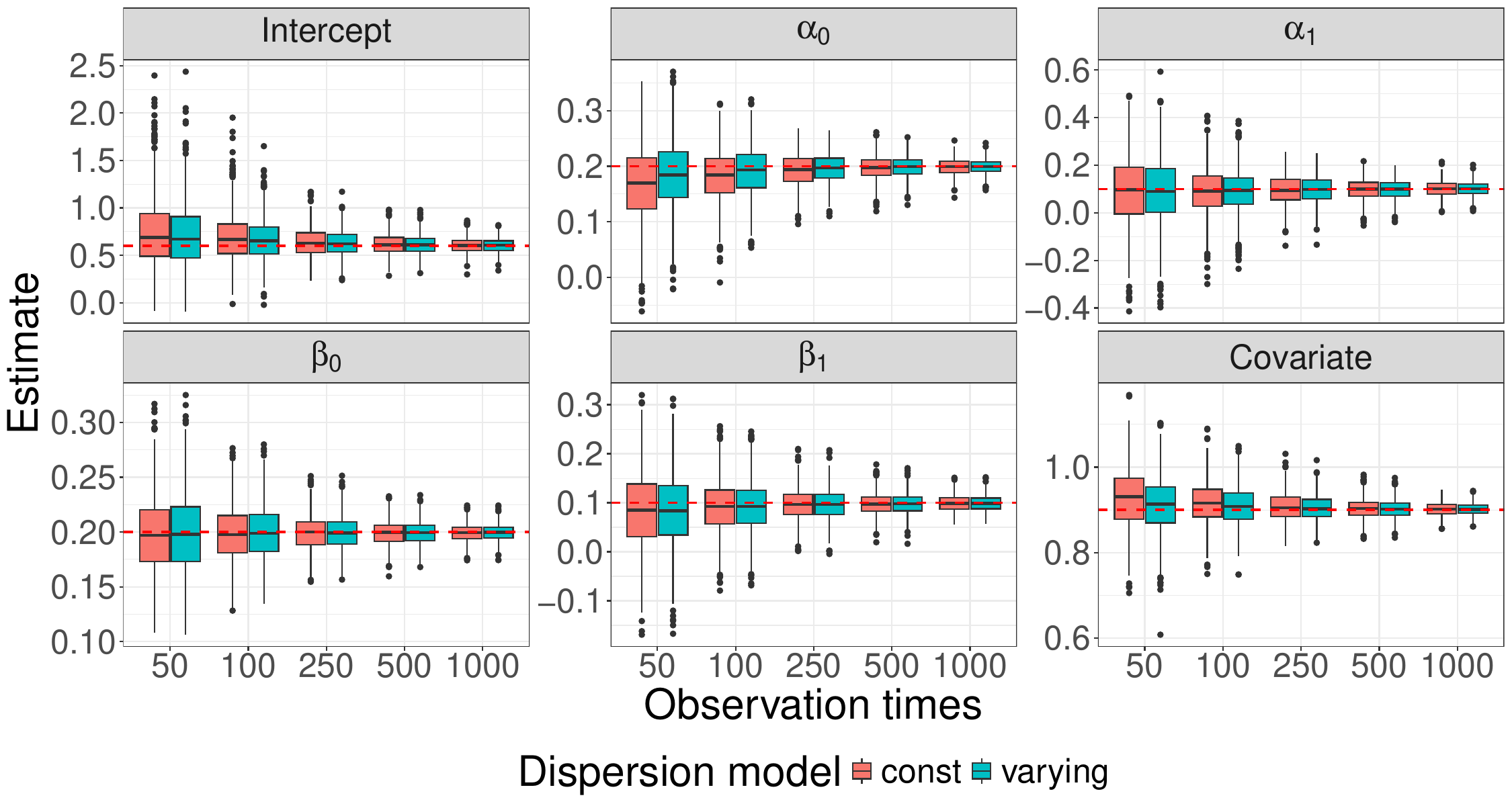}
		\caption{Parameter estimates of the mean model~\eqref{eq:simulation_mean_with_feedback} under the constant dispersion specification~\eqref{eq:simulation_dispersion_constant} and the varying dispersion specification~\eqref{eq:simulation_dispersion_without_feedback}.}
	\end{subfigure}
	
	\vspace{3em}
	
	\begin{subfigure}{\textwidth}
		\centering
		\includegraphics[width=\textwidth, keepaspectratio]{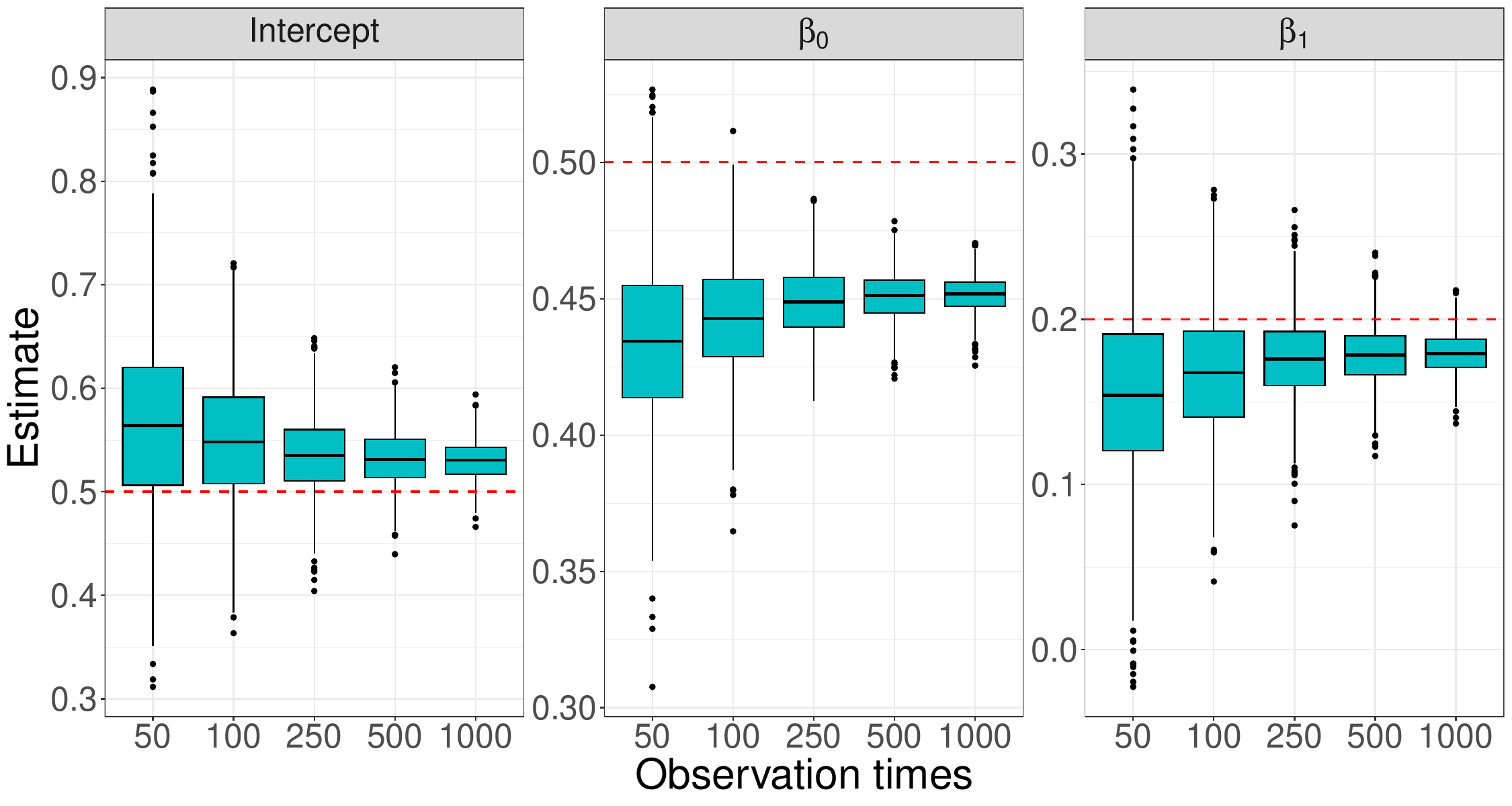}
		\caption{Parameter estimates of the varying dispersion model~\eqref{eq:simulation_dispersion_without_feedback} combined with the mean model~\eqref{eq:simulation_mean_with_feedback}.}
        \label{subfig:boxplot_poisson_with_without}
	\end{subfigure}
	\caption{Box plots of parameter estimates under alternative dispersion specifications.  Panel~(a) shows estimates of the mean model under constant and varying dispersion, whereas panel~(b) displays estimates of the varying dispersion model. The data-generating process with generalized Poisson marginals corresponds to \eqref{eq:simulation_mean_with_feedback} and~\eqref{eq:simulation_dispersion_without_feedback}, i.e., the mean model includes a feedback mechanism and the dispersion process not.}
	\label{fig:boxplot_poisson_with_without}
\end{figure}

\begin{figure}[p]
	\centering
	\begin{subfigure}{\textwidth}
		\centering
		\includegraphics[width=\textwidth, keepaspectratio]{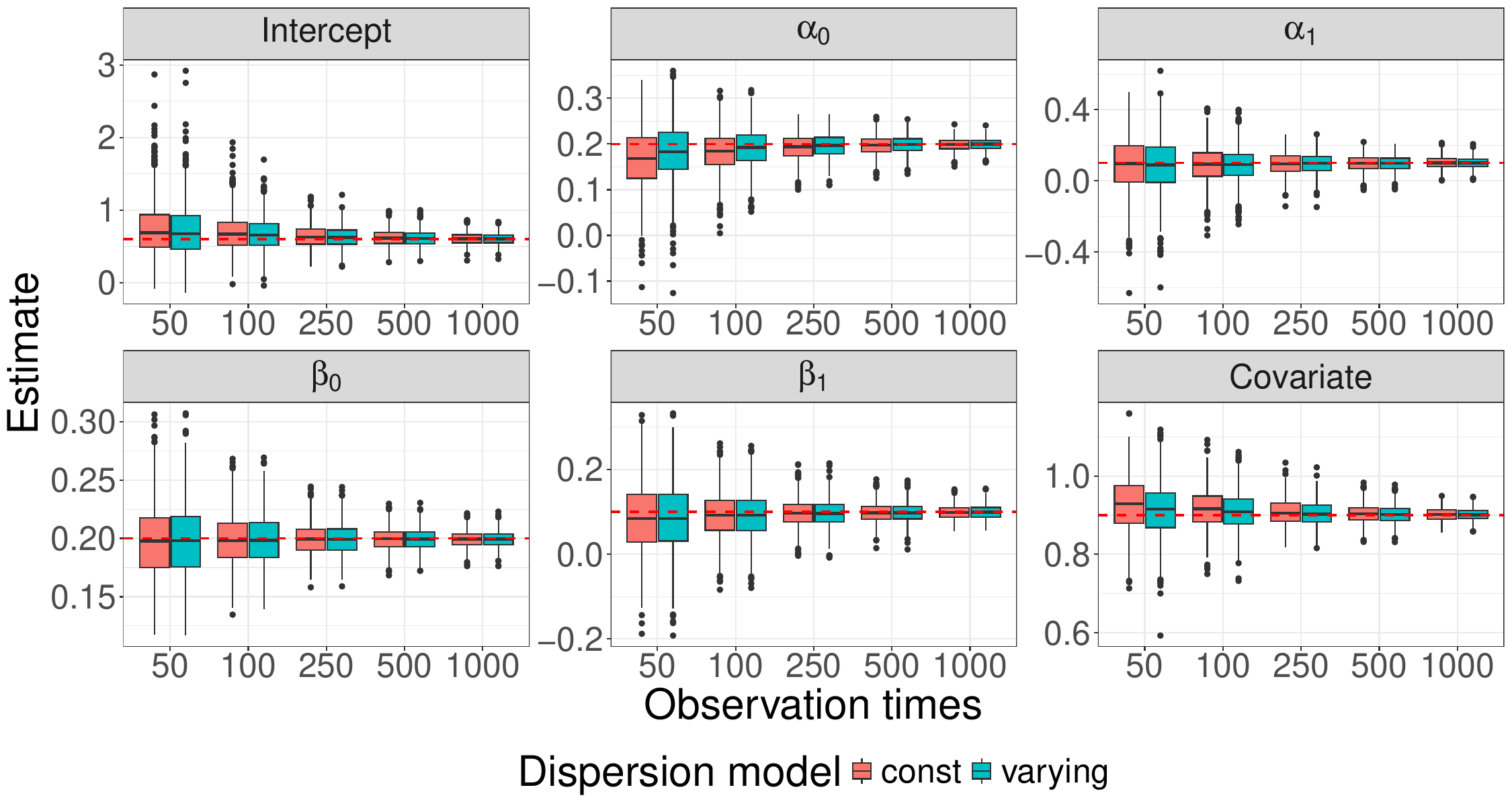}
		\caption{Parameter estimates of the mean model~\eqref{eq:simulation_mean_with_feedback} under the constant dispersion specification~\eqref{eq:simulation_dispersion_constant} and the varying dispersion specification~\eqref{eq:simulation_dispersion_with_feedback}.}
	\end{subfigure}
	
	\vspace{3em}
	
	\begin{subfigure}{\textwidth}
		\centering
		\includegraphics[width=\textwidth, keepaspectratio]{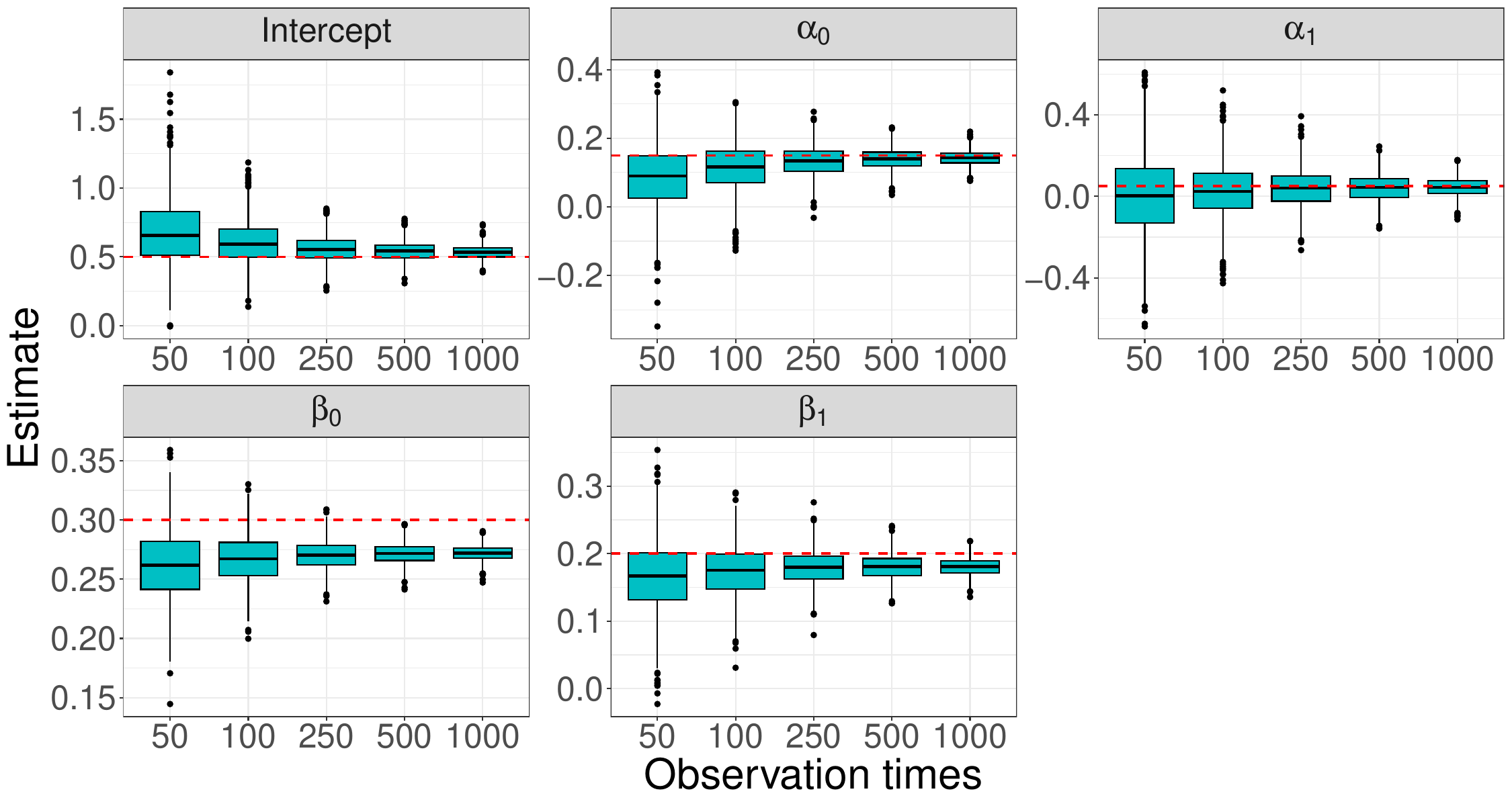}
		\caption{Parameter estimates of the varying dispersion model~\eqref{eq:simulation_dispersion_with_feedback} combined with the mean model~\eqref{eq:simulation_mean_with_feedback}.}
        \label{subfig:boxplot_poisson_with_with}
	\end{subfigure}
	\caption{Boxplots of parameter estimates under alternative dispersion specifications.  Panel~(a) shows estimates of the mean model under constant and varying dispersion, whereas panel~(b) displays estimates of the varying dispersion model. The data-generating process with generalized Poisson marginals corresponds to \eqref{eq:simulation_mean_with_feedback} and~\eqref{eq:simulation_dispersion_with_feedback}, i.e., both models include a feedback mechanism}
	\label{fig:boxplot_poisson_with_with}
\end{figure}

For the dispersion model, greater uncertainty is observed in the parameter estimates, which arises from the model being estimated on an error process. In the case of marginally normal distributions, the parameter estimates converge to the true values as the time series length increases. For generalized Poisson marginals, however, the estimates of~$\beta_0$ and~$\beta_1$ exhibit a notable bias from the true parameters (see Figures~\ref{subfig:boxplot_poisson_without_without}, \ref{subfig:boxplot_poisson_without_with}, \ref{subfig:boxplot_poisson_with_without}, and~\ref{subfig:boxplot_poisson_with_with}), which does not fully vanish as the length~$T$ of the time series increases. This phenomenon is more pronounced in the purely autoregressive dispersion model~\eqref{eq:simulation_dispersion_without_feedback} than in model~\eqref{eq:simulation_dispersion_with_feedback}, which additionally incorporates a feedback mechanism. In particular, the strength of the autoregressive dependence tends to be underestimated. Crucially, however, the estimates stabilize at values between 0 and the true parameter, i.e. they still capture some of the dependence structure of the dispersion process. This stabilization might explain why misspecifying the dispersion as constant entails only a moderate increase in the variability of the mean model estimates, as described above: even an imprecise dispersion model suffices to preserve the quality of mean estimation, provided the estimated parameters do not degenerate. We further analyzed this behavior in the next Subsection.

QQ-plots of the parameter estimates for both the mean and dispersion models at~$T = 500$ confirm convergence to a normal distribution, with only minor deviations when a feedback mechanism is included in the respective model component.

Table~\ref{tab:simulation_empirical_size} reports the empirical sizes of an approximate Wald test at the~$5\%$ significance level, derived from the normal approximation~\eqref{eq:asymptotics} of the estimates. We individually test the null hypotheses $\text{H}_0\colon \beta_{0,1} = 0$ and $\text{H}_0\colon \beta_{1,1} = 0$ for the first two scenarios listed in Table~\ref{tab:settings}. Rejection of either hypothesis indicates that a constant global dispersion parameter can no longer be assumed within the model framework. Even for small sample sizes, i.e., $T = 50$, the empirical size closely approximates the nominal~$5\%$ level, and this approximation is maintained as the time series length increases.

\begin{table}[ht]
    \centering
    \begin{tabular}{llc*{5}{c}}
        \toprule
        \multirow{2}{*}{Mean model} & \multirow{2}{*}{Distribution} & \multirow{2}{*}{Parameter} & \multicolumn{5}{c}{Observations ($T$)} \\
        \cmidrule(lr){4-8}
        & & & 50 & 100 & 250 & 500 & 1000 \\
        \midrule
        \multirow{4}{*}{\eqref{eq:simulation_mean_without_feedback}} 
        & \multirow{2}{*}{Normal} 
        & $\beta_{0,1}$ & 0.066 & 0.068 & 0.044 & 0.046 & 0.052 \\
        & & $\beta_{1,1}$ & 0.065 & 0.063 & 0.059 & 0.052 & 0.048 \\ \cmidrule(lr){2-8}
        & \multirow{2}{*}{Gen. Poisson} 
        & $\beta_{0,1}$ & 0.069 & 0.071 & 0.051 & 0.045 & 0.049 \\
        & & $\beta_{1,1}$ & 0.061 & 0.053 & 0.055 & 0.054 & 0.056 \\
        \midrule
        \multirow{4}{*}{\eqref{eq:simulation_mean_with_feedback}} 
        & \multirow{2}{*}{Normal} 
        & $\beta_{0,1}$ & 0.070 & 0.072 & 0.045 & 0.047 & 0.052 \\
        & & $\beta_{1,1}$ & 0.069 & 0.063 & 0.061 & 0.052 & 0.050 \\ \cmidrule(lr){2-8}
        & \multirow{2}{*}{Gen. Poisson} 
        & $\beta_{0,1}$ & 0.072 & 0.066 & 0.045 & 0.043 & 0.047 \\
        & & $\beta_{1,1}$ & 0.063 & 0.050 & 0.049 & 0.059 & 0.050 \\
        \bottomrule
    \end{tabular}
    \caption{Empirical size of the Wald test at the 5\% significance level for the spatio-temporal dependence parameters of \eqref{eq:simulation_dispersion_without_feedback} under constant dispersion, i.e., \eqref{eq:simulation_dispersion_constant}.}
    \label{tab:simulation_empirical_size}
\end{table}

\subsubsection{Increasing the observation level}

To investigate the phenomenon of underestimated autoregressive dependencies in more detail, we slightly modified the setting of the previous scenario. We restrict our attention to a $5 \times 5$ grid with the same dependence structure as before, focusing on the purely autoregressive models without a feedback term, i.e., the combination of \eqref{eq:simulation_mean_without_feedback} and \eqref{eq:simulation_dispersion_without_feedback}, which exhibited the most notable behavior in the previous simulations.

We now vary the true intercept of the mean model, thereby shifting the expected value of the observations upward. In the case of marginal normal distributions, we consider $\delta_0 \in [3, 10]$, whereas for marginal generalized Poisson distributions the data are simulated with $\delta_0 \in [0.1, 2]$. The dispersion model is left unchanged as~\eqref{eq:simulation_dispersion_without_feedback}.

\begin{figure}[p]
\centering
	\begin{subfigure}{\textwidth}
		\centering
		\includegraphics[width=0.8\textwidth, keepaspectratio]{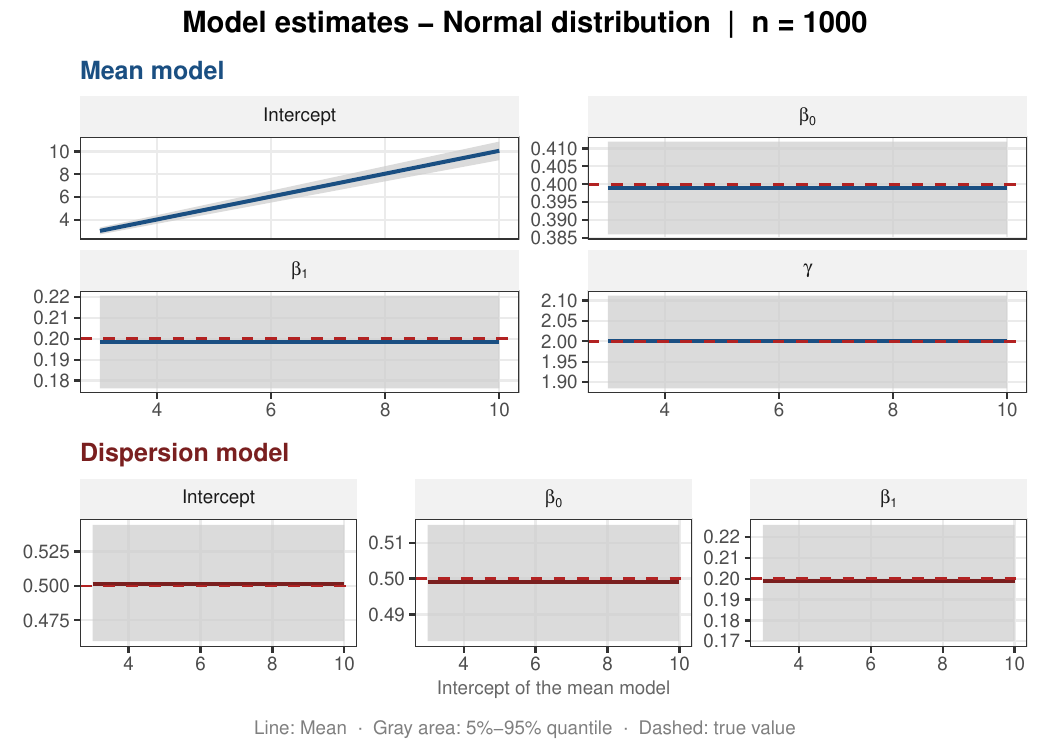}
		\caption{Marginals generated by a normal distribution}
        \label{fig:simulation_increasing_intercept_vnormal}
	\end{subfigure}
	
	\vspace{3em}
	
	\begin{subfigure}{\textwidth}
		\centering
		\includegraphics[width=0.8\textwidth, keepaspectratio]{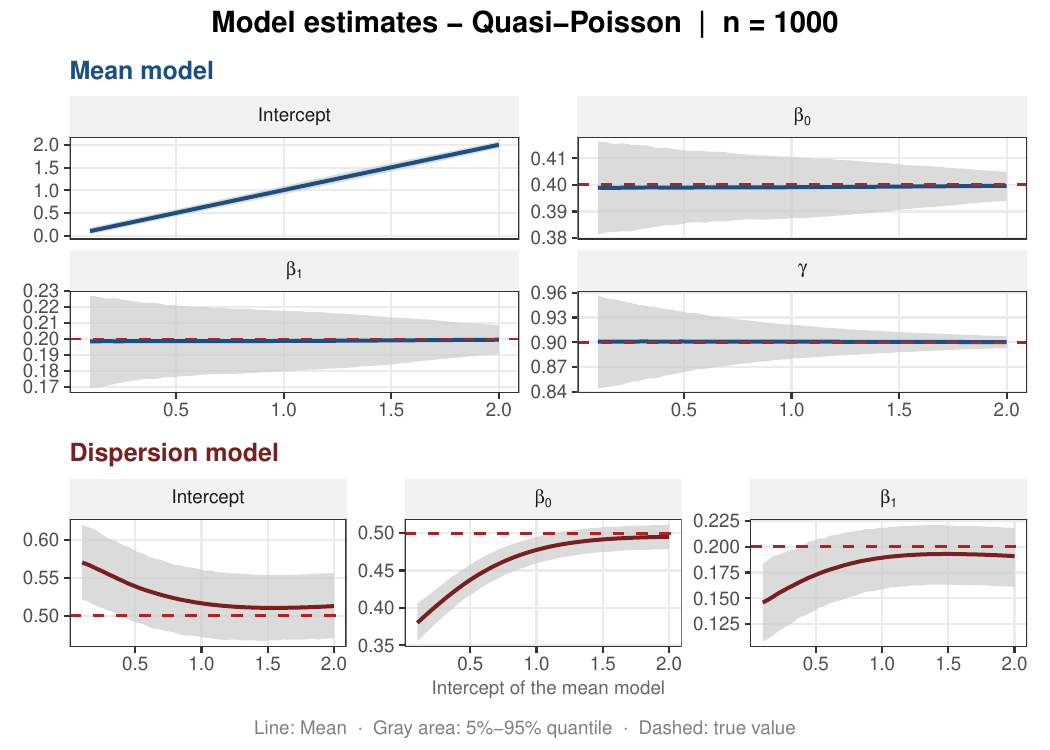}
		\caption{Marginals generated by a generalized Poisson distribution}
        \label{fig:simulation_increasing_intercept_vpoisson}
	\end{subfigure}
	\caption{Mean parameter estimates of mean model~\eqref{eq:simulation_mean_without_feedback} and dispersion model~\eqref{eq:simulation_dispersion_without_feedback} as the intercept of the mean model varies, based on a $5 \times 5$ grid with $T = 1000$. Shaded regions indicate pointwise empirical 90\% intervals.}
	\label{fig:simulation_increasing_intercept}
\end{figure}

Figure~\ref{fig:simulation_increasing_intercept_vnormal} shows the mean parameter estimates for the case of marginal normal distributions with a time series length of $T = 1000$. The shaded gray region encloses the pointwise empirical 5\% and 95\% quantiles of the parameter estimates. Figure~\ref{fig:simulation_increasing_intercept_vpoisson} presents the results for the marginal generalized Poisson distribution. Whereas the magnitude of the intercept has no notable effect on parameter estimates in the normal distribution setting, the count data setting exhibits increased precision in estimating the mean model parameters as the intercept increases. Furthermore, the parameter estimates of the dispersion model converge on average to the true parameters as the intercept of the mean model increases, though this does not affect the variance of the estimates.

The intercept governs the overall level of the observations, which in turn affects the Gamma approximation of the pseudo-observations employed in the estimation of the dispersion model. While this approximation is known to be exact for the normal distribution \citep{smyth_generalized_1989}, count data require a sufficiently high observation level. In the count data setting with $T = 1000$, the simulated observations vary on average between approximately $4.492$ ($\delta_0 = 0.1$) and $492.879$ ($\delta_0 = 2$). From $\delta_0 \approx 1$, corresponding to an average of approximately $38.83$, the true parameter value is contained within the empirical 90\% confidence interval of the parameter estimates.

In summary, the combination of the mean model and dispersion in our framework is able to detect spatial and temporal dependencies in the dispersion, but the effect might be systematically underestimated in some cases. Further studies are needed to investigate under which conditions the relationships can be well estimated or improved. For this purpose, REML based estimation procedures analogous to the DGLMs for independent data \citep{smyth_adjusted_1999} may be used. In the mean model, the dependencies are already captured very well on average.

% No Feedback at all:
\begin{figure}[p]
	\centering
	\begin{subfigure}{\textwidth}
		\centering
		\includegraphics[width=\textwidth, keepaspectratio]{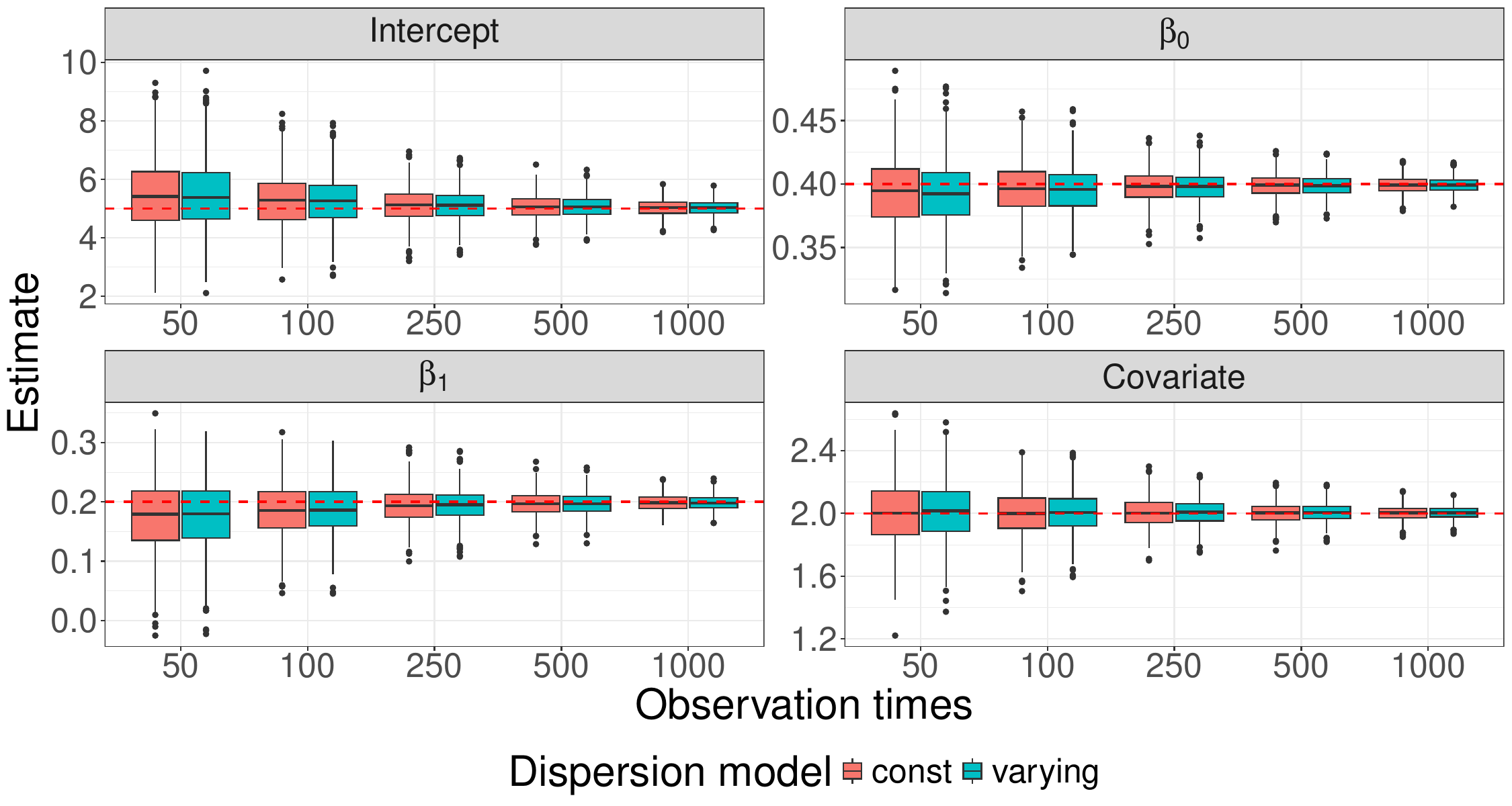}
		\caption{Parameter estimates of the mean model~\eqref{eq:simulation_mean_without_feedback} under the constant dispersion specification~\eqref{eq:simulation_dispersion_constant} and the varying dispersion specification~\eqref{eq:simulation_dispersion_without_feedback}.}
	\end{subfigure}
	
	\vspace{3em}
	
	\begin{subfigure}{\textwidth}
		\centering
		\includegraphics[width=\textwidth, keepaspectratio]{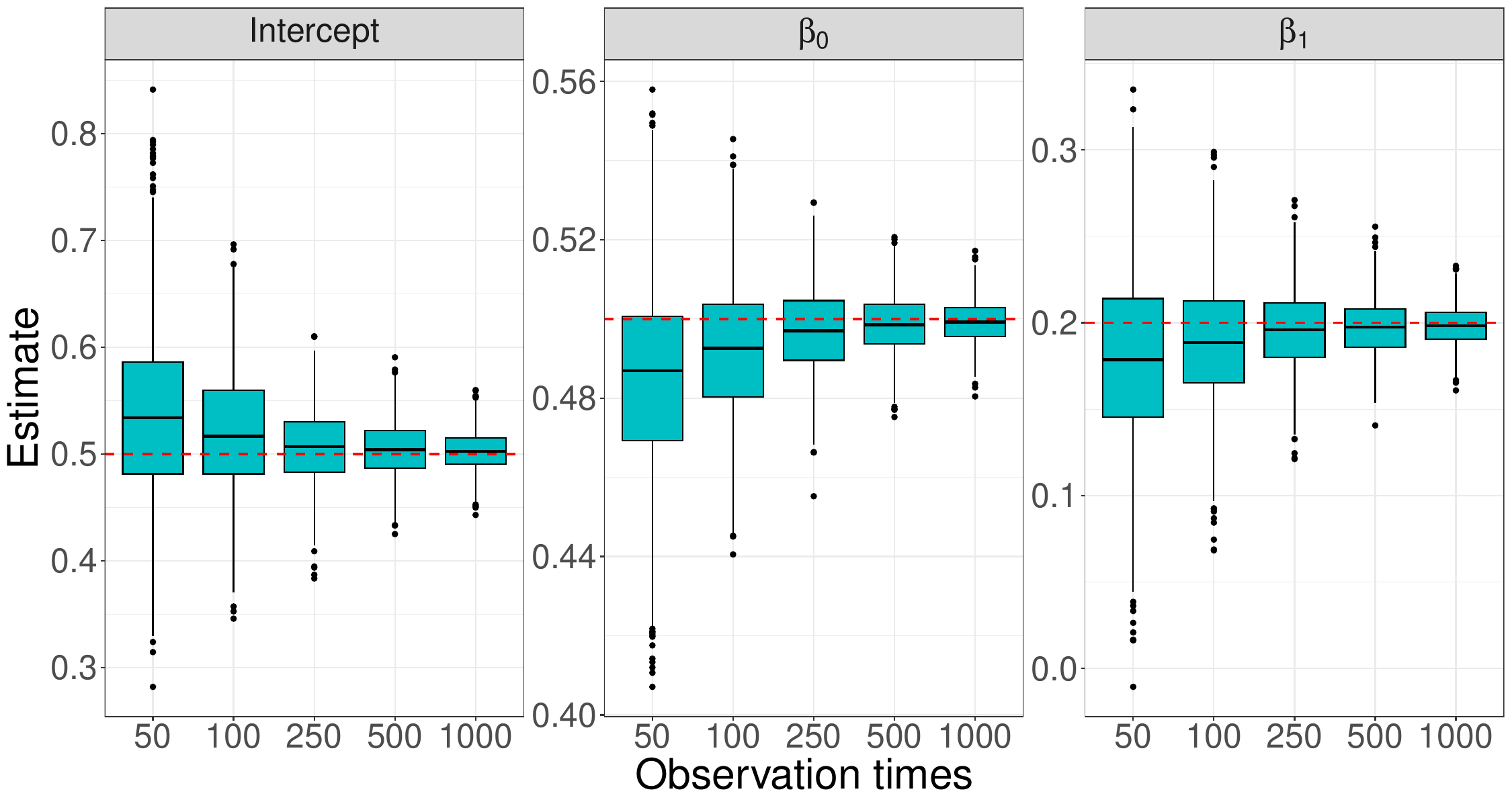}
		\caption{Parameter estimates of the varying dispersion model~\eqref{eq:simulation_dispersion_without_feedback} combined with the mean model~\eqref{eq:simulation_mean_without_feedback}.}
	\end{subfigure}
	\caption{Box plots of parameter estimates under alternative dispersion specifications.  Panel~(a) shows estimates of the mean model under constant and varying dispersion, whereas panel~(b) displays estimates of the varying dispersion model. The data-generating process with marginal normal distributions corresponds to \eqref{eq:simulation_mean_without_feedback} and~\eqref{eq:simulation_dispersion_without_feedback}, i.e., both models without feedback mechanism.}
	\label{fig:boxplot_normal_without_without}
\end{figure}

\begin{figure}[p]
	\centering
	\begin{subfigure}{\textwidth}
		\centering
		\includegraphics[width=\textwidth, keepaspectratio]{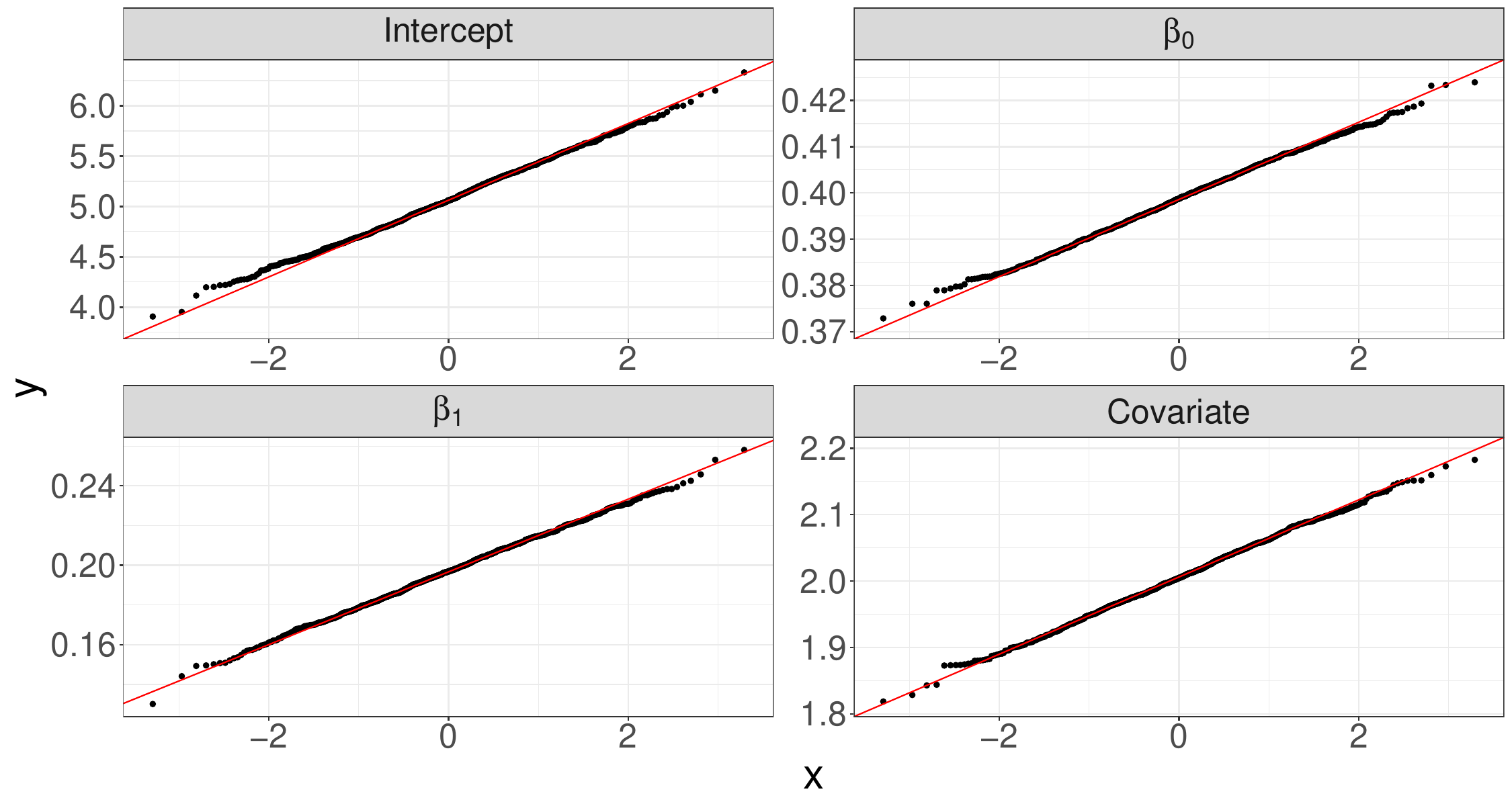}
		\caption{Mean model~\eqref{eq:simulation_mean_without_feedback}, i.e. without feedback mechanism.}
	\end{subfigure}
	
	\vspace{3em}
	
	\begin{subfigure}{\textwidth}
		\centering
		\includegraphics[width=\textwidth, keepaspectratio]{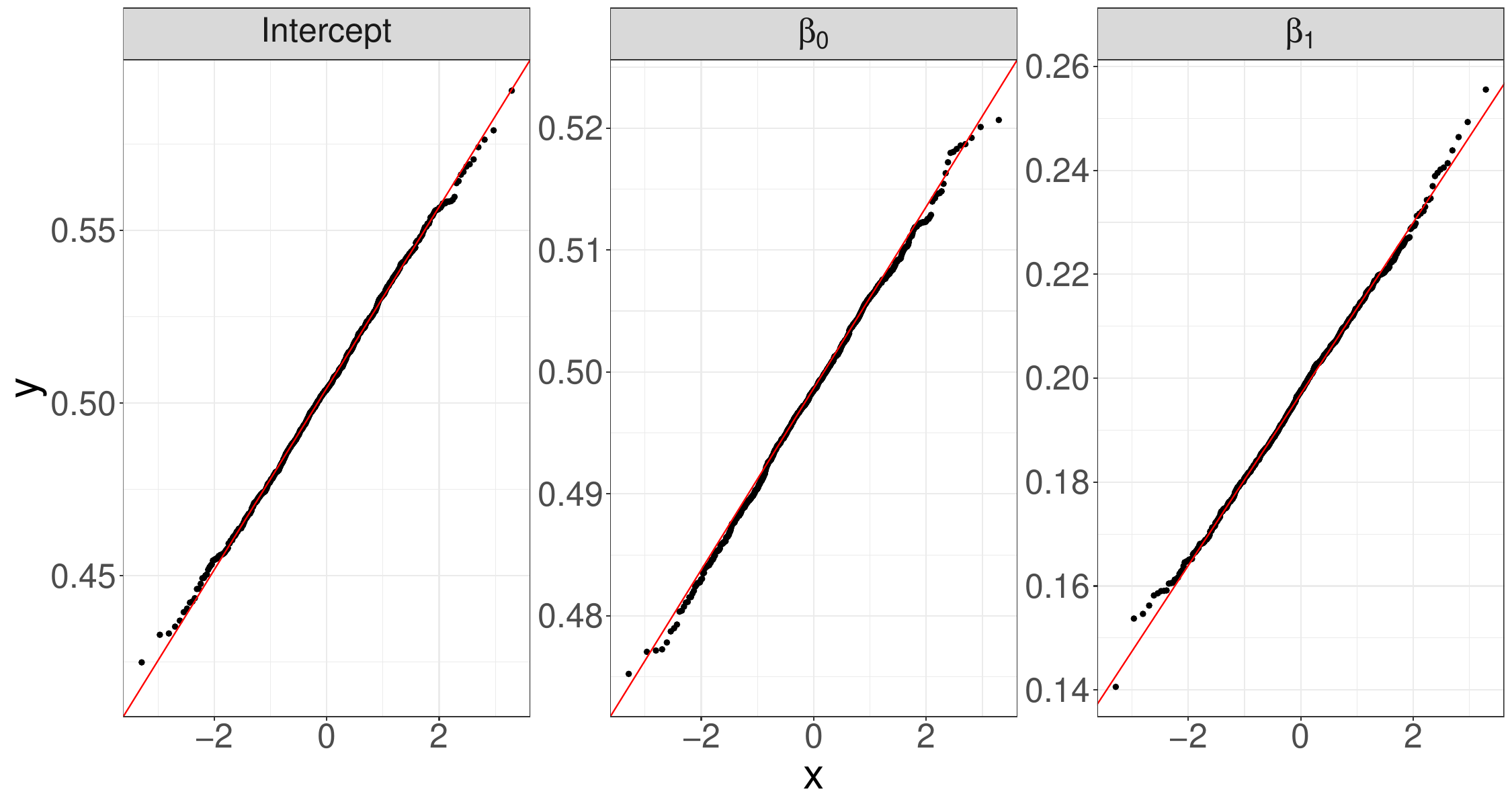}
		\caption{Dispersion model~\eqref{eq:simulation_dispersion_without_feedback}, i.e. without feedback mechanism.}
	\end{subfigure}
	\caption{Normal Q--Q plots of parameter estimates under the true data-generating process \eqref{eq:simulation_mean_without_feedback} and~\eqref{eq:simulation_dispersion_without_feedback} with marginal normal distributions, based on $T=500$ observation times.}
	\label{fig:qqplot_normal_without_without}
\end{figure}

% No Feedback mechanism in mean model
% Feedback mechanism in dispersion model

\begin{figure}[p]
	\centering
	\begin{subfigure}{\textwidth}
		\centering
		\includegraphics[width=\textwidth, keepaspectratio]{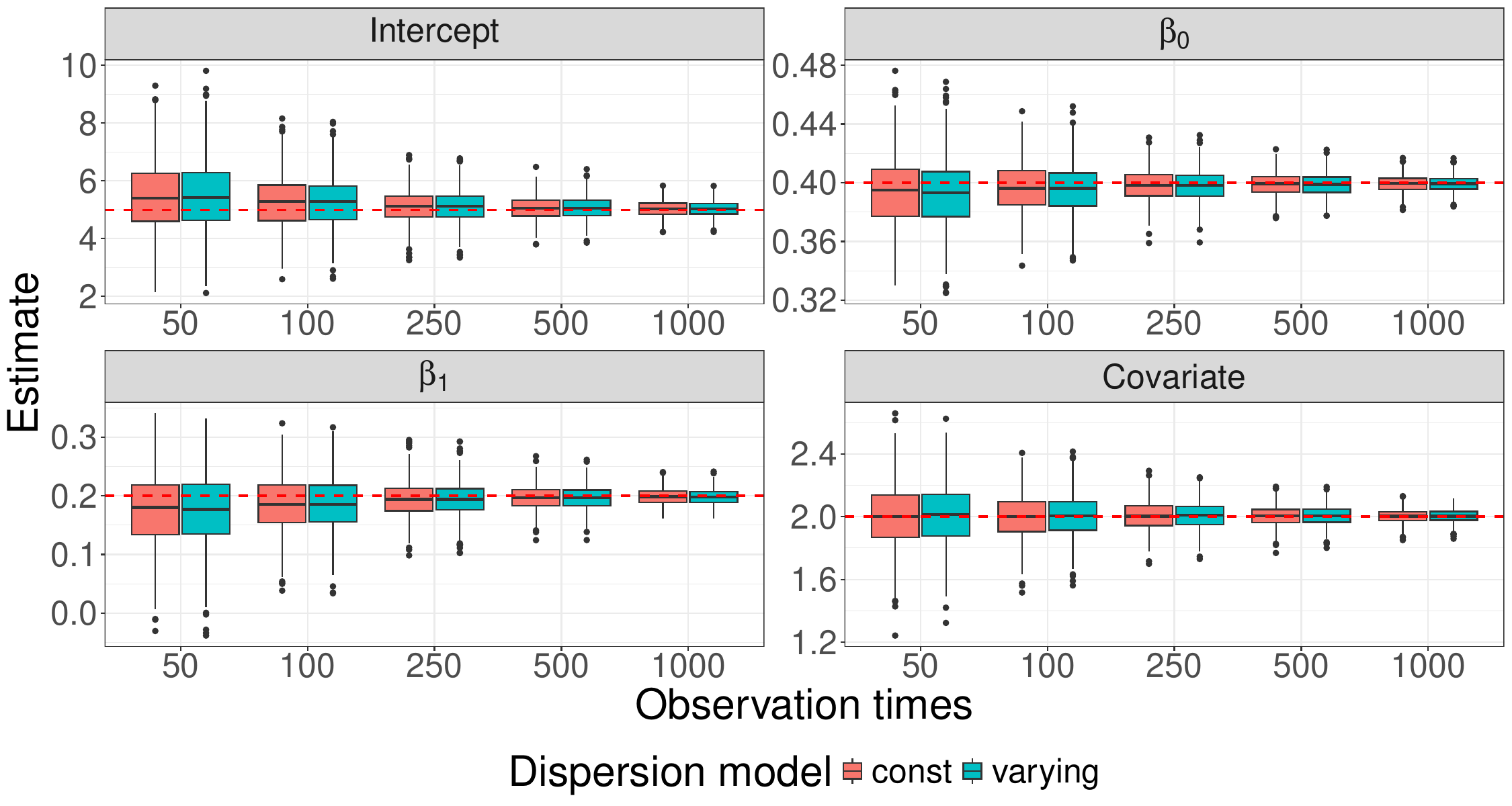}
		\caption{Parameter estimates of the mean model~\eqref{eq:simulation_mean_without_feedback} under the constant dispersion specification~\eqref{eq:simulation_dispersion_constant} and the varying dispersion specification~\eqref{eq:simulation_dispersion_with_feedback}.}
	\end{subfigure}
	
	\vspace{3em}
	
	\begin{subfigure}{\textwidth}
		\centering
		\includegraphics[width=\textwidth, keepaspectratio]{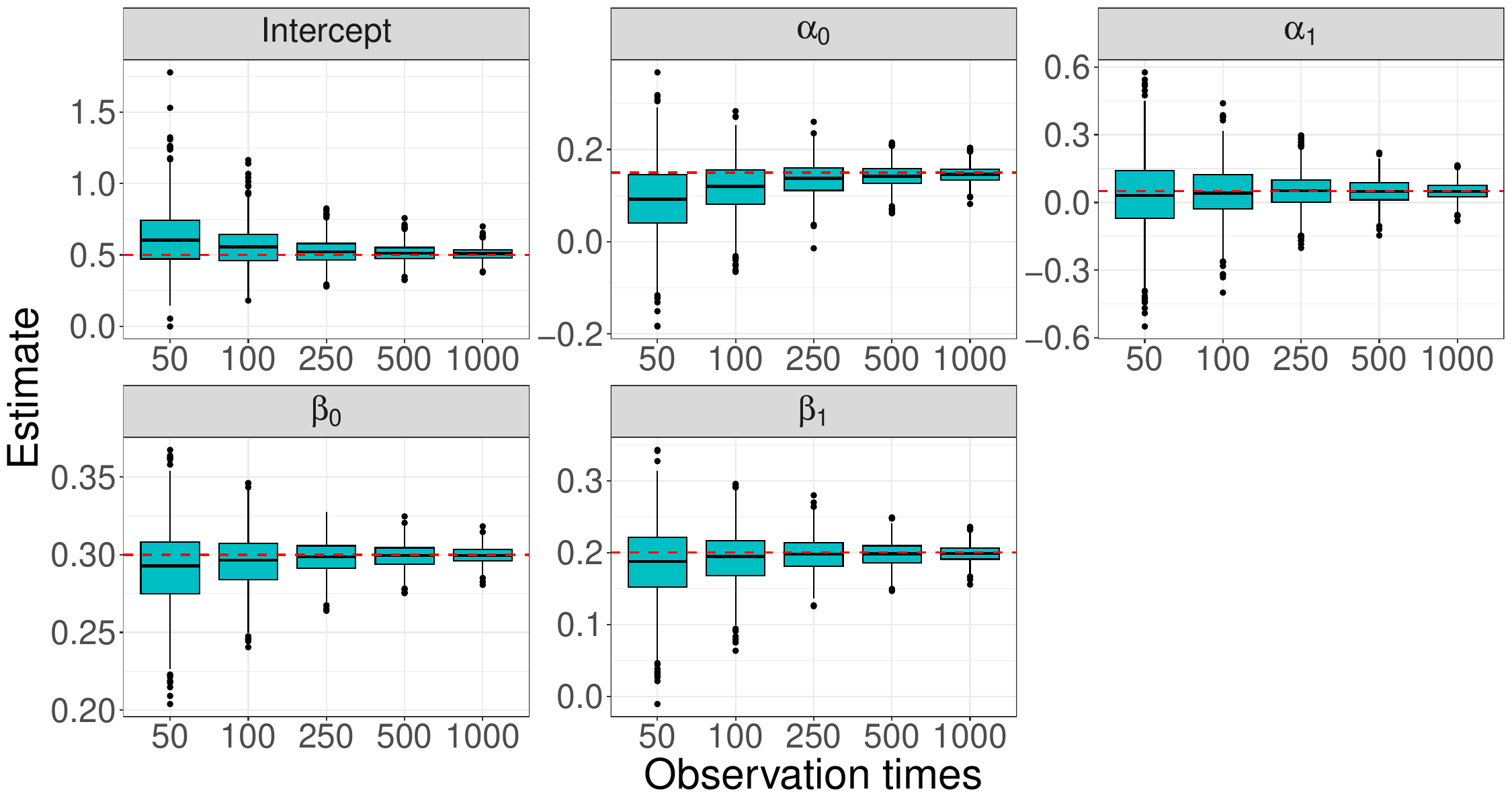}
		\caption{Parameter estimates of the varying dispersion model~\eqref{eq:simulation_dispersion_with_feedback} combined with the mean model~\eqref{eq:simulation_mean_without_feedback}.}
	\end{subfigure}
	\caption{Box plots of parameter estimates under alternative dispersion specifications.  Panel~(a) shows estimates of the mean model under constant and varying dispersion, whereas panel~(b) displays estimates of the varying dispersion model. The data-generating process with marginal normal distributions corresponds to \eqref{eq:simulation_mean_without_feedback} and~\eqref{eq:simulation_dispersion_with_feedback}, i.e., the mean model does not include a feedback mechanism while the dispersion process does.}
	\label{fig:boxplot_normal_without_with}
\end{figure}

\begin{figure}[p]
\centering
	\begin{subfigure}{\textwidth}
		\centering
		\includegraphics[width=\textwidth, keepaspectratio]{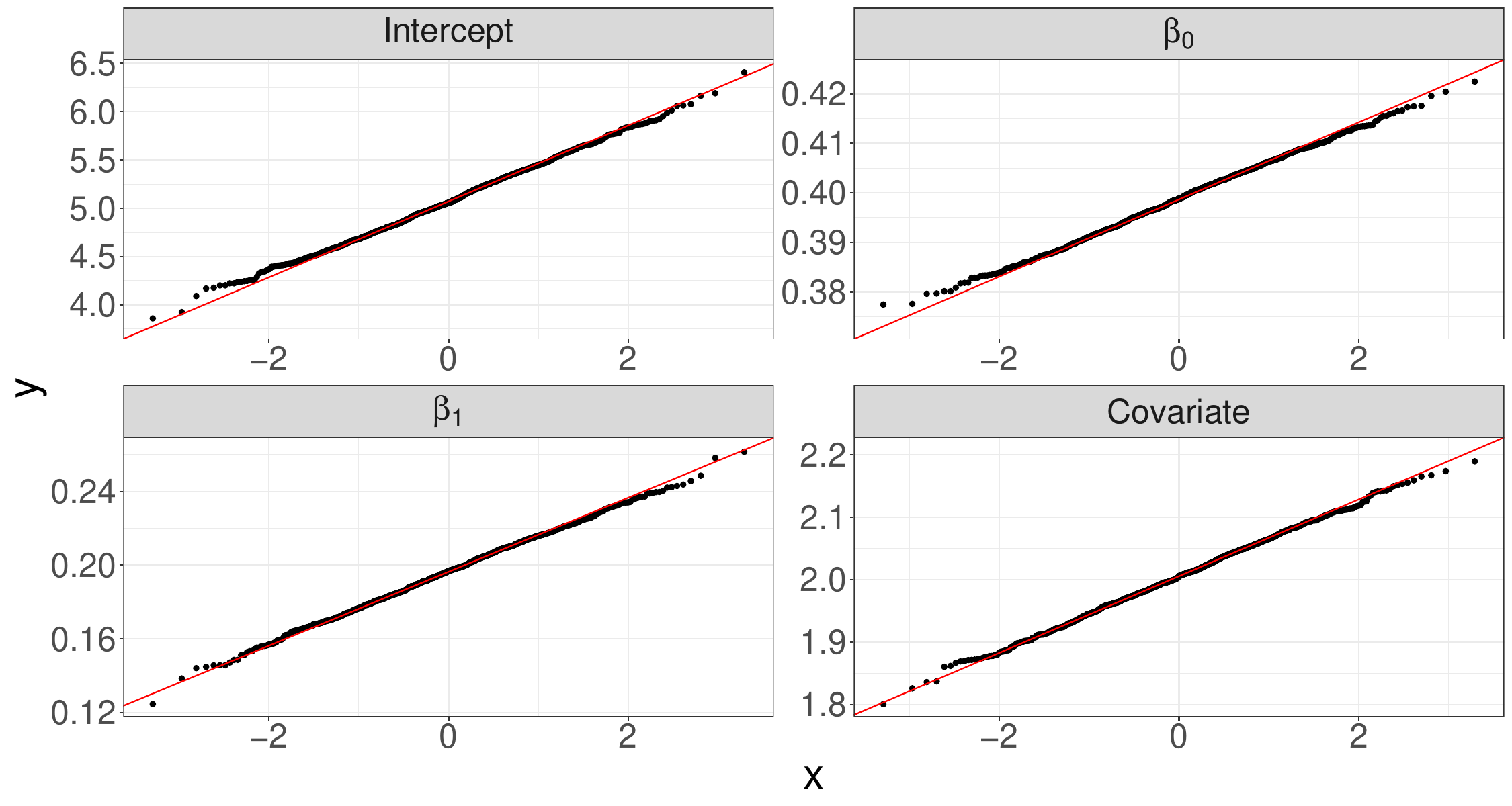}
		\caption{Mean model~\eqref{eq:simulation_mean_without_feedback}, i.e., without feedback mechanism.}
	\end{subfigure}
	
	\vspace{3em}
	
	\begin{subfigure}{\textwidth}
		\centering
		\includegraphics[width=\textwidth, keepaspectratio]{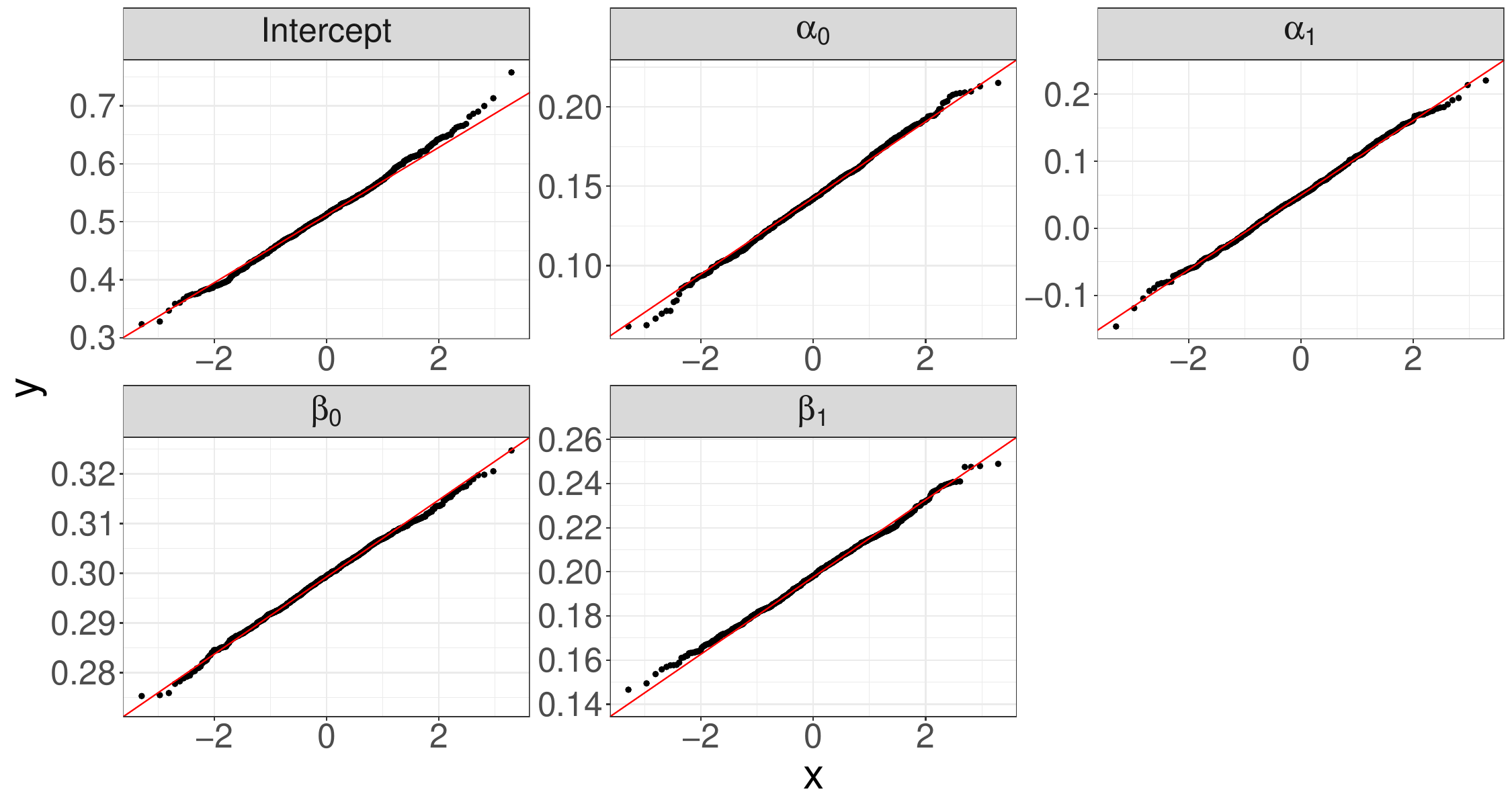}
		\caption{Dispersion model~\eqref{eq:simulation_dispersion_with_feedback}, i.e., with feedback mechanism.}
	\end{subfigure}
	\caption{Normal Q--Q plots of parameter estimates under the true data-generating process \eqref{eq:simulation_mean_without_feedback} and~\eqref{eq:simulation_dispersion_with_feedback} with marginal normal distributions, based on $T=500$ observation times.}
	\label{fig:qqplot_normal_without_with}
\end{figure}

% Feedback mechanism in Meanmodel
% No Feedback in Dispersion model

\begin{figure}[p]
	\centering
	\begin{subfigure}{\textwidth}
		\centering
		\includegraphics[width=\textwidth, keepaspectratio]{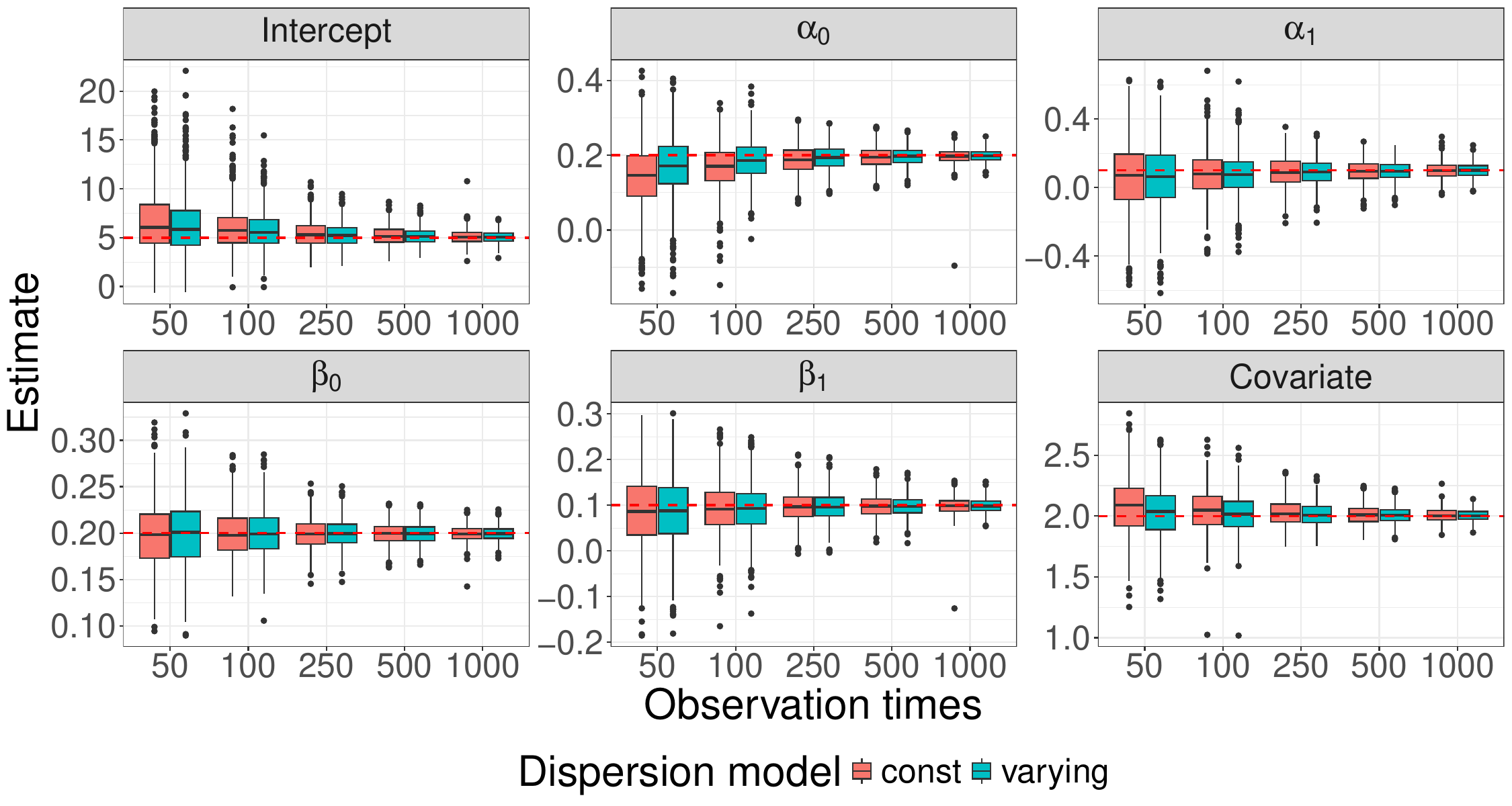}
		\caption{Parameter estimates of the mean model~\eqref{eq:simulation_mean_with_feedback} under the constant dispersion specification~\eqref{eq:simulation_dispersion_constant} and the varying dispersion specification~\eqref{eq:simulation_dispersion_without_feedback}.}
	\end{subfigure}
	
	\vspace{3em}
	
	\begin{subfigure}{\textwidth}
		\centering
		\includegraphics[width=\textwidth, keepaspectratio]{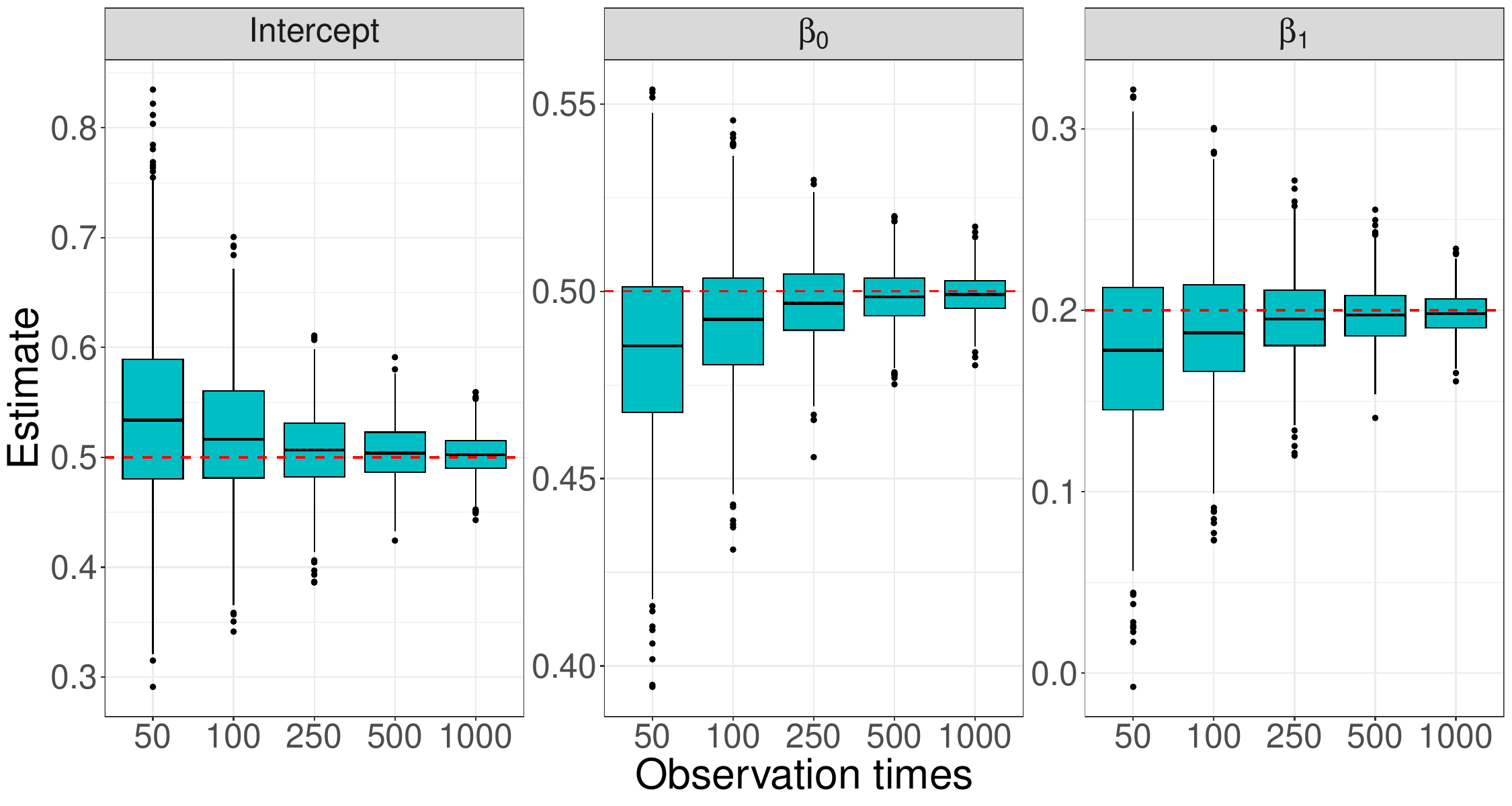}
		\caption{Parameter estimates of the varying dispersion model~\eqref{eq:simulation_dispersion_without_feedback} combined with the mean model~\eqref{eq:simulation_mean_with_feedback}.}
	\end{subfigure}
	\caption{Box plots of parameter estimates under alternative dispersion specifications.  Panel~(a) shows estimates of the mean model under constant and varying dispersion, whereas panel~(b) displays estimates of the varying dispersion model. The data-generating process with marginal normal distributions corresponds to \eqref{eq:simulation_mean_with_feedback} and~\eqref{eq:simulation_dispersion_without_feedback}, i.e., the mean model includes a feedback mechanism and the dispersion process not.}
	\label{fig:boxplot_normal_with_without}
\end{figure}

\begin{figure}[p]
\centering
	\begin{subfigure}{\textwidth}
		\centering
		\includegraphics[width=\textwidth, keepaspectratio]{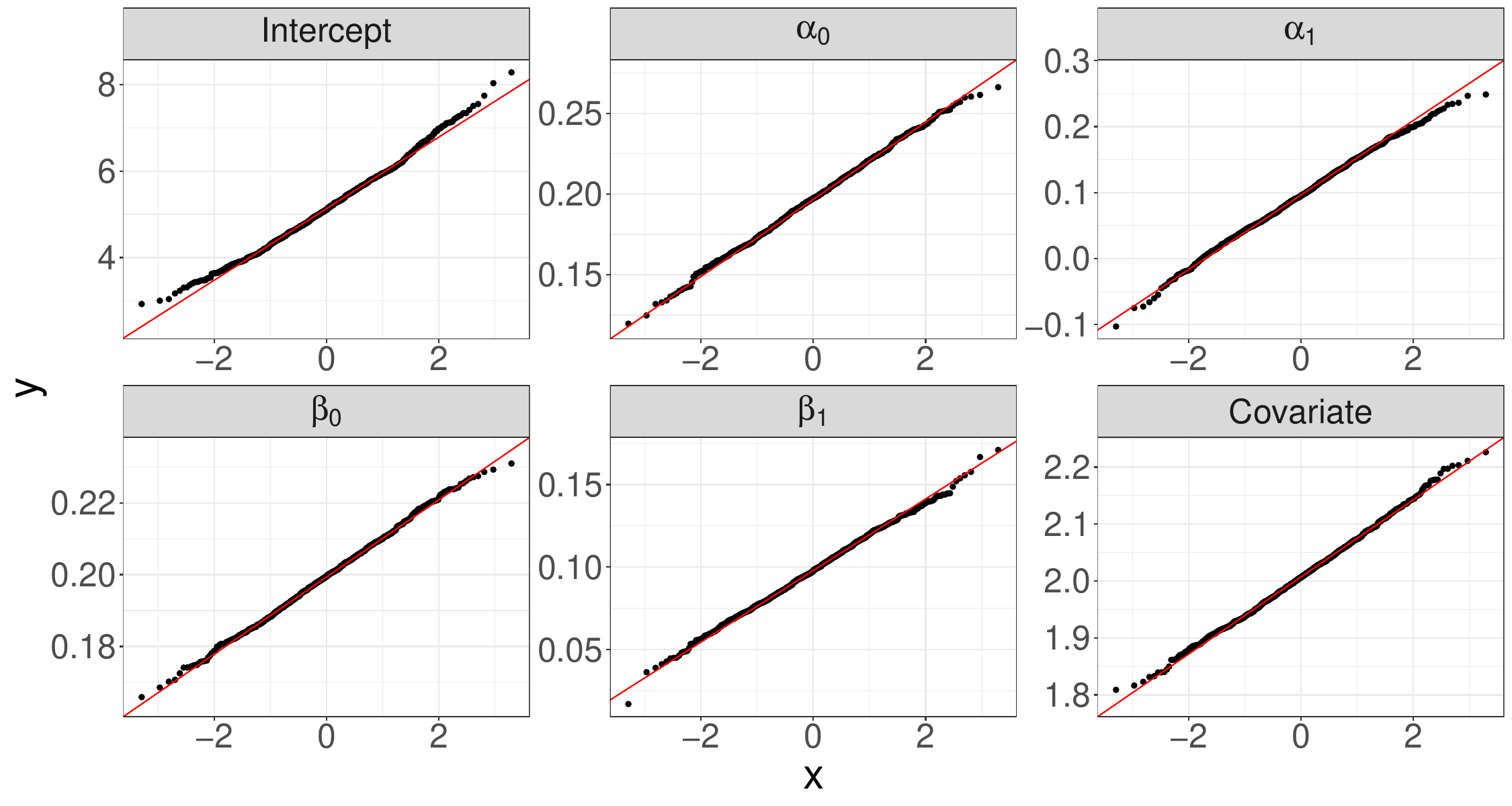}
		\caption{Mean model~\eqref{eq:simulation_mean_with_feedback}, i.e., with feedback mechanism.}
	\end{subfigure}
	
	\vspace{3em}
	
	\begin{subfigure}{\textwidth}
		\centering
		\includegraphics[width=\textwidth, keepaspectratio]{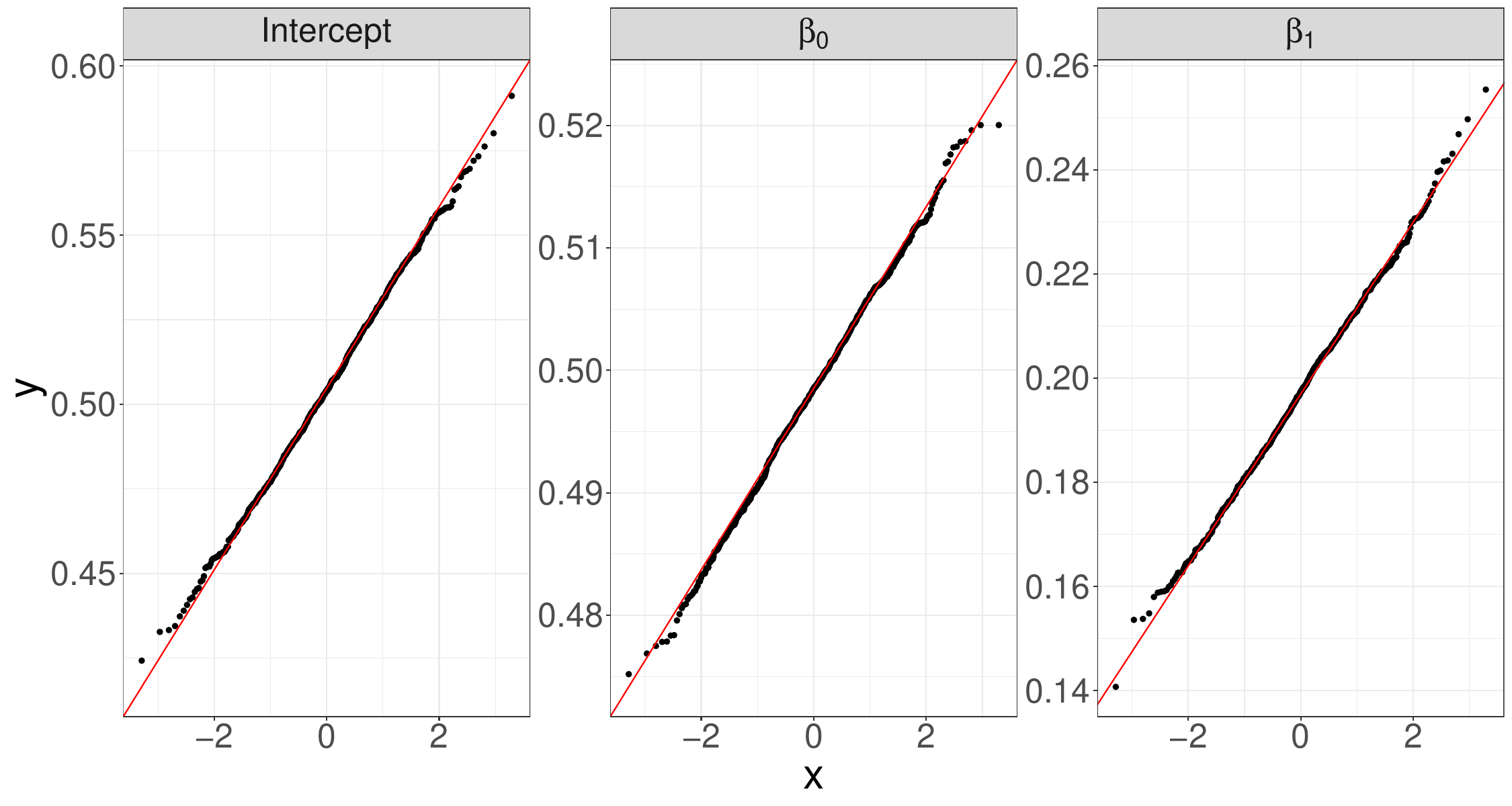}
		\caption{Dispersion model~\eqref{eq:simulation_dispersion_without_feedback}, i.e., without feedback mechanism.}
	\end{subfigure}
	\caption{Normal Q--Q plots of parameter estimates under the true data-generating process \eqref{eq:simulation_mean_with_feedback} and~\eqref{eq:simulation_dispersion_without_feedback} with marginal normal distributions, based on $T=500$.}
	\label{fig:qqplot_normal_with_without}
\end{figure}

% Feedbackmechanism in mean and dispersion

\begin{figure}[p]
	\centering
	\begin{subfigure}{\textwidth}
		\centering
		\includegraphics[width=\textwidth, keepaspectratio]{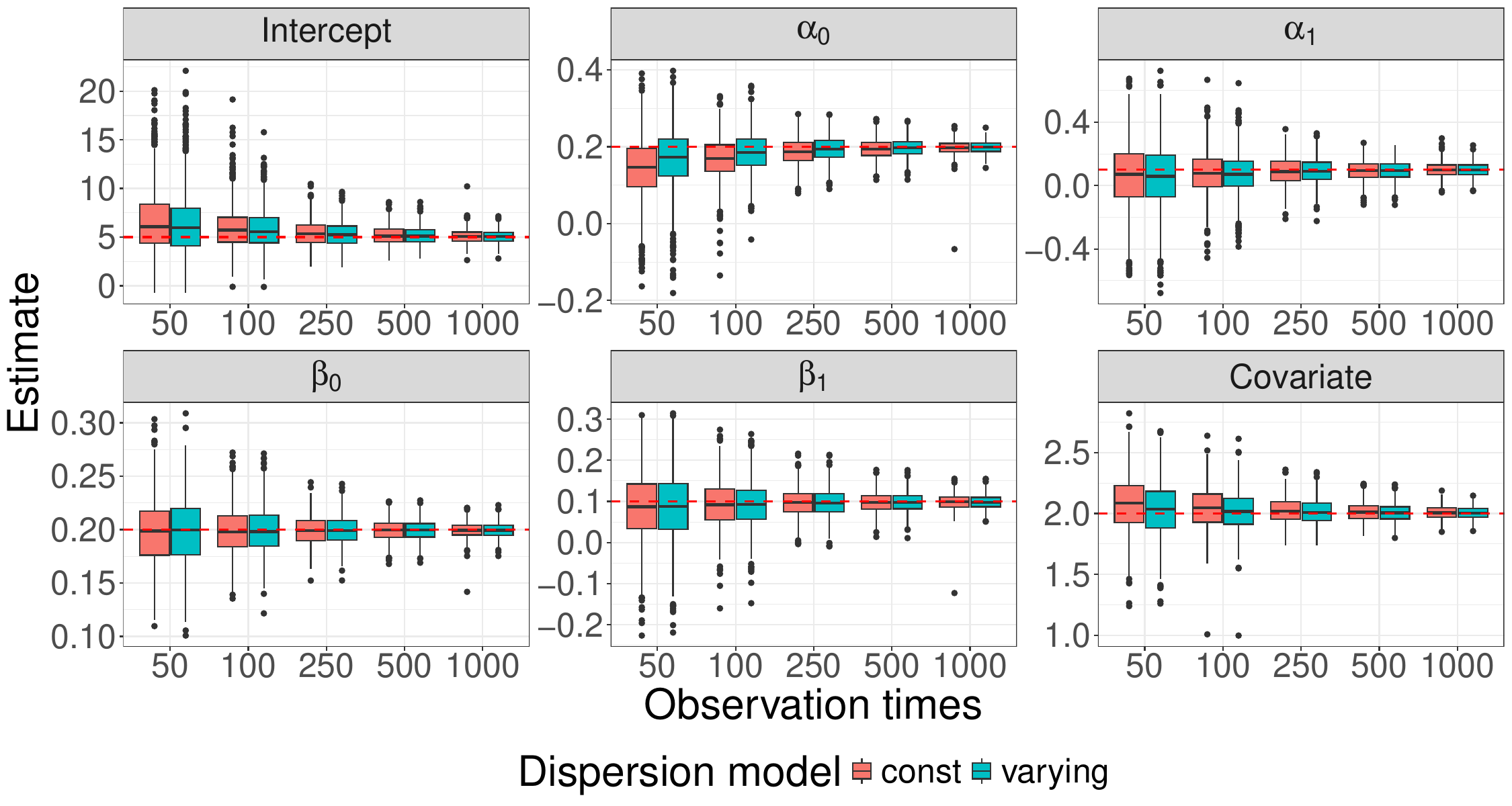}
		\caption{Parameter estimates of the mean model~\eqref{eq:simulation_mean_with_feedback} under the constant dispersion specification~\eqref{eq:simulation_dispersion_constant} and the varying dispersion specification~\eqref{eq:simulation_dispersion_with_feedback}.}
	\end{subfigure}
	
	\vspace{3em}
	
	\begin{subfigure}{\textwidth}
		\centering
		\includegraphics[width=\textwidth, keepaspectratio]{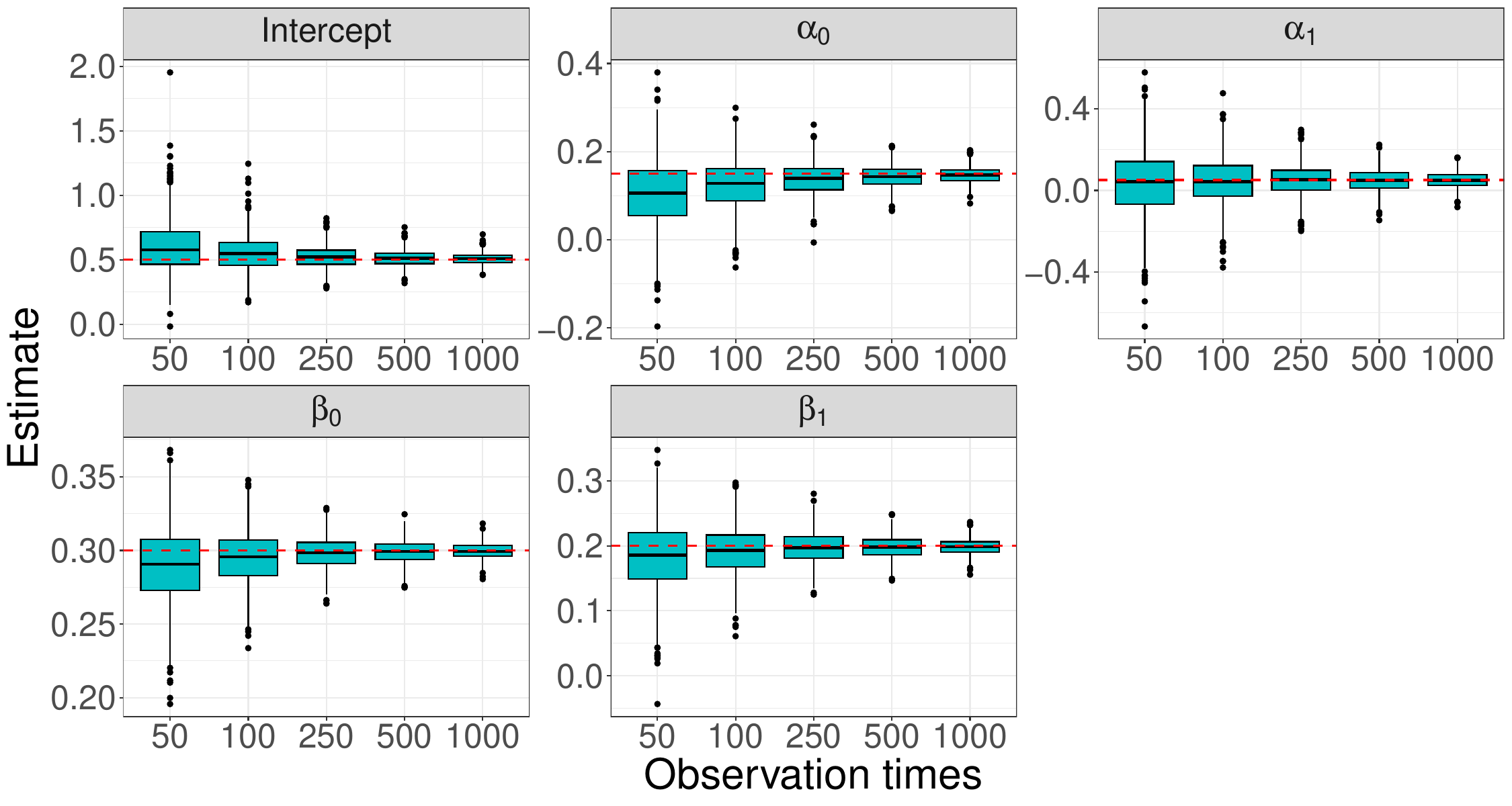}
		\caption{Parameter estimates of the varying dispersion model~\eqref{eq:simulation_dispersion_with_feedback} combined with the mean model~\eqref{eq:simulation_mean_with_feedback}.}
	\end{subfigure}
	\caption{Box plots of parameter estimates under alternative dispersion specifications.  Panel~(a) shows estimates of the mean model under constant and varying dispersion, whereas panel~(b) displays estimates of the varying dispersion model. The data-generating process with marginal normal distributions corresponds to \eqref{eq:simulation_mean_without_feedback} and~\eqref{eq:simulation_dispersion_with_feedback}, i.e., both models include a feedback mechanism}
	\label{fig:boxplot_normal_with_with}
\end{figure}

\begin{figure}[p]
\centering
	\begin{subfigure}{\textwidth}
		\centering
		\includegraphics[width=\textwidth, keepaspectratio]{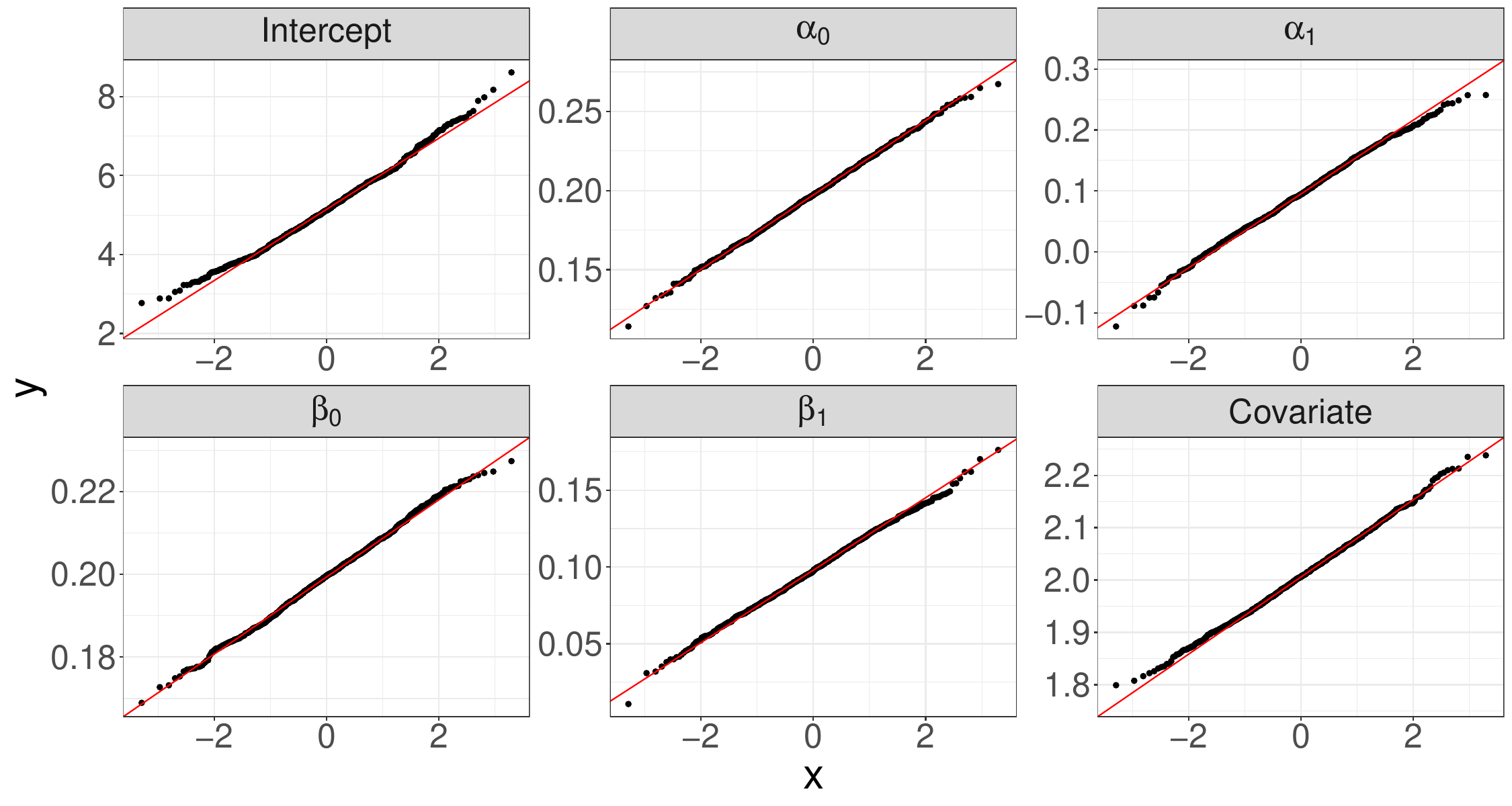}
		\caption{Mean model~\eqref{eq:simulation_mean_with_feedback}, i.e., with feedback mechanism.}
	\end{subfigure}
	
	\vspace{1em}
	
	\begin{subfigure}{\textwidth}
		\centering
		\includegraphics[width=\textwidth, keepaspectratio]{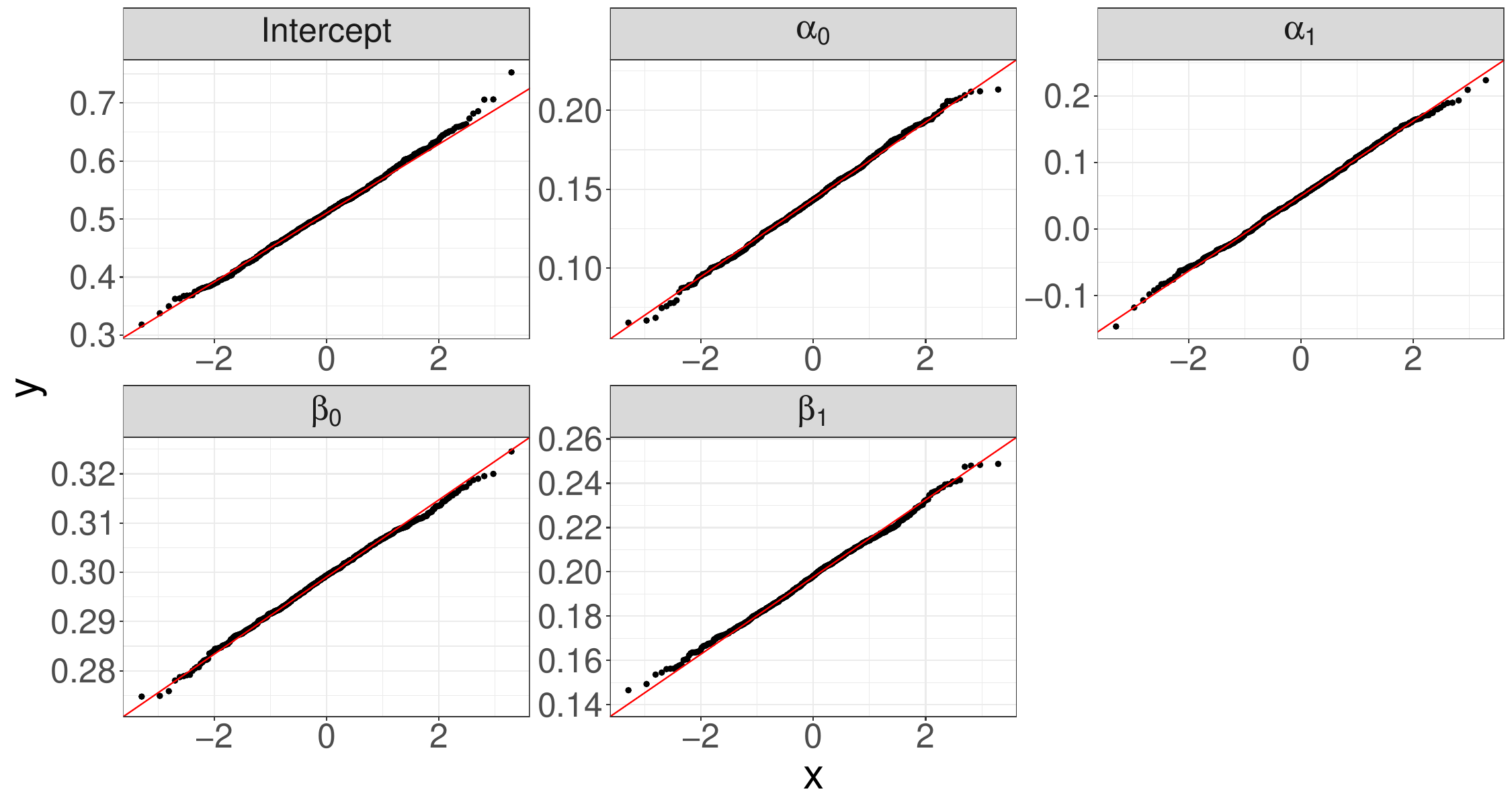}
		\caption{Dispersion model~\eqref{eq:simulation_dispersion_with_feedback}, i.e., with feedback mechanism.}
	\end{subfigure}
	\caption{Normal Q--Q plots of parameter estimates under the true data-generating process \eqref{eq:simulation_mean_with_feedback} and~\eqref{eq:simulation_dispersion_with_feedback} with marginal normal distributions, based on $T=500$.}
	\label{fig:qqplot_normal_with_with}
\end{figure}

% No Feedback at all:

\begin{figure}[p]
	\centering
	\begin{subfigure}{\textwidth}
		\centering
		\includegraphics[width=\textwidth, keepaspectratio]{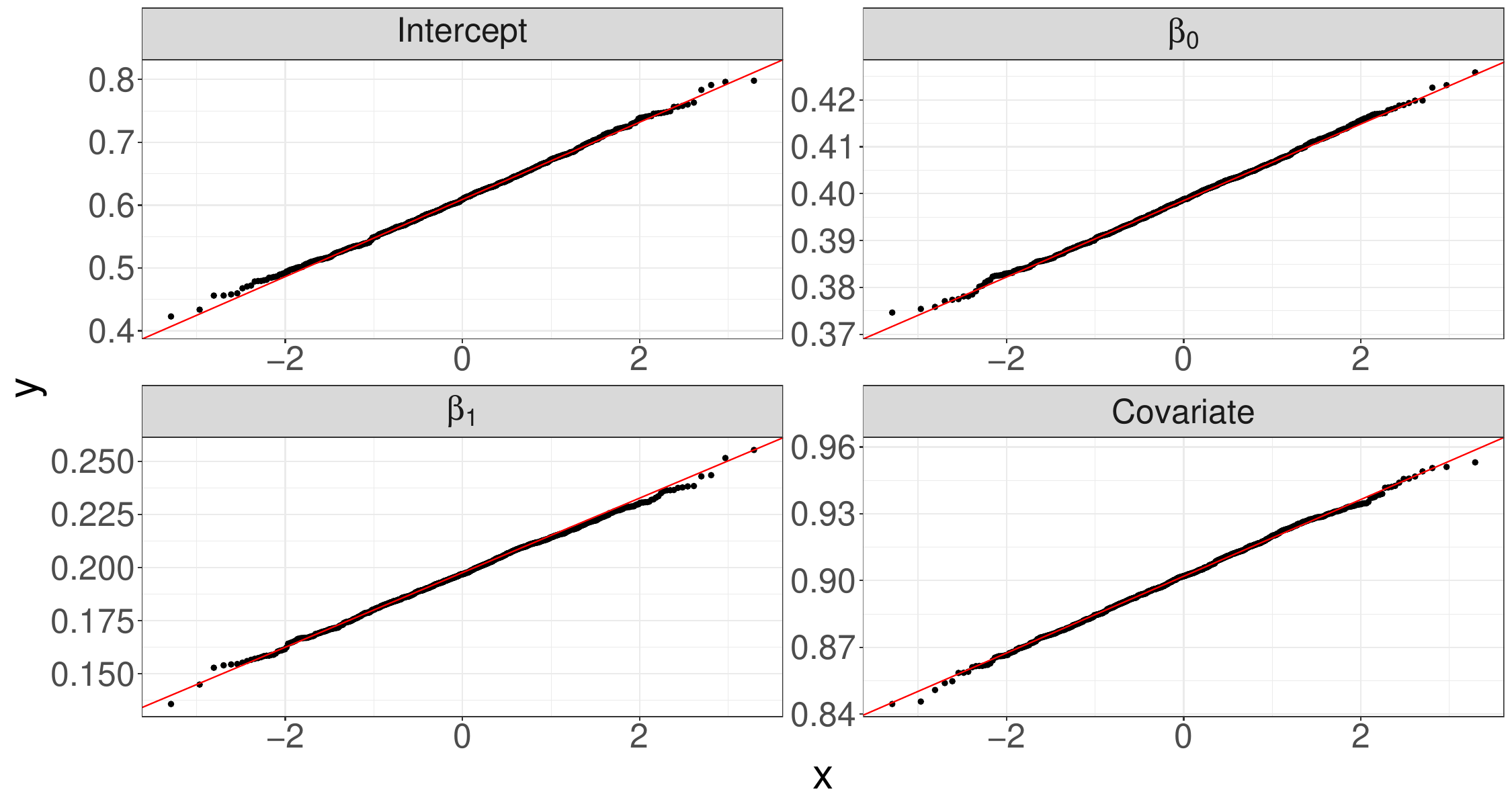}
		\caption{Mean model~\eqref{eq:simulation_mean_without_feedback}, i.e., without feedback mechanism.}
	\end{subfigure}
	
	\vspace{3em}
	
	\begin{subfigure}{\textwidth}
		\centering
		\includegraphics[width=\textwidth, keepaspectratio]{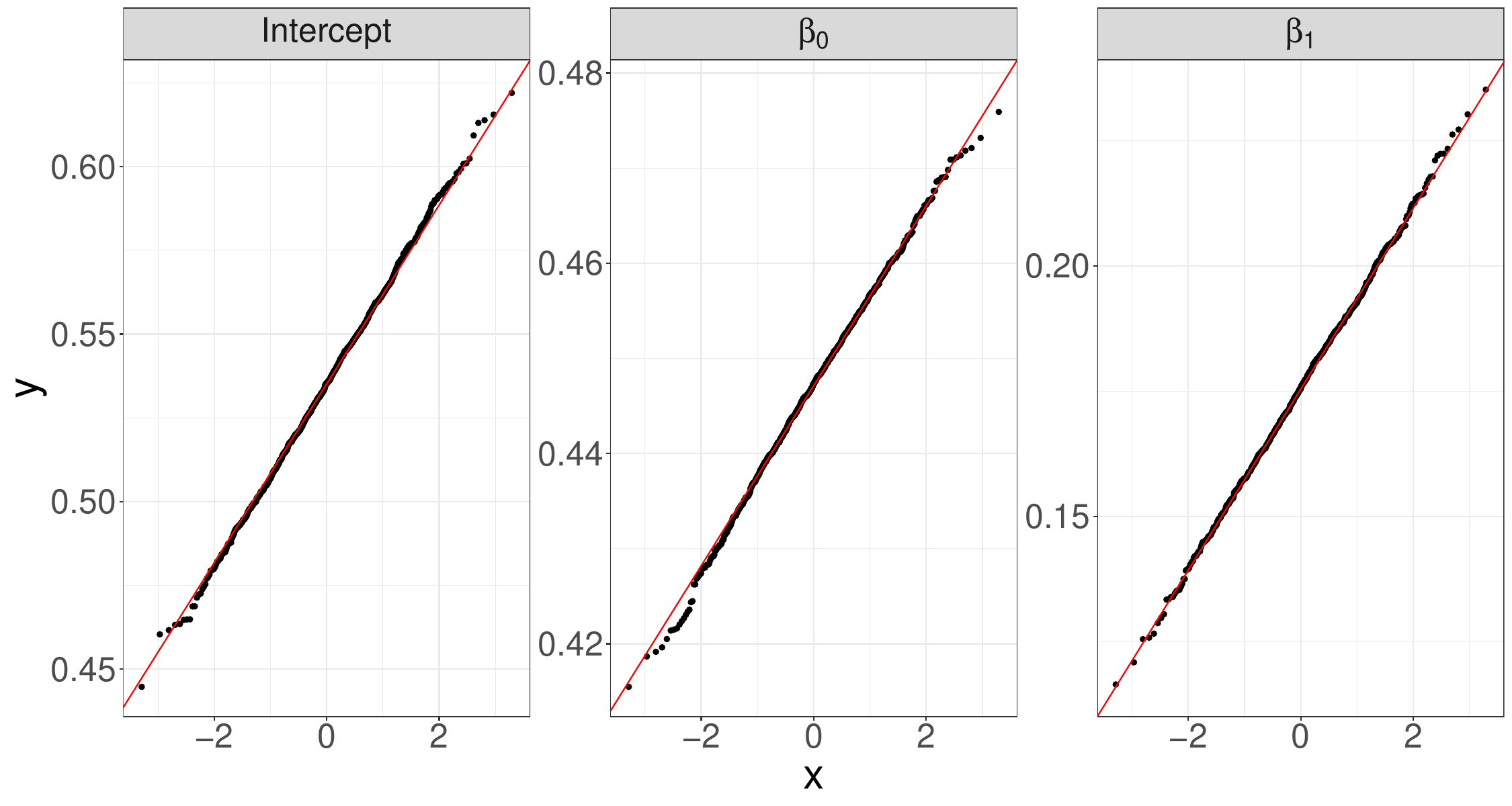}
		\caption{Dispersion model~\eqref{eq:simulation_dispersion_without_feedback}, i.e., without feedback mechanism.}
	\end{subfigure}
	\caption{Normal Q--Q plots of parameter estimates under the true data-generating process \eqref{eq:simulation_mean_without_feedback} and~\eqref{eq:simulation_dispersion_without_feedback} with generalized Poisson marginals, based on $T=500$.}
	\label{fig:qqplot_poisson_without_without}
\end{figure}

% No Feedback mechanism in mean model
% Feedback mechanism in dispersion model

\begin{figure}[p]
\centering
	\begin{subfigure}{\textwidth}
		\centering
		\includegraphics[width=\textwidth, keepaspectratio]{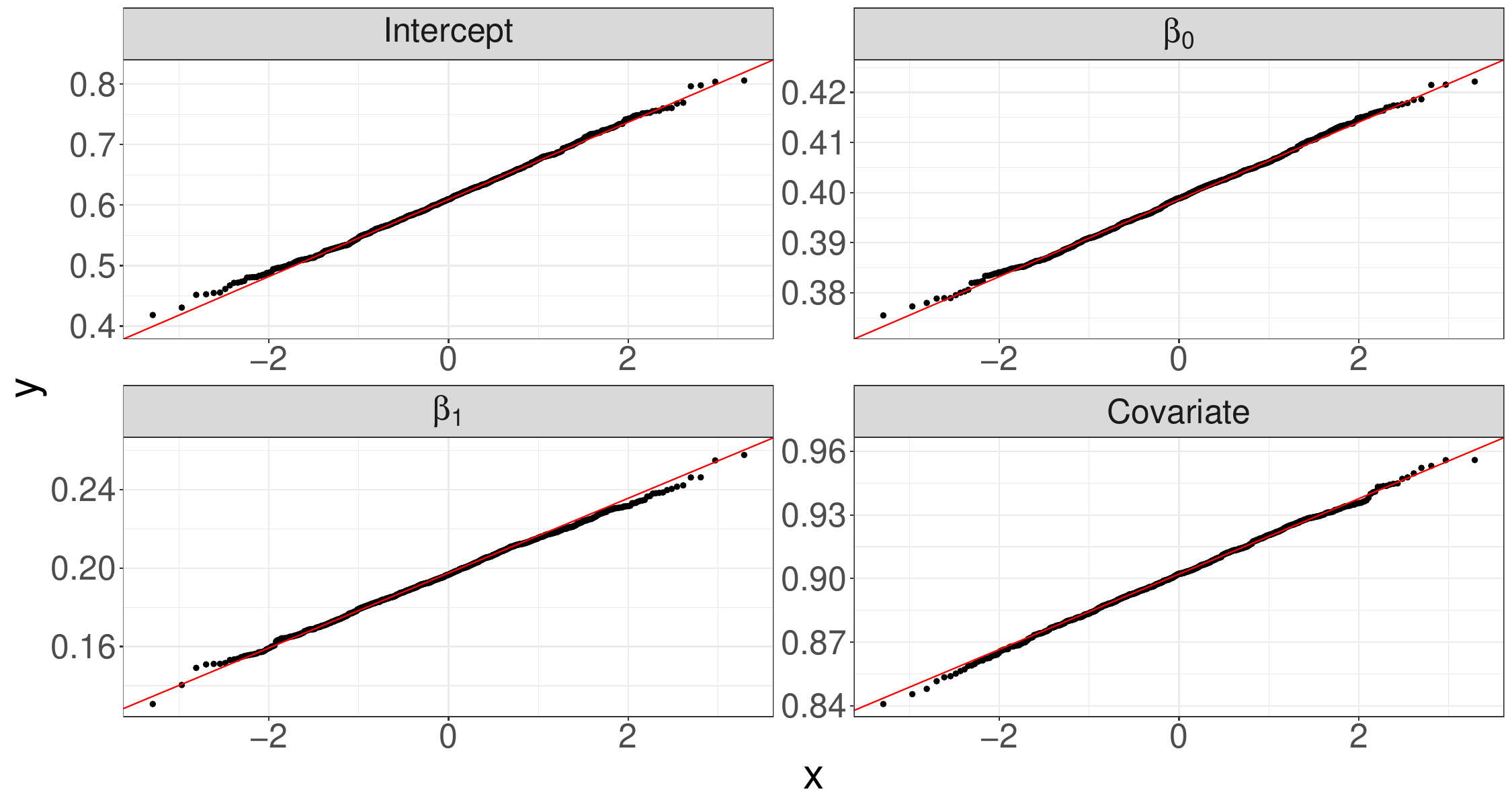}
		\caption{Mean model~\eqref{eq:simulation_mean_without_feedback}, i.e., without feedback mechanism.}
	\end{subfigure}
	
	\vspace{3em}
	
	\begin{subfigure}{\textwidth}
		\centering
		\includegraphics[width=\textwidth, keepaspectratio]{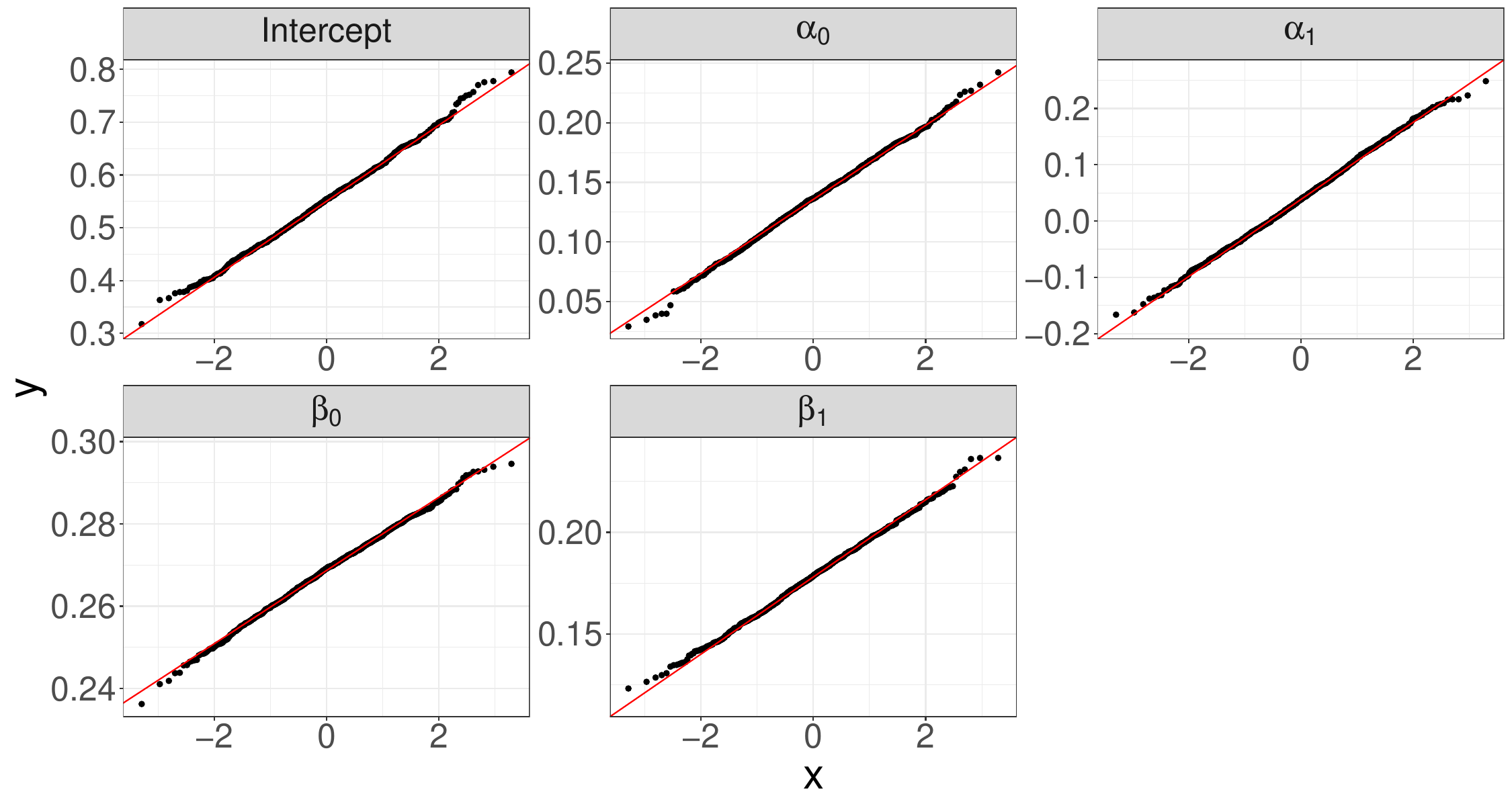}
		\caption{Dispersion model~\eqref{eq:simulation_dispersion_with_feedback}, i.e., with feedback mechanism.}
	\end{subfigure}
	\caption{Normal Q--Q plots of parameter estimates under the true data-generating process \eqref{eq:simulation_mean_without_feedback} and~\eqref{eq:simulation_dispersion_with_feedback} with generalized Poisson marginals, based on $T=500$.}
	\label{fig:qqplot_poisson_without_with}
\end{figure}

% Feedback mechanism in Meanmodel
% No Feedback in Dispersion model

\begin{figure}[p]
	\centering
	\begin{subfigure}{\textwidth}
		\centering
		\includegraphics[width=\textwidth, keepaspectratio]{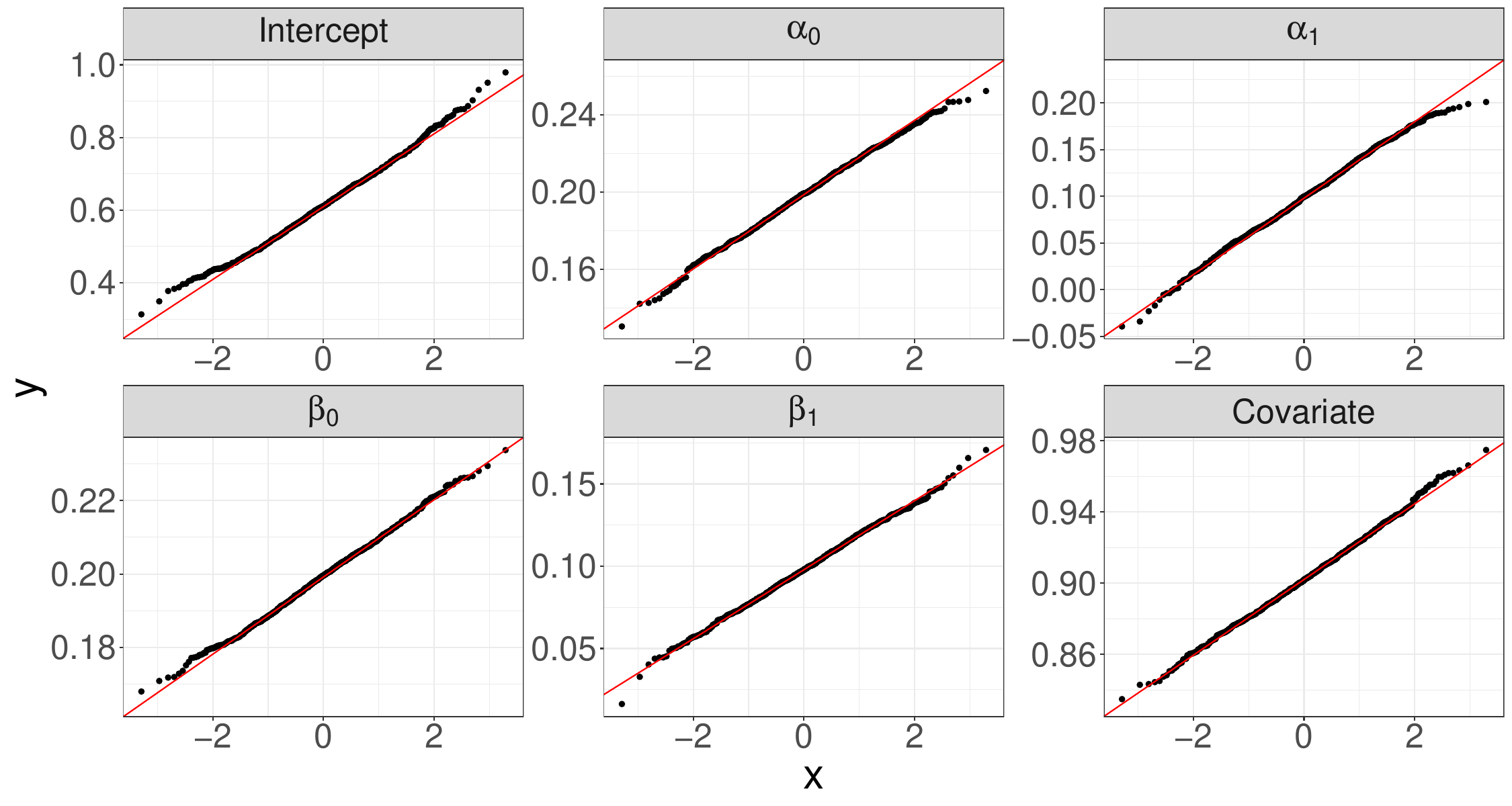}
		\caption{Mean model~\eqref{eq:simulation_mean_with_feedback}, i.e., with feedback mechanism.}
	\end{subfigure}
	
	\vspace{1em}
	
	\begin{subfigure}{\textwidth}
		\centering
		\includegraphics[width=\textwidth, keepaspectratio]{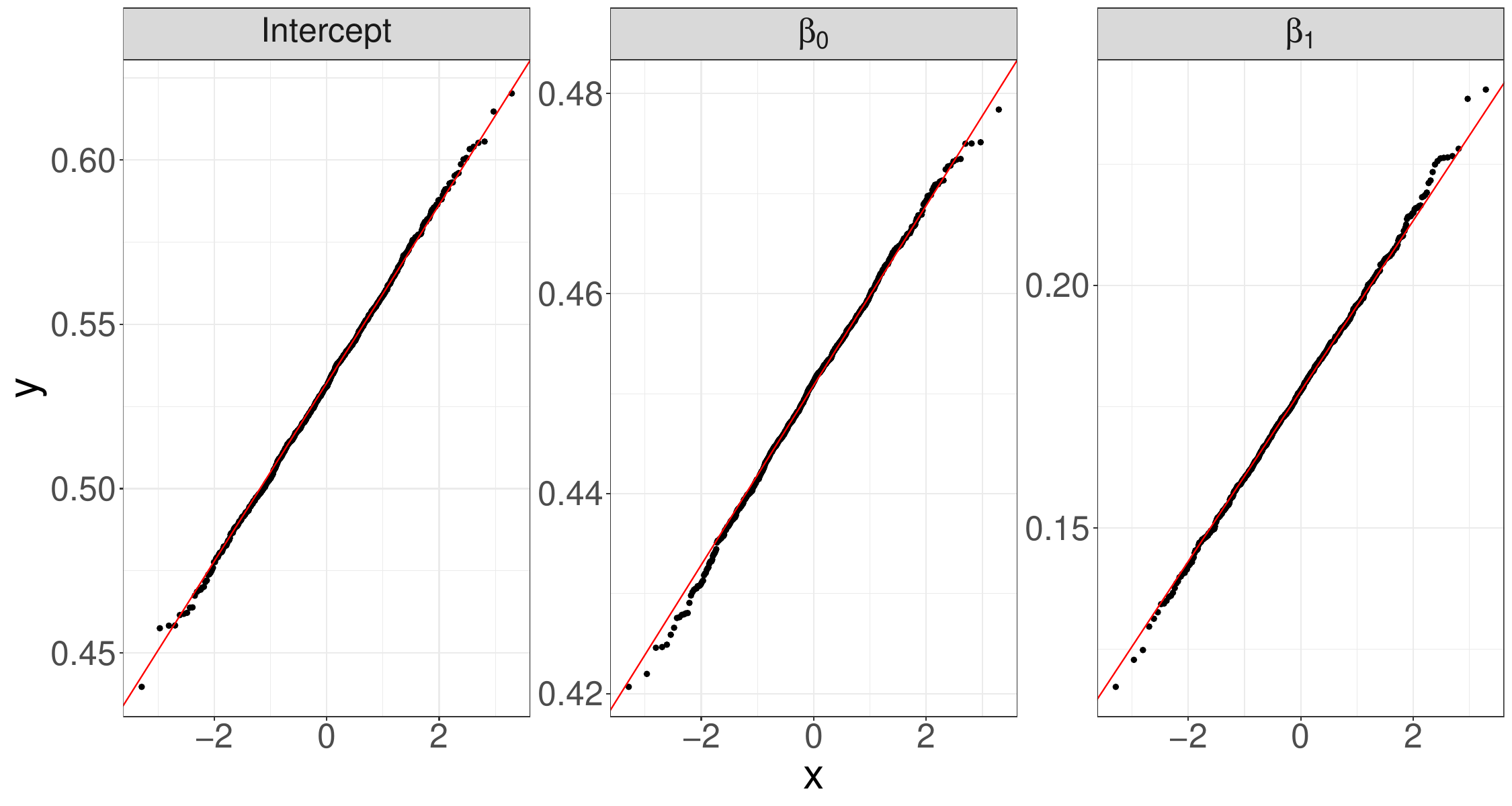}
		\caption{Dispersion model~\eqref{eq:simulation_dispersion_without_feedback}, i.e., without feedback mechanism.}
	\end{subfigure}
	\caption{Normal Q--Q plots of parameter estimates under the true data-generating process \eqref{eq:simulation_mean_with_feedback} and~\eqref{eq:simulation_dispersion_without_feedback} with generalized Poisson marginals, based on $T=500$.}
	\label{fig:qqplot_poisson_with_without}
\end{figure}

% Feedbackmechanism in mean and dispersion

\begin{figure}[p]
\centering
	\begin{subfigure}{\textwidth}
		\centering
		\includegraphics[width=\textwidth, keepaspectratio]{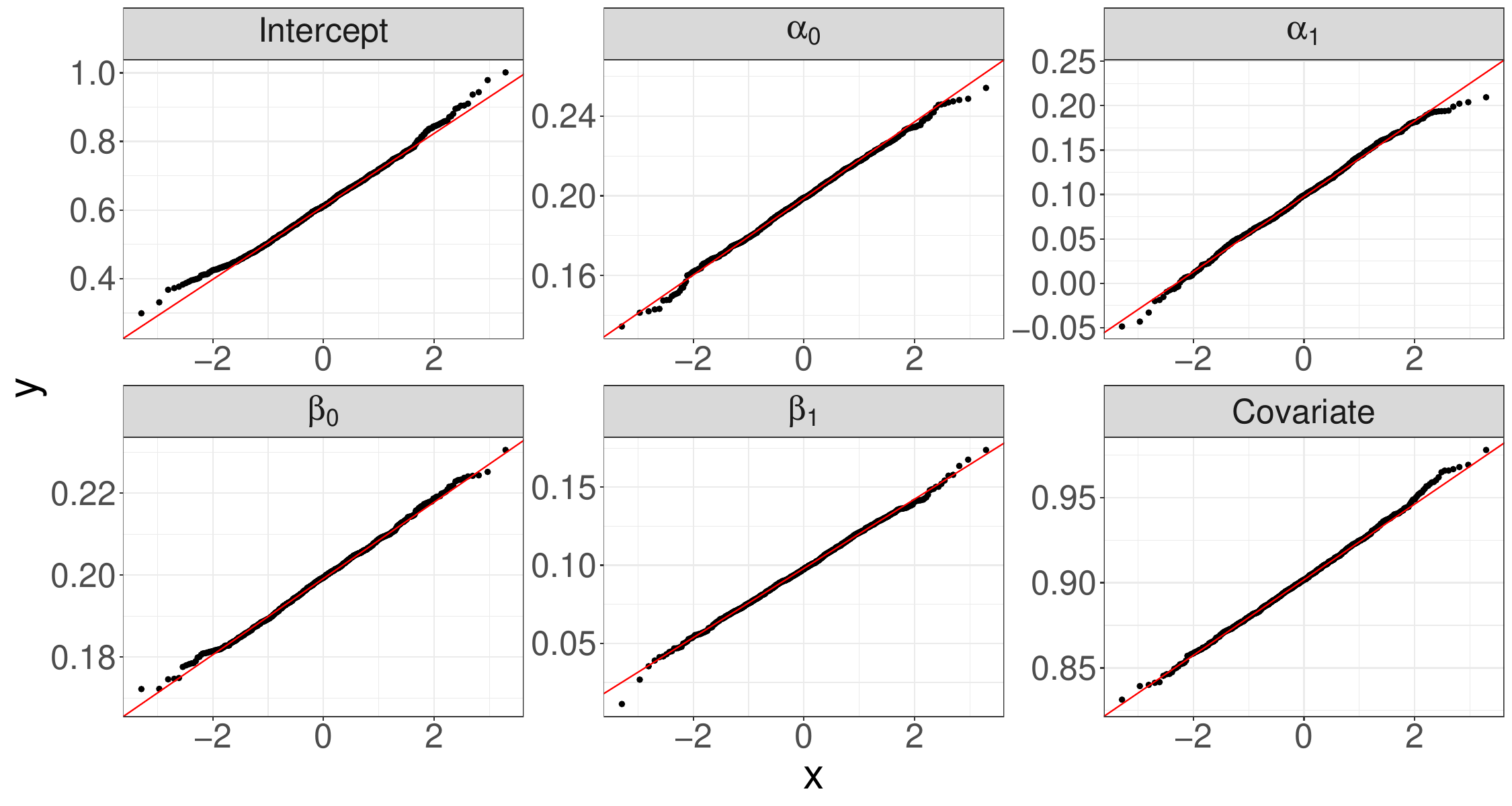}
		\caption{Mean model~\eqref{eq:simulation_mean_with_feedback}, i.e., with feedback mechanism.}
	\end{subfigure}
	
	\vspace{3em}
	
	\begin{subfigure}{\textwidth}
		\centering
		\includegraphics[width=\textwidth, keepaspectratio]{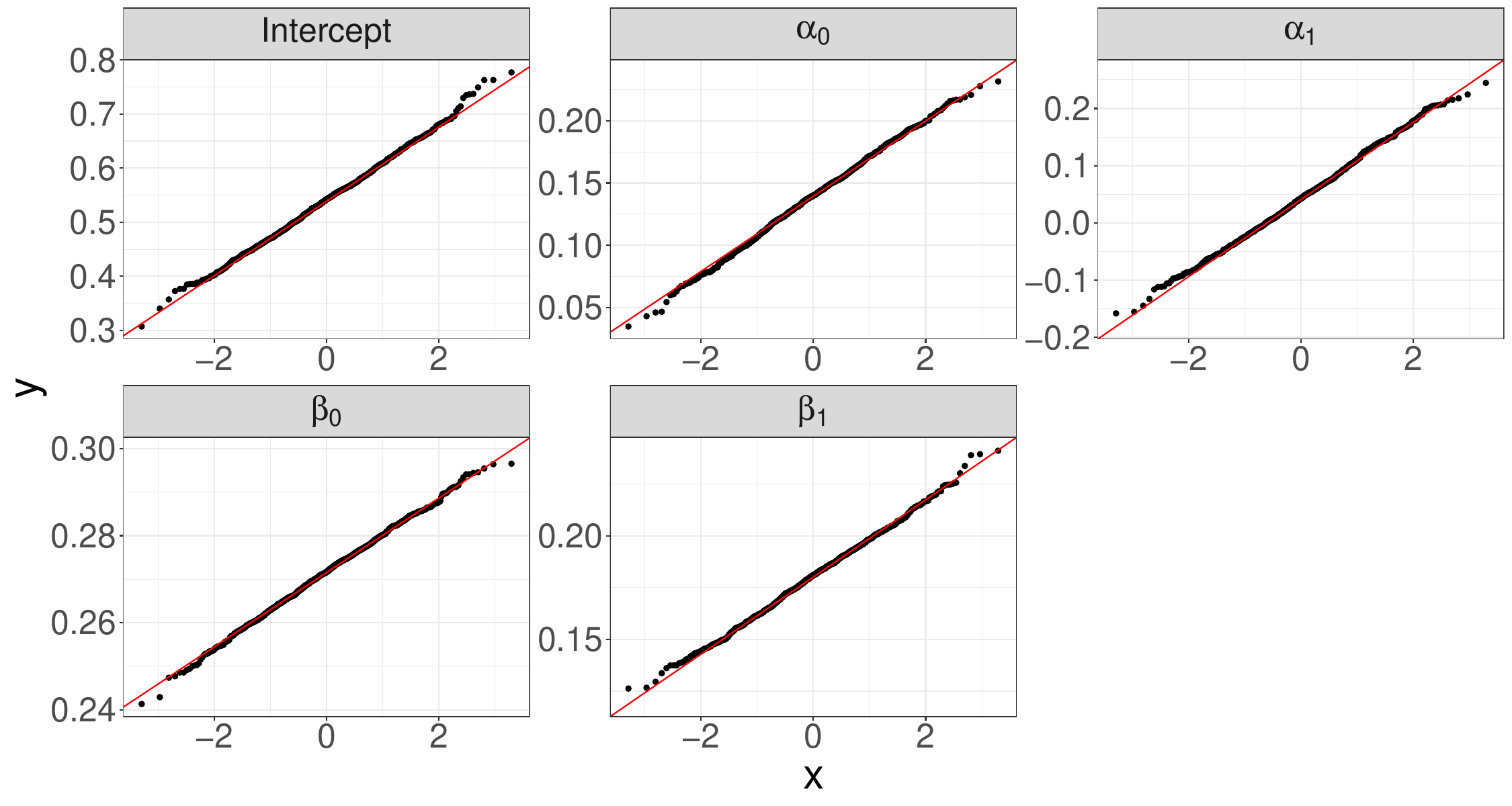}
		\caption{Dispersion model~\eqref{eq:simulation_dispersion_with_feedback}, i.e., with feedback mechanism.}
	\end{subfigure}
	\caption{Normal Q--Q plots of parameter estimates under the true data-generating process \eqref{eq:simulation_mean_with_feedback} and~\eqref{eq:simulation_dispersion_with_feedback} with generalized Poisson marginals, based on $T=500$.}
	\label{fig:qqplot_poisson_with_with}
\end{figure}

